\documentclass[twocolumn]{aastex631}
\usepackage{svg}
\usepackage{subfloat}
\usepackage{rotating}

\received{October 2, 2023}
\revised{April 5, 2024}
\accepted{April 18, 2024}

\shorttitle{Stellar Abundances of ETGs with Fine Structure}
\shortauthors{Barth et al.}
\graphicspath{{./}}

\begin{document}

\title{Stellar Abundances at the Center of Early Type Galaxies with Fine Structure}

\author[0000-0003-0602-5287]{Nicholas Barth}
\affiliation{Department of Astronomy, University of Florida,
1772 Stadium Rd, Gainesville, FL, 32611, US}

\author[0000-0003-3474-1125]{George C. Privon}
\affiliation{National Radio Astronomy Observatory,
520 Edgemont Road, Charlottesville, VA 22903, US}
\affiliation{Department of Astronomy, University of Florida,
1772 Stadium Rd, Gainesville, FL, 32611, US}
\affiliation{Department of Astronomy, University of Virginia, 530 McCormick Road, Charlottesville, VA 22904, US}

\author[0000-0002-8504-8470]{Rana Ezzeddine}
\affiliation{Department of Astronomy, University of Florida,
1772 Stadium Rd, Gainesville, FL, 32611, US}

\author[0000-0003-2638-1334]{Aaron S. Evans}
\affiliation{National Radio Astronomy Observatory,
520 Edgemont Road, Charlottesville, VA 22903, US}
\affiliation{Department of Astronomy, University of Virginia, 530 McCormick Road, Charlottesville, VA 22904, US}

\author[0000-0001-7568-6412]{Ezequiel Treister}
\affiliation{Instituto de Astrof{\'{\i}}sica, Facultad de F{\'{i}}sica, Pontificia Universidad Cat{\'{o}}lica de Chile, Campus San Joaquín, Av. Vicu{\~{n}}a Mackenna 4860, Macul Santiago, Chile, 7820436}



\begin{abstract}

Our understanding of early-type galaxies (ETGs) has grown in the past decade with the advance of full-spectrum fitting techniques used to infer the properties of the stellar populations that make-up the galaxy. We present ages, central velocity dispersions, and abundance ratios relative to Fe of C, N, O, Mg, Si, Ca, Ti, Cr, Mn, Co, Ni, Cu, Sr, Ba, and Eu, derived using full-spectrum fitting techniques for three ETGs NGC 2865, NGC 3818, and NGC 4915. Each of these three galaxies were selected because they have optical, disturbed structures (fine structure) that are linked to major merger events that occurred 1, 7, and 6 Gyr ago, respectively. Two of the ETGs, NGC 3818 and NGC 4915, show chemical signatures similar to ETGs without fine structure, which is consistent with a gas-poor  merger of elliptical galaxies in which substantial star formation is not expected. For NGC 2865, we find a statistically higher abundance of Ca (an $\alpha$-element) and Cr and Mn (Fe-peak elements). We show that for NGC 2865, a simple gas-rich merger scenario fails to explain the larger abundance ratios  compared to ETGs without fine structure. These three early-type galaxies with fine structure exhibit a range of abundances, suggesting ETGs with fine structure can form via multiple pathways and types of galaxy mergers.

\end{abstract}



\section{Introduction} \label{sec:intro}



The chemical abundance ratios of galaxies are powerful tools for constraining the evolution processes and formation channels that galaxies experience over their cosmic lifetime (\citealt{2019A&ARv..27....3M}, \citealt{2021ApJ...923...65F}). Early-type galaxies (ETGs) have two main stages of formation; an early ($z\sim2$) collapse of gas, forming stars in a compact core, or an ongoing second stage where satellite galaxies are accreted and ex situ formed stars (\citealt{Naab2009:0903.1636v4}, \citealt{Oser2010:1010.1381v2}). The fraction of ETGs that show recent star formation is low ($\sim10$\%), and it has been suggested that the increases in recent star formation could be a result of galaxy interactions and merger activity in the past few Gyrs (\citealt{Greif2010:1003.0472v2}, \citealt{2017A&A...606A..77U}).

With little-to-no star formation occurring in ETGs, the integrated light from these systems are dominated by old stellar populations. Through full spectral fitting, we could learn about how the ETGs were assembled and what the early star formation period looked like \citep{2021ApJ...923...65F}. The chemical abundance ratios derived for the ETGs' stellar populations point to chemical enrichment events, and are a powerful tool in untangling the chemical evolution within the galaxy. Beginning with the $\alpha$ elements (e.g. C, N, O, Mg, Si, Ca, Ti), these are produced via massive ($> 8M_{\odot}$) stars undergoing core-collapse supernovae (CCSNe). Fe peak elements are formed later, as Type Ia SNe start to occur within the stellar populations. The delay between when core-collapse supernovae dominate, and when Type Ia SNe dominate, is represented by the abundance ratios of elements formed in each event (e.g. [$\alpha$/Fe], where $\alpha$-elements are formed in CCSNe, and Fe is formed in Type Ia SNe). The abundance ratios thus can point to characteristics of the star formation history of a galaxy \citep{2021ApJ...923...65F}.

Some spectroscopic galactic surveys, such as the ATLAS3D and MASSIVE surveys, have widely improved our knowledge of the chemical composition of ETGs through the study of absorption and emission features in their spectra (\citealt{Cappellari2010:1012.1551v1}, \citealt{Ma2014:1407.1054v2}). Such surveys have defined trends showing how ETGs chemical abundances correlate with their ages and masses, with more massive ETGs showing higher metallicities, higher $\alpha$-element abundances, and older stellar populations. However, many of these surveys specifically target ETGs with unperturbed light profiles (i.e. lacking stellar features such as tails, shells, or streams, also known as \textit{fine structure}) to study the early Universe and to remove dynamical and chemical effects that took place at later epochs. The trends in chemodynamical properties of ETGs with fine structure have not been as widely studied as a result, and it is not obvious that ETGs with fine structure would share the same characteristics of ETGs without fine structure. The type of fine structure present in ETGs is indicative of the type of merger (gas-rich versus gas-poor merger, major versus minor merger, etc.) (\citealt{1990ApJ...364L..33S}, \citealt{1992AJ....104.1039S}, \citealt{2023MNRAS.520.5870G}), and thus with our study of stellar populations within ETGs with fine structure we are able to link the type of merger to any chemical evolution history of the galaxies. 




Our analysis of early-type galaxies with fine structure begins with a description of our targets and observations in Section \ref{sec:data}. We describe the full-spectrum fitting method used to determine chemodynamical parameters for each galaxy in Section \ref{sec:fitting}. We present our results of the fitting in Section \ref{sec:results}, which include ages, velocity dispersions, and chemical abundance ratios for each ETG. We describe how our results compare to ETG's without fine structure in Section \ref{sec:comp}.  Finally, we discuss and summarize our conclusions in Section \ref{sec:disc} and \ref{sec:conclusion}, respectively.

\section{Targets and Observations} \label{sec:data}

\subsection{Target Selection}

\begin{deluxetable*}{cccccccc} \label{tab:one}

\tablecaption{Our program target ETGs: (1) Galaxy ID, (2) Redshift \citep{2000MNRAS.313..469S}, (3) Hubble Morphological Type \citep{1991rc3..book.....D}, (4) Fine Structure Parameter (\citealt{1990ApJ...364L..33S}, \citealt{1992AJ....104.1039S}), (5) Time since last merger in Gyr (\citealt{1990ApJ...364L..33S}, \citealt{1992AJ....104.1039S}), (6) Estimated stellar mass in $M_\odot$ \citep{2006ApJ...636..698F}, (7) and (8) observed colors \citep{1990ApJ...364L..33S}.}

\tablehead{\colhead{Galaxy} & \colhead{Redshift} & \colhead{Morph. Type} & \colhead{$\Sigma$} & \colhead{$\tau_{\rm merger}$} & {$\log_{10} M_{*}$}  & \colhead{$(U-B)$} & \colhead{$(B-V)$} \\ \colhead{(1)} & \colhead{(2)} & \colhead{(3)} & \colhead{(4)} & \colhead{(5)} & \colhead{(6)} & \colhead{(7)} & \colhead{(8)}} 

\startdata
NGC 2865 & 0.008 & E3 & 6.4 & $\sim$1 Gyr & 10.5 & 0.42 & 0.91\\
NGC 3818 & 0.005 & E5 & 1.30 & $\sim$7 Gyr & 10.65 & 0.48 & 0.96\\
NGC 4915 & 0.01 & E0 & 5.48 & $\sim$6 Gyr & 10.52 & 0.47 & 0.94 \\
\enddata


\end{deluxetable*}


Three ETGs were selected for high resolution observations, NGC 2865, NGC 3818, and NGC 4915. Each target shows evidence of a major merger through photometric observations of fine structure. NGC 3818 and NGC 4915 were studied in \citet{1990ApJ...364L..33S} and \citet{1992AJ....104.1039S}, who found a correlation between fine structure and the line strengths and $UBV$ colors of ETGs. The $UBV$ colors, used to derive a heuristic merger age for the ETGs, show that galaxies with a larger amount of fine structure have had more recent (within $1-7$Gyr) major mergers. \citet{1990ApJ...364L..33S} define the \textit{fine-structure parameter}, $\Sigma$, as

\begin{equation}
    \Sigma = S + \log (1 + n) + J + B + X
\end{equation}

\noindent where $S$ is the visual estimate of the strength of the ETGs ripples, $n$ is the number of detected ripples, $J$ is the number of plumes, streams, or tails of luminous material, $B$ is the visual estimate of the boxiness of the isophote fit to the ETGs light profile, and $X$ indicates the absence or presence of an $X-$structure. $\Sigma$ was defined as a rough measurement of dynamical youth or rejuvenation \citep{1992AJ....104.1039S}. Further studies have found that the type of fine structure found in a merger remnant is indicative of the type of merger it underwent \citep{2023MNRAS.520.5870G}.

NGC 2865 is an E3 galaxy originally studied in \citet{1992A&AS...95..489J}. Further observations conducted by \citet{2007MNRAS.381..245R}, found fine structure that indicated a recent merger history. It has the highest fine structure content ($\Sigma=6.4$) among the 3 ETGs, with extended tidal tails and shells surrounding the galaxy. \citet{1986ApJ...306..110F} reports that NGC 2865's shells contribute 11-22\% of the total luminosity and have colors that resemble disk galaxies. NGC 2865 is thought to have a young, central stellar component formed within the last Gyr and a kinematically distinct core (\citealt{1999astro.ph.10032L}, \citealt{2006MNRAS.371..633H}, \citealt{2010MNRAS.401..852R}).

NGC 3818 is a bulge-dominated, E5 galaxy, with a disk fully embedded in a boxy bulge and the least amount of fine structure out of the three ETGs \citep{1998MNRAS.300..469S}. NGC 3818 is located in a low-density environment (density of galaxies brighter than $M_B = -16$ is 0.2) in the Crater Cloud group of galaxies \citep{1988ngc..book.....T}. \citet{2011A&A...528A..10P} studied mid-infrared (MIR) spectra of NGC 3818 and determined it as a \textit{passive} ETG, with no emission lines in its spectra, indicating little-to-no current activity within the galaxy. 

NGC 4915 is a relatively isolated, E0 galaxy thought to have formed $<$7 Gyr ago \citep{1992AJ....104.1039S}. At its center, NGC 4915 hosts a young stellar population that has a strong H$\beta$ absorption feature \citep{1992AJ....104.1039S}. Strong Balmer features, such as the H$\beta$ line, point to A or F stars added to an old stellar population, which is consistent with an aging starburst population that formed during the merger. \citet{1992AJ....104.1039S} describes the fine structure around NGC 4915 to be extremely boxy, with an apparent X-structure. 

\subsection{Observations}

\begin{figure*}
    \includegraphics[scale=0.48]{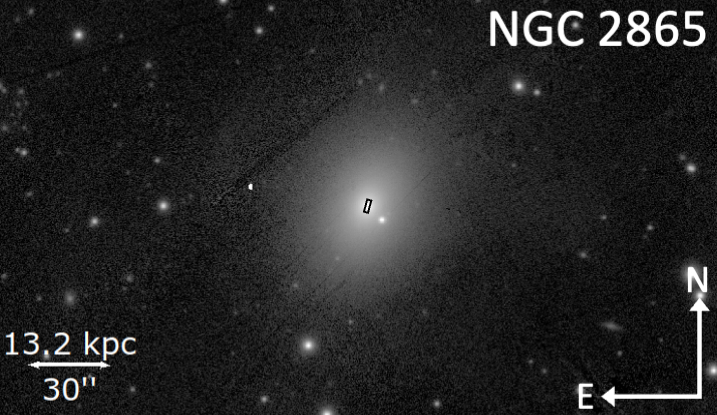}
    \includegraphics[scale=0.48]{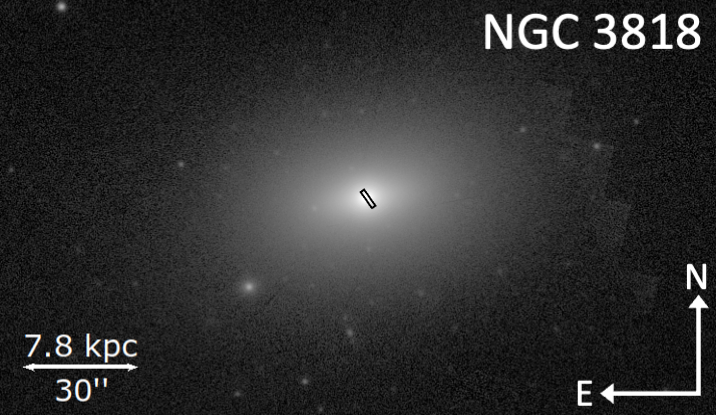}
    \includegraphics[scale=0.48]{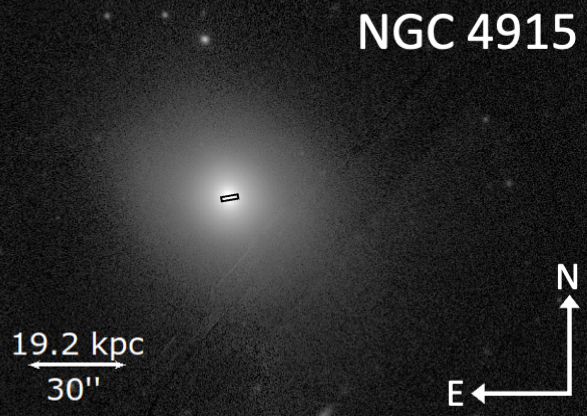}
    \caption{Slit coverage for NGC 2865, NGC 3818, and NGC 4915. The black box at the center of each galaxy represents the $1''\times5''$ slit used in the observations taken with the MIKE spectrograph. Images of each galaxy were taken from the PanSTARRS DR1 survey (z band) \citep{2016arXiv161205560C}.}
    \label{fig:slitImages}
\end{figure*}

The observations of NGC 2865, NGC 3818, and NGC 4915 were obtained on May 3rd, 2016 using the Magellan Inamori Kyocera Echelle (MIKE) double echelle spectrograph located on the MAGELLAN 6.5m Clay telescope \citep{2003SPIE.4841.1694B}.  Each exposure was taken with 3x2 binning using a $1''\times5''$ single slit resulting in a resolving power of $R \sim 28,000$ in the blue arm of MIKE and a $R \sim 22,000$ in the red arm. Figure \ref{fig:slitImages} shows the central slit coverage for each galaxy with respect to the galaxy as a whole. With a $2.2''$ extraction aperture, using the distances derived from the redshift listed in Table \ref{tab:one}, we obtain a spatial coverage of $0.2\times0.4$, $0.1\times0.2$, and $0.2\times0.5$ kpc for NGC 2865, NGC 3818, and NGC 4915, respectively. Each galaxy had 3 1500 second exposures, with the exception of NGC 4915, whose last exposure was cut short at 800 seconds due to time constraints.

We used the \texttt{CarPy} data reduction pipeline to combine, correct, and wavelength calibrate the observations of each galaxy (\citealt{2000ApJ...531..137K}, \citealt{2003PASP..115..688K}). Each spectrum was shifted to the rest frame from the observed frame by using the galaxies' recorded redshift listed in Table \ref{tab:one}. The resulting spectra covered a rest frame wavelength range of $\sim$ 3300 \AA$\;$to 9200 \AA. 


\section{Chemical Analysis} \label{sec:fitting}


\subsection{Full Spectral Fitting} \label{sec:fullfit}

We fit the spectra of the three ETGs using the Absorption Line Fitter (ALF), a simple stellar population (SSP) fitting method developed by \citet{2012ApJ...747...69C}. ALF utilizes both synthetic and empirical stellar libraries to create a grid of SSP models used to fit integrated light spectra of galaxies \citep{2012ApJ...760...71C}. In order to explore the large parameter space defined by the prior bounds, ALF uses a Monte Carlo Markov Chain (MCMC) fitting routine that produces models with various parameter combinations and uses a $\chi^2$ minimization to determine the best-fitting model.

ALF's full capabilities include the ability to determine the abundance ratios of 19 elements, kinematic properties such as recessional velocity and velocity dispersion, and an age and mass fraction of a sub-dominant, younger stellar population. We apply ALF to our target galaxies to determine the parameters of the light-dominating stellar population at the center of each galaxy with priors for each parameter listed in Table \ref{tab:priors}, and we fix the IMF to a Kroupa IMF \citep{2001MNRAS.322..231K}. After each fitting run by ALF, we check the posterior distributions of each parameter to make sure that the MCMC walkers are not settling on a value at the prior edge. 


\begin{deluxetable}{l|cc} \label{tab:priors}

\tablecaption{Prior limits set in this work for the ALF fitting routine.}


\tablehead{\colhead{Parameter} & \colhead{Low Prior} & \colhead{High Prior}} 

\startdata
$\sigma$ (km s$^{-1}$) &  20.0 & 600.0\\
Age (Gyr) &  2.0 & 13.5\\
Age of Young SSP (Gyr) &  0.5 & 3.0\\
$\log$ (Mass \% of Young SSP) &  -6.0 & -0.5\\
$\left[{\rm Fe/H} \right]$ &  -0.4 & 0.5\\
$\left[{\rm O/Fe} \right]$ &  -1.0 & 0.6\\
$\left[{\rm C/Fe} \right]$ &  -1.0 & 0.5\\
$\left[{\rm N/Fe} \right]$ &  -1.0 & 0.5\\
$\left[{\rm Mg/Fe} \right]$ &  -0.4 & 0.5\\
$\left[{\rm Si/Fe} \right]$ &  -0.5 & 0.5\\
$\left[{\rm Ca/Fe} \right]$ &  -0.5 & 0.5\\
$\left[{\rm Ti/Fe} \right]$ &  -1.0 & 0.5\\
$\left[{\rm Cr/Fe} \right]$ &  -0.5 & 0.5\\
$\left[{\rm Mn/Fe} \right]$ &  -0.3 & 1.0\\
$\left[{\rm Co/Fe} \right]$ &  -0.5 & 0.5\\
$\left[{\rm Ni/Fe} \right]$ &  -1.0 & 0.5\\
$\left[{\rm Cu/Fe} \right]$ &  -0.5 & 0.5\\
$\left[{\rm Sr/Fe} \right]$ &  -1.0 & 1.0\\
$\left[{\rm Ba/Fe} \right]$ &  -1.0 & 1.0\\
$\left[{\rm Eu/Fe} \right]$ &  -1.0 & 1.0\\
\enddata

\end{deluxetable}

\citet{Conroy2013:1303.6629v2} show the residuals to their fitting procedure of stacked, high SNR, SDSS galaxy spectra. They note that certain wavelength regions of their spectra show a correlation between galaxys' velocity dispersions and the residuals, indicating that there are unknown factors that affect those regions that are not accounted for by ALF. As a result of this uncertainty, we chose to mask regions that showed correlations with the velocity dispersion in the \citet{Conroy2013:1303.6629v2} stacked, SDSS data when fitting our galaxies with ALF. 

\begin{subfigures}
\begin{figure*}
    \centering
    \includegraphics[scale=0.875]{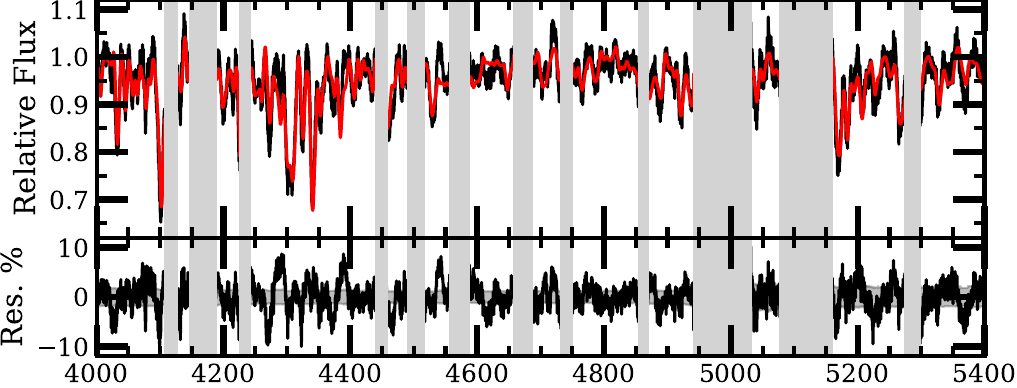}
    \includegraphics[scale=0.875]{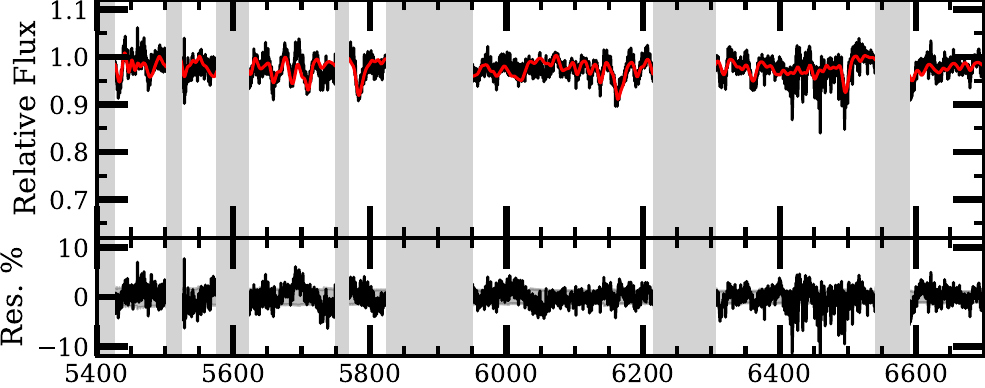}
    \includegraphics[scale=0.875]{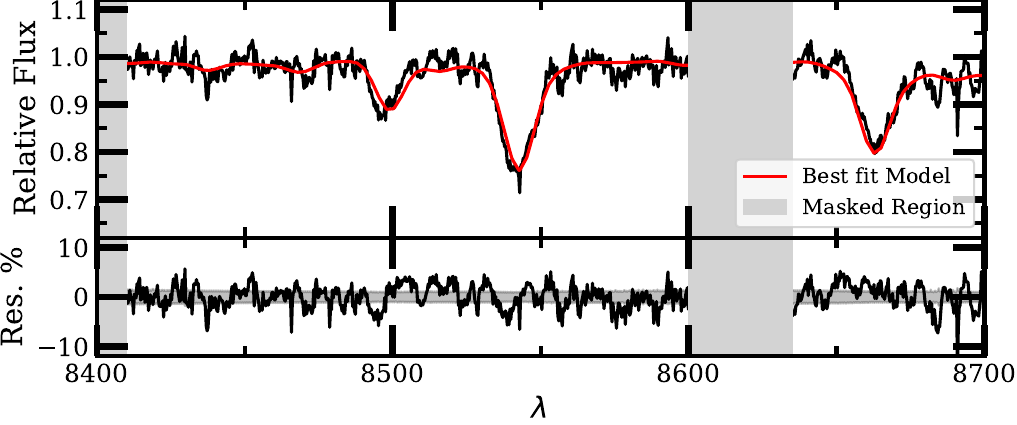}
    \caption{\textit{Top:} Comparison of the fitted model (red) to the observed spectrum (black) of NGC 2865. The data have been binned using a bin-width of 0.3$\AA$ for ease of visibility. Regions masked from the fit are the vertical areas, shaded grey. \textit{Bottom}: Residual percentage between the model and data. The 1$\sigma$ uncertainty region is shown as the shaded grey region.}
    \label{fig:n2modelData}
\end{figure*}

\begin{figure*}
    \centering
    \includegraphics[scale=0.875]{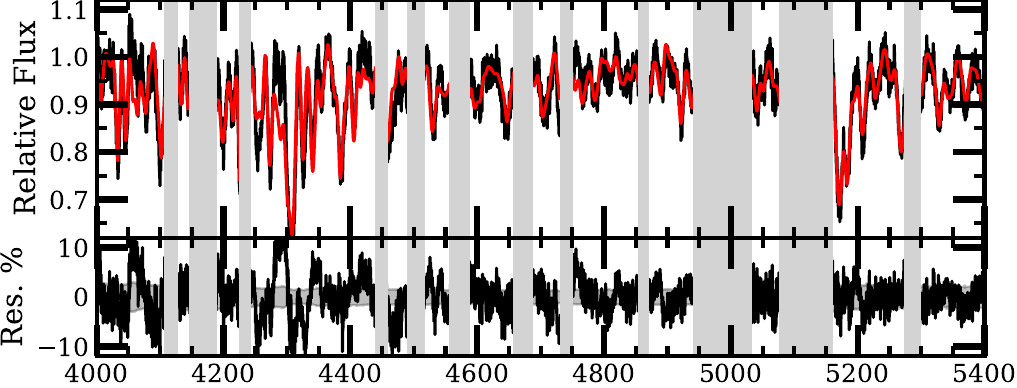}
    \includegraphics[scale=0.875]{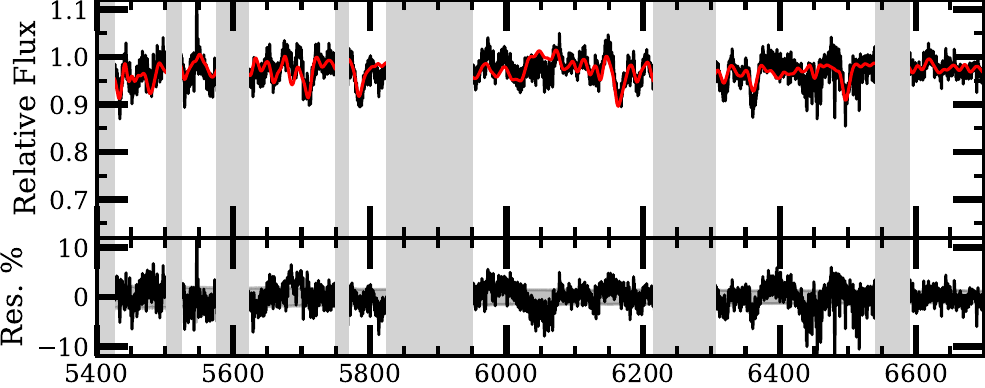}
    \includegraphics[scale=0.875]{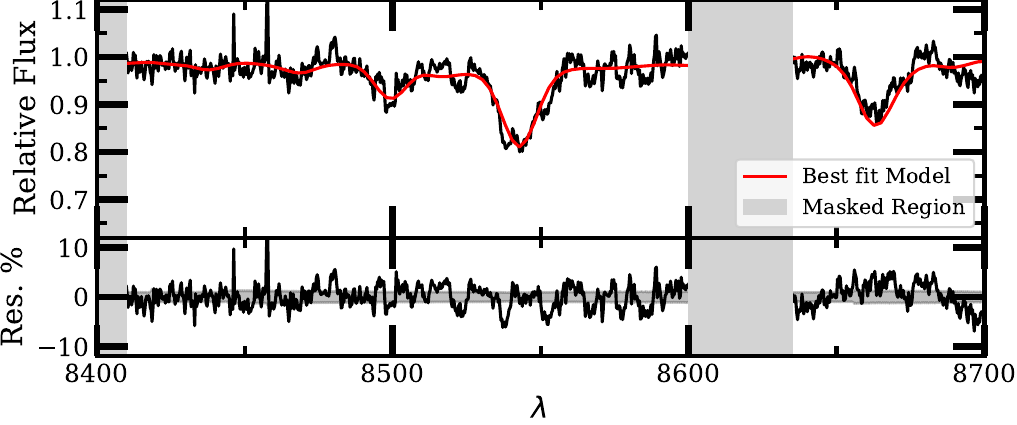}
    \caption{Same as Figure \ref{fig:n2modelData} but for NGC 3818.}
    \label{fig:n3modelData}
\end{figure*}

\begin{figure*}
    \centering
    \includegraphics[scale=0.875]{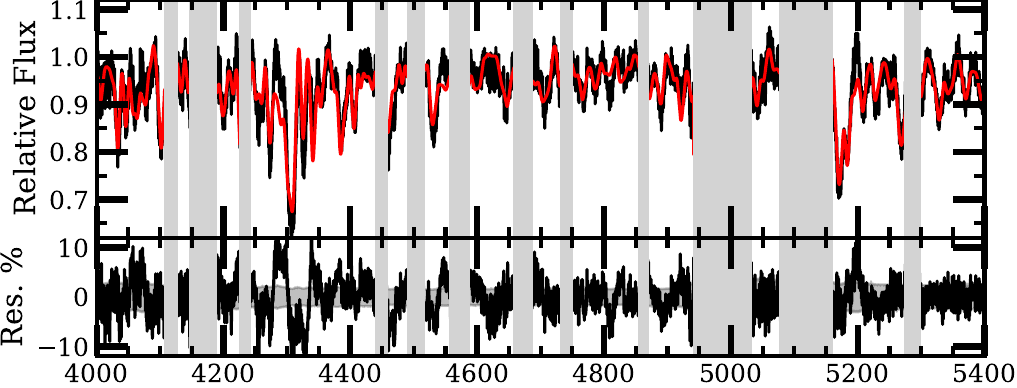}
    \includegraphics[scale=0.875]{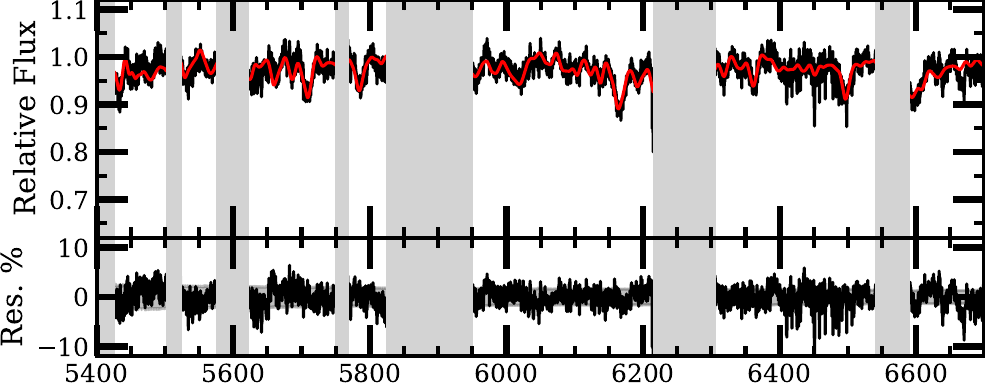}
    \includegraphics[scale=0.875]{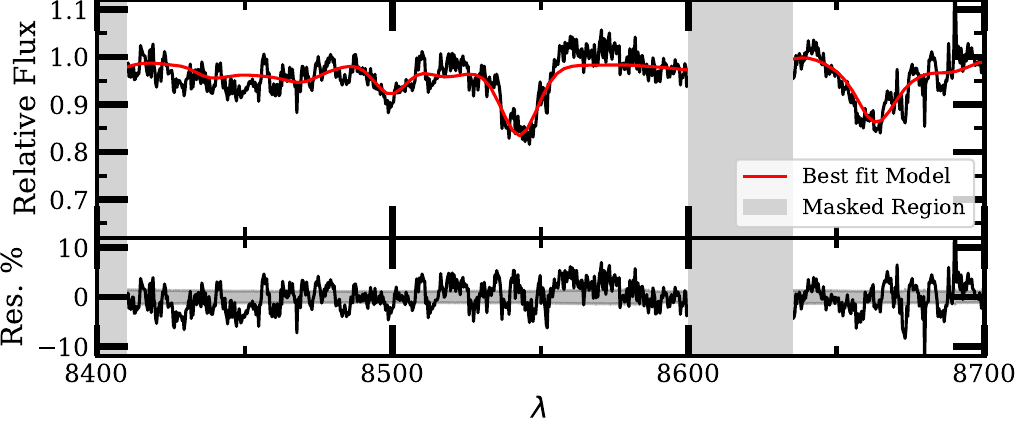}
    \caption{Same as Figure \ref{fig:n2modelData} but for NGC 4915.}
    \label{fig:n4modelData}
\end{figure*}
\end{subfigures}

Additionally, the major Na lines used in the fitting routine is the NaD doublet at 5900$\AA$, which is heavily influenced by ISM Na. Since the NaD line is the only strong Na source in our spectra, we risk fitting models with inflated Na abundances due to contributions from both stellar photospheres and ISM absorption. As such, we mask the NaD line from our fit and do not report a stellar Na abundance here.

Figure \ref{fig:n2modelData} - \ref{fig:n4modelData} show our best fit models to each ETG. We fit the wavelength regions 4000\AA-6800\AA, and 8400\AA-8700\AA, taking care to mask regions  affected by atmospheric and galaxy emission lines. The residual percentage between our best fit model and the data is also shown, with a range of $\pm10\%$.




\section{Results} \label{sec:results}

\begin{deluxetable*}{l|rrr} \label{tab:results}




\tablecaption{Parameters of the best-fit model shown in Figure \ref{fig:n2modelData}-\ref{fig:n4modelData} for NGC 2865, NGC 3818, and NGC 4915. We report values and uncertainties corresponding to the median and standard deviation of parameter values determined through our bootstrap fits.}

\tablenum{3}

\tablehead{\colhead{Parameter} & \colhead{NGC 2865} & \colhead{NGC 3818} & \colhead{NGC 4915}} 

\startdata
$\sigma$ (km s$^{-1}$) & $150\pm9$ & $179\pm8$ & $190\pm17$\\
Age (Gyr) & $3.9\pm1.3$ & $7.3\pm1.4$ & $6.0\pm1.5$\\
Age of Young SSP (Gyr) & $1.4\pm1.5$ & $1.4\pm1.5$ & $1.3\pm1.6$\\
$\log$ (Mass \% of Young SSP) & $-3.2\pm1.3$ & $-3.1\pm1.2$ & $-3.1\pm1.3$\\
$\left[{\rm Fe/H} \right]$ & $-0.1\pm0.1$ & $0.0\pm0.1$ & $0.0\pm0.1$\\
$\left[{\rm O/Fe} \right]$ & $0.2\pm0.3$ & $0.0\pm0.3$ & $0.1\pm0.3$\\
$\left[{\rm C/Fe} \right]$ & $0.0\pm0.1$ & $0.0\pm0.1$ & $0.0\pm0.2$\\
$\left[{\rm N/Fe} \right]$ & $0.1\pm0.2$ & $0.2\pm0.2$ & $0.0\pm0.3$\\
$\left[{\rm Mg/Fe} \right]$ & $0.4\pm0.2$ & $0.4\pm0.1$ & $0.3\pm0.2$\\
$\left[{\rm Si/Fe} \right]$ & $0.2\pm0.2$ & $0.0\pm0.2$ & $0.0\pm0.3$\\
$\left[{\rm Ca/Fe} \right]$ & $0.3\pm0.1$ & $-0.1\pm0.1$ & $0.0\pm0.1$\\
$\left[{\rm Ti/Fe} \right]$ & $-0.2\pm0.2$ & $-0.4\pm0.2$ & $0.0\pm0.3$\\
$\left[{\rm Cr/Fe} \right]$ & $0.2\pm0.1$ & $0.0\pm0.1$ & $0.0\pm0.2$\\
$\left[{\rm Mn/Fe} \right]$ & $0.3\pm0.2$ & $0.2\pm0.2$ & $0.1\pm0.2$\\
$\left[{\rm Co/Fe} \right]$ & $0.1\pm0.2$ & $0.1\pm0.2$ & $0.0\pm0.3$\\
$\left[{\rm Ni/Fe} \right]$ & $0.0\pm0.2$ & $0.0\pm0.2$ & $-0.1\pm0.3$\\
$\left[{\rm Cu/Fe} \right]$ & $0.1\pm0.3$ & $0.0\pm0.3$ & $0.0\pm0.3$\\
$\left[{\rm Sr/Fe} \right]$ & $0.1\pm0.3$ & $0.1\pm0.3$ & $0.0\pm0.4$\\
$\left[{\rm Ba/Fe} \right]$ & $0.0\pm0.3$ & $0.0\pm0.3$ & $0.0\pm0.4$\\
$\left[{\rm Eu/Fe} \right]$ & $0.1\pm0.3$ & $0.1\pm0.3$ & $0.0\pm0.4$\\
\enddata



\end{deluxetable*}

\subsection{Stellar Population Parameters} \label{sec:abund}

\begin{figure*}
    \centering
    \includegraphics[scale=0.4]{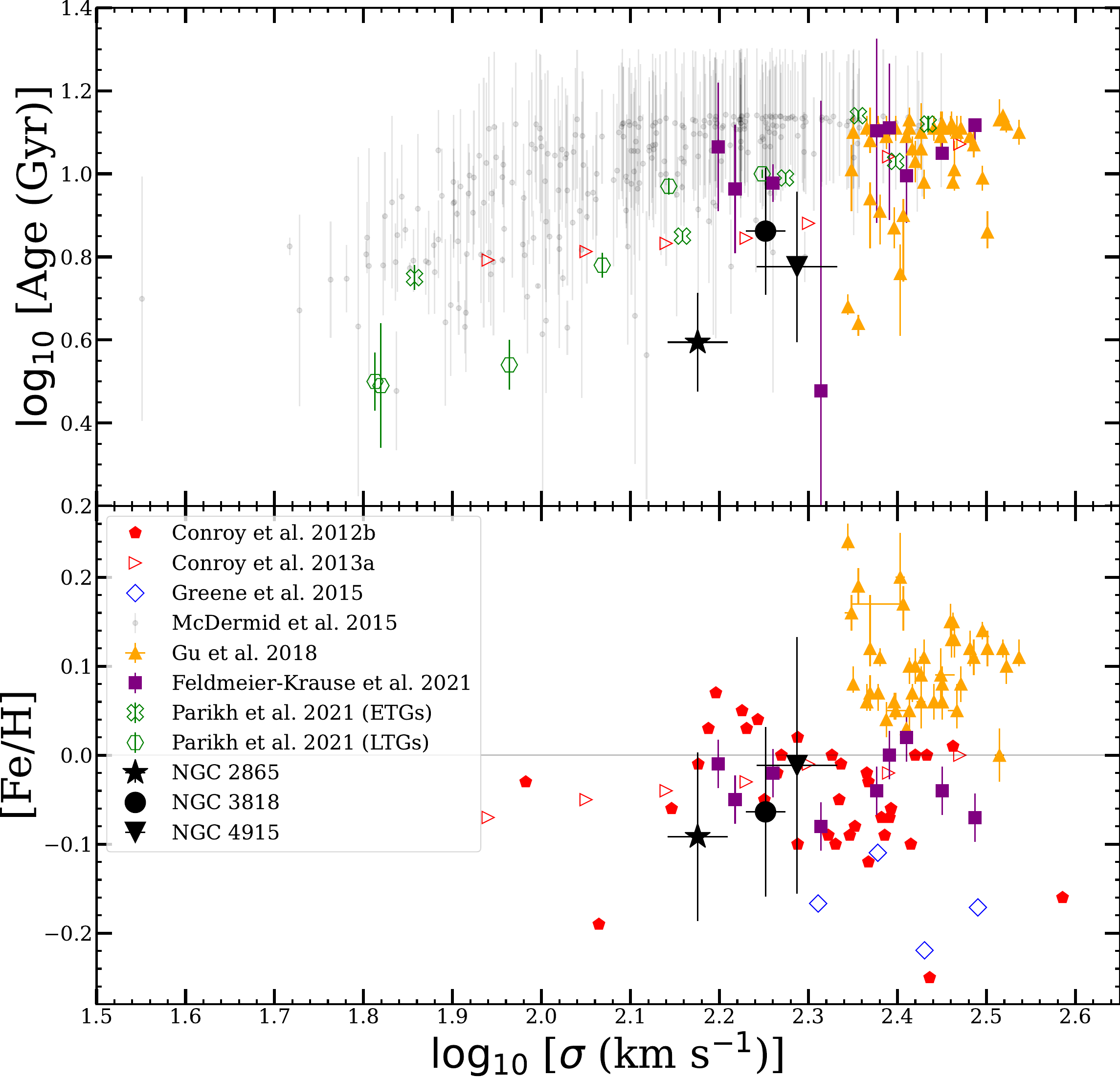}
    \caption{Relationship between the velocity dispersion and the age and [Fe/H] for NGC 2865 (\textit{black star}), NGC 3818 (\textit{black circle}), and NGC 4915 (\textit{black square}). Ages and metallicities determined from \citet{2012ApJ...760...71C} (filled, red pentagons), \citet{Conroy2013:1303.6629v2} (unfilled, red triangles), \citet{2015ApJ...807...11G} (unfilled, blue diamonds), \citet{2015MNRAS.448.3484M} (grey dots), \citet{2018ApJ...862L..18G} (filled, orange triangles), \citet{2021ApJ...923...65F} (filled, purple squares), \citet{2021MNRAS.502.5508P} (unfilled X's for ETGs, unfilled hexagons for LTGs), are included for comparison. Filled in markers represent results determined for individual galaxies, while open markers correspond to results derived from stacked data over discrete mass bins. A horizontal, black line is shown for a solar metallicity ([Fe/H] = 0) in the bottom plot. The central regions of our 3 ETGs show younger ages than most other ETGs of the same $\sigma$, especially in the case of NGC 2865.}
    \label{fig:sigAge}
\end{figure*}

The stellar population parameter of our analysis of NGC 2865, NGC 3818, and NGC 4915 are listed in Table \ref{tab:results}. Figure \ref{fig:sigAge} shows our fitted age and [Fe/H] values as a function of central velocity dispersion compared to other literature samples of ETGs. ALF fits an element $X$ with respect to hydrogen, and to transform this ratio with respect to iron, we subtract the fitted [Fe/H] measurement ($[X/{\rm Fe}] = [X/{\rm H}] - [{\rm Fe/H}]$). Additionally, certain elements must be corrected with theoretical response functions. These response function abundance corrections come from synthetic stellar libraries and supplement areas of the spectra where the empirical model library fails to capture the contributions from various elements (\citealt{2012ApJ...760...71C}, \citealt{2018ApJ...854..139C}). As a result, the elements O, Mg, Ca, Si, and Ti, are corrected based off relations derived in \citet{2007ApJS..171..146S}, \citet{2014A&A...562A..71B}, and \citet{2011MNRAS.414.1227M}. \citet{2018ApJ...854..139C} found that the empirical library abundances for C, N, Na, V, Cr, Mn, Co, Cu, Ba, Sr, and Eu were close to 0 for a wide metallicity range (${\rm X/Fe}]_{\rm lib} ([{\rm Z/H}])=0)$, indicating that no additional abundance was needed to add to the theoretical response function for these elements. The bottom panels of the figures in Appendix \ref{sec:app} show where each element can have an effect on the spectra in the observed wavelength region. By fitting the entire spectrum, we are able to constrain abundances of elements using their combined impact on the integrated light. In the case of O, which does not have individual lines in our spectra, we derive an O abundance using the contributions from molecules that include O (such as H$_2$O, TiO, MgO, AlO, VO, CO, OH, and SiO) and produce an absorption features in the response function.

As shown in Figure \ref{fig:sigAge}, we derive a younger stellar population for NGC 2865 at $\sim$3.9 Gyr, while we obtain relatively older derived stellar populations of $\sim$7.3 and $\sim$6.0 Gyr for NGC 3818 and NGC 4915, respectively. We also show comparisons of ages to stacked SDSS galaxies \citep{2012ApJ...760...71C}, LRIS galaxies \citep{2012ApJ...760...71C}, the ATLAS3D galaxies \citep{2015MNRAS.448.3484M}, the MaNGA survey galaxies (\citealt{2018ApJ...862L..18G}, \citealt{2021MNRAS.502.5508P}), and the 9 galaxies studied in \citet{2021ApJ...923...65F}. We find that as compared to these ETGs without fine structure, NGC 2865 has a stellar population that is younger relative to its determined velocity dispersion. None of the three ETGs, however,  show any significant mass fraction ($>15\%$) of a younger ($<0.5$ Gyr) stellar population.

\begin{figure*}
    \centering
    \includegraphics[scale=0.4,width=\textwidth]{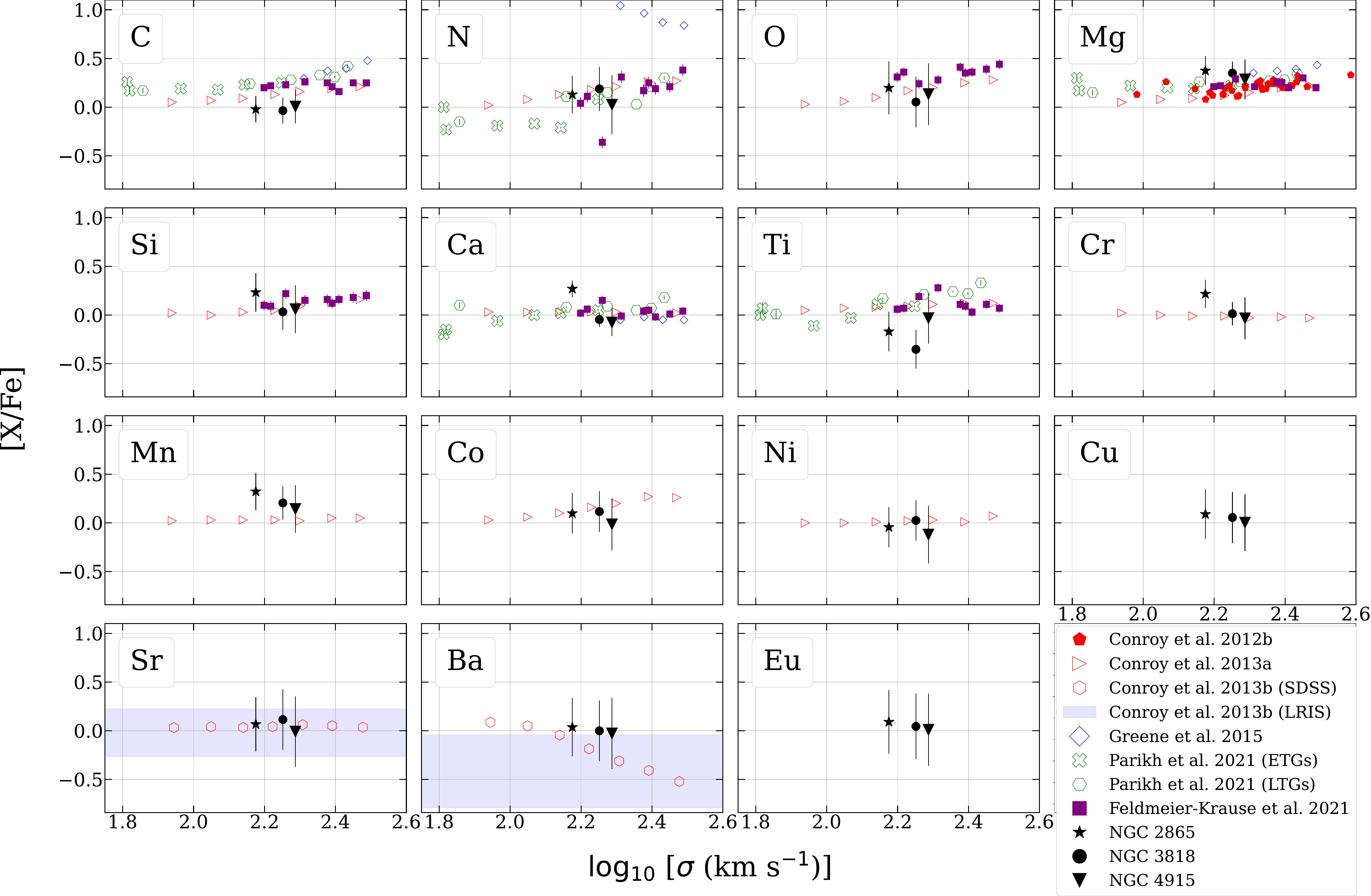}
    \caption{[X/Fe] ratio for NGC 2865 (\textit{black stars}), NGC 3818 (\textit{black circles}), and NGC 4915 (\textit{black squares}) are plotted against the galaxies' central velocity dispersion and compared to literature abundances of \citet{2012ApJ...760...71C} (filled, red pentagons), \citet{Conroy2013:1303.6629v2} (unfilled, red triangles), \citet{Conroy2013:1301.0321v2} (unfilled, red hexagons), \citet{2015ApJ...807...11G} (unfilled, blue diamonds), \citet{2021ApJ...923...65F} (filled, purple squares), \citet{2021MNRAS.502.5508P} (unfilled X's for ETGs, unfilled hexagons for LTGs). Sr and Ba are determined for individual galaxies in \citet{Conroy2013:1301.0321v2}, however the velocity dispersions are not reported. We instead show the range of abundances covered by their results as the light-blue shaded region. The abundances plotted correspond to the values and uncertainties reported in Table 3. Filled in markers represent results determined for individual galaxies, while open markers correspond to results derived from stacked data over discrete mass bins.}
    \label{fig:allEls}
\end{figure*}

We obtain a slightly subsolar to solar in metallicity for all three ETGs as can be seen in Figure \ref{fig:sigAge}. The abundance ratios of our reported elemental abundances are shown in Figure \ref{fig:allEls} as a function of velocity dispersion.

\subsubsection{Carbon, Nitrogen, and Oxygen}

Carbon, nitrogen, and oxygen are light elements formed in massive stars (O) and AGB stars (C and N), and are primarily determined in galaxies by ALF using molecular lines (e.g. CH, CN, TiO) \citep{2018ApJ...854..139C}. N is formed in intermediate mass AGB stars (unlike C, which is produced in low-mass AGB stars), allowing stellar populations to be enriched in N earlier than C \citep{2011MNRAS.414.3231K}.

All three ETGs are fit with a solar C abundance ([C/Fe]$\sim0.0\pm0.1$), while the N and O median abundance ratios are slightly super-solar at [N,O/Fe] of $0.2-0.3\pm0.2$ dex. 

\subsubsection{$\alpha$-elements} \label{sec:alpha_abund}

Mg, Si, Ca, and Ti (along with O), are produced in massive stars, and are ejected through CCSNe (\citealt{1995ApJS...98..617T}, \citealt{2006ApJ...653.1145K}, \citealt{2020ApJ...900..179K}). Mg, Ti, and O affect large wavelength regions of the integrated light spectra due to the TiO and MgH molecular absorption features present in the cooler, and older stellar spectra that typically dominate in ETGs. The effect of these elements on the integrated light spectra can be seen in the lower panels of the graphs in the Appendix figure set. The [$\alpha$/Fe] ratio is used as a proxy for the star formation duration, with higher $\sigma$ (i.e. more massive) galaxies showing larger [$\alpha$/Fe] ratios \citep{2005ApJ...621..673T}. However, outliers from the Mg-$\sigma$ relationship have been shown to be caused by contribution from intermediate/young stellar populations within the ETGs \citep{2005ApJ...621..673T}.

The $\alpha$-elements vary slightly from C, N, and O in that NGC 2865 shows significant super-solar ([X/Fe] $>0.2$) abundances for Mg, Si, and Ca, while the other two ETGs, NGC 3818 and NGC 4915, are super-solar in Mg, but are determined to have solar Si and Ca abundances. Ti is determined to have significantly sub-solar abundances for our 3 studied ETGs, particularly for NGC 3818. While Ti is expected to follow $\alpha$ elements like Si and Ca in abundances for ETGs, we do not observe this trend for our galaxies, in particularly for NGC 3818. NIR coverage of the TiO lines, which are unfortunately not covered by our spectra,  are likely to allow a more robust test of the low Ti abundance, as described in \citet{2018ApJ...854..139C}.



\subsubsection{Fe-peak elements}

The second group of elements that we fit using ALF are the Fe-peak elements, Fe, Cr, Mn, Co, Ni, and Cu. The Fe-peak elements generally follow the same trends as Fe, and often have many lines that affect the entire spectrum \citep{Conroy2013:1303.6629v2}. Cr and Mn are produced in the incomplete Si-burning regions of Type Ia SNe. Mn is produced in larger quantities than Fe leading to a positive [Mn/Fe] ratio with more Type Ia SNe, while Cr is produced near the same amount as Fe \citep{2020ApJ...900..179K}. Moreover, the increase of [Co/Fe] at higher velocity dispersion in the \cite{2012ApJ...760...71C} data is speculated to be from Co enrichment from massive stars. Ni is produced in similar quantities to Fe \citep{2020ApJ...900..179K}, however large abundance changes in Ni and Co do not produce significant spectral line changes, making both elements difficult to detect (see Figure 4 in \citealt{Conroy2013:1303.6629v2}).

The Fe-peak elements produce weaker features in integrated light spectra than the lighter CNO and $\alpha$-elements. In NGC 2865 we find significant super-solar values of Cr ([Cr/Fe] $=0.22\pm0.14$) and Mn ([Mn/Fe] $=0.32\pm0.19$), and solar abundances within uncertainties for Co, Ni, and Cu. NGC 3818 and NGC 4915 have solar abundance ratios of Cr, Mn, Ni, and Cu. Interestingly, NGC 2865 shows the largest abundance of Cr out of the 3 ETGs, and while the dominant Cr yield is from Type Ia Sne, there is a non-negligible amount of Cr produced in Type II SNe \citep{2011MNRAS.414.3231K}. NGC 2865 is enriched in both Mn and Cr, while NGC 3818 and NGC 4915 show no enrichment of Cr.

\subsubsection{Neutron capture elements}

Additionally, we use ALF to determine abundances for the neutron capture elements Sr, Ba, and Eu. Elements such as Ba, are produced in intermediate and low-mass AGB stars through the slow neutron capture process ($s-$process) \citep{Conroy2013:1301.0321v2}. Rapid neutron capture process ($r-$process) elements, such as Eu, are produced in dense neutron environments such as neutron star mergers (NSM) and Type II CCSNe \citep{2020ApJ...900..179K}. Determining the abundances of the $r-$process element, Eu, while difficult to detect due to the faint nature of its spectral absorption features in stellar spectra, would allow for constraints on the rate of NSM/CCSNe within ETGs, and the impact these events have on a galaxy.

The neutron capture elements produce subtle and isolated responses in the integrated spectra of ETGs due to the weak nature of the lines in stellar spectra, unlike some of the lighter elements that affect broad wavelength regions \citep{Conroy2013:1301.0321v2}. As a result, the abundances of these elements are difficult to fit as the response is often lost in the noise of the spectra or in the broadening of nearby, stronger spectral features. In Figure \ref{fig:ncap}, we show a sample of the distribution of fits from our bootstrap analysis to areas of the spectrum that are affected by C, Mg, Ba, Sr, and Eu. As stated in Section \ref{sec:uncert}, the uncertainty in our determined abundance corresponds to the spread of fitted abundances ratios from our bootstrap analysis. In Appendix \ref{sec:app} we show the regions that each element affects for our observed wavelength range. For lighter elements such as C and Mg, we report smaller uncertainties because of the elements' strong contribution to the spectra and reliability of models with respect to those elements. For each of the three ETGs, the median abundance of the bootstrap fitting for Sr, Ba, and Eu was solar, which suggests that the slightest changes to the regions of the integrated light spectra that are affected by Sr, Ba, and Eu can produce large changes in the determined abundances. However, we discuss the implications of our determined abundances for Sr, Ba, and Eu to show how they relate to our interpretation of ETG chemical evolution.

We present a first order determination of Eu, for the first time, from integrated light spectra of distant galaxies. We note that the Eu lines are very weak in our spectra, which is reflected in the large error bars $\sim 0.5$\,dex. Each of our three ETGs with fine structure were determined to have an [Eu/Fe]$\sim0$. Such values were predicted in simulation work done by \citet{2018MNRAS.477.1206N} which showed that Milky Way type galaxies at [Fe/H]$\sim0$ have an [Eu/Fe] abundance of $\sim0$. We note that their simulations include merger events. The lack of enhancement or depletion post-merger as signified by a [Eu/Fe]$\sim0$, indicate that either Eu and Fe were both equally produced during the merger, or that one of the elements was overproduced in a manner that brought the [Eu/Fe] closer to 0.


\subsection{Parameter Uncertainty} \label{sec:uncert}

ALF provides uncertainties for each fit parameter using the shape of the posterior probability distribution created during the MCMC fitting routine. However, we found that the output from ALF provided uncertainties that did not reflect the true magnitudes of the total uncertainties in fitting the integrated light spectra. In order to provide a more robust uncertainty for each parameter, we determine uncertainties for each of the fit parameters using a bootstrap method. 

We generate 1000 spectra by drawing from a normal distribution centered on the original data spectrum for each galaxy, with a standard deviation equal to the uncertainty obtained in the data reduction pipeline. Each spectrum is then fit using the same fitting routine to obtain 1000 sets of parameter values for each galaxy. Figure \ref{fig:ncap} shows a sample of fitted models for different regions of the galaxy spectra, which demonstrates the spread in model fits with our spectra. In Table \ref{tab:results}, we report the median value and the uncertainty corresponds to the standard deviation of the distribution of the set of parameter values obtained in the bootstrap analysis. For the chemical abundances, our uncertainties range from $\sim0.1-0.4$ dex, with the lighter elements (C, N, O, Mg, Si, Ca, Ti) generally having smaller uncertainties than the heavier elements (Cr, Mn, Co, Ni, Cu, Sr, Ba, and Eu). 

\section{Comparison to ETGs without Fine Structure} \label{sec:comp}

We compare our results to other spectroscopic galaxy surveys that use full-spectrum fitting in order to study the relation between fine structure content and the chemical abundance ratios in ETGs. The first of our literature sources is the work done in \cite{2012ApJ...760...71C}, \cite{Conroy2013:1303.6629v2}, \cite{Conroy2013:1301.0321v2}, and \cite{2018ApJ...854..139C}, from which the ALF fitting routine originated. Their work uses stacked SDSS observation to obtain high SNR spectra for discrete velocity dispersion bins of ETGs. In \cite{2012ApJ...760...71C}, they examine 34 ETGs from the SAURON survey (\citealt{2001MNRAS.326...23B}, \citealt{2002MNRAS.329..513D}). We also compare our work to the ATLAS3D survey of ETGs \citep{Cappellari2010:1012.1551v1}, which includes a star formation history and stellar population study done in \citet{2015MNRAS.448.3484M}. \cite{2018arXiv181013242G} also uses ALF to study stacked spectra taken from the MaNGA survey (\citealt{2015IAUS..311..100B}, \citealt{2016AJ....152..197Y}, \citealt{2015AJ....149...77D}, \citealt{2017AJ....154...86W}, \citealt{2018ApJS..235...42A}), specifically ETGs in the Coma Cluster. Additionally from the MaNGA survey, \cite{2021MNRAS.502.5508P} uses stacked spectra to study gradients using line indice models to obtain ages, chemical abundances, and kinematic information. Their work uniquely studies not only ETGs, but select late-type galaxies as well, which allows comparisons of our sample of recently merged galaxies with other galaxy types. We compare to \cite{2015ApJ...807...11G} which uses galaxies from the MASSIVE survey \citep{Ma2014:1407.1054v2}, which is a volume-limited survey that \textit{excludes} large spiral and interacting galaxies. Finally, we include results from \cite{2021ApJ...923...65F}, that obtained abundance ratios for 8 ETGs and 1 low surface brightness spiral galaxy using full spectrum fitting methods. Many of the surveys explore gradients, and thus report abundances for various radii from the galactic center. For the purpose of this study, we only compare to the literature results at the center of each system.

\subsection{Velocity Dispersions and Ages}

The velocity dispersion, $\sigma$, is a proxy for the galaxy mass, with higher $\sigma$ correlating with more massive galaxies \citep{1976ApJ...204..668F}. As can be seen in Figure \ref{fig:sigAge}, more massive galaxies tend to be dominated by older stellar populations (\citealt{2015MNRAS.448.3484M}, \citealt{2018ApJ...862L..18G}, \citealt{2021ApJ...923...65F}). As shown in Figure \ref{fig:sigAge}, NGC 2865 shows a younger stellar populations than most of the galaxies in its mass range (with the exception of a few other outliers), while NGC 3818 and NGC 4915 are determined to have ages that are similar to other ETGs of similar mass ranges, within uncertainties. It is possible that the merger events of NGC 2865 produced periods of star formation from which there are more young stars contributing to the integrated light \citep{2008ApJ...682L..13H}.

\subsection{Chemical Abundance Comparisons}

Elements that produce strong features in ETG spectra, such as C, N, Mg, Ca, and Ti, have been measured in many surveys (\citealt{2012ApJ...760...71C}, \citealt{2015ApJ...807...11G}, \citealt{2021MNRAS.502.5508P}, \citealt{2021ApJ...923...65F}). To reach the SNR needed (SNR $>100$, \citealt{2012ApJ...760...71C}) to measure elements such as Cr, Mn, Co, Ni, Cu, Sr, Ba, and Eu, surveys often bin ETGs into discrete mass bins and coadd the low SNR spectra. As a result, Fe-peak elements are less reported than the lighter elements, and we are limited to comparisons to \cite{2012ApJ...760...71C}. There are no galaxies with reported abundances for some heavy elements, such as Cu and Eu.

We find that while the three ETGs have median [C/Fe] derived abundances of about 0.1-0.2 dex lower than literature ETGs. While our C abundances are lower than the trend established by the literature ETGs, the result may be unsurprising as we see a slightly super-solar N abundances. At metallicities of [Fe/H]$>-0.2$ dex, \citet{2019A&ARv..27....3M} show that N may be produced as a secondary element from intermediate stars already enhanced in C and O, which could result in a higher N abundance and a lower C abundance following a burst of star formation. Furthermore, \citet{Conroy2013:1303.6629v2} show that N tracks O very closely, while C does not at higher velocity dispersions ($\log \sigma > 2.1$ km s$^{-1}$). Similar to N, all three ETGs are determined to have [O/Fe] ratios that lay in the same abundance range as the literature ETGs.

Mg is a well-studied element in ETGs, as it produces strong and localized spectral lines that contribute to the integrated light spectrum (as can be seen in the bottom panel of Figure 8.4 in Appendix \ref{sec:app}).  As can be seen in Figure \ref{fig:allEls}, all three of our ETGs lie close to or on the observed literature ETG [Mg/Fe] abundance values. We also find similar abundnaces within uncertainty for Si to the literature ETGs. However, with the [Ca/Fe] abundance, we find that NGC 2865 has a significantly higher [Ca/Fe] abundance when compared to ETGs of a similar velocity dispersion, while NGC 3818 and NGC 4915 have abundances that lie near the literature ETG abundances. As mentioned in Section \ref{sec:alpha_abund}, the $\alpha$-elements are formed through the stellar evolution of massive stars \citep{2020ApJ...900..179K}. The [$\alpha$/Fe] is observed to decrease at [Fe/H]$>-1$ due to increased Fe contributions from Type Ia SNe. 

As \citet{Conroy2013:1303.6629v2} uses stacked $\sigma$ bins, we do not know the scatter for Cr, Mn, Co, and Ni in each $\sigma$ bin, however their measurements show trends as the galaxies grow in mass for the Fe-peak elements. Cr, Mn, Ni remain near solar ([X/Fe]$\sim 0$) for all mass bins, as these elements are produced in the same processes that produce Fe. Co, an element thought to form in massive stars, does break this trend by increasing in [Co/Fe] with increasing $\sigma$. The 3 ETGs studied in this work generally lie around the trends seen in \citet{Conroy2013:1303.6629v2}, however we find that NGC 2865 lies above the results of \citet{Conroy2013:1303.6629v2} in Cr and Mn. As noted in Section \ref{sec:abund}, NGC 2865 being enhanced in Cr and Mn could be a result of a rapid enrichment, possibly due to the merger event, where Type Ia SNe could not occur quick enough to inject Fe into the ISM and lower the abundance ratios with respect to Fe. We find a Solar [Ni/Fe]$\sim 0$ ratio for NGC 2865, which does not follow the the other Fe-peak elements (like Cr, Mn, Co). The solar value of [Ni/Fe] is consistent with chemical evolution models and observations of stars at a [Fe/H] $>-1$. Ni and Fe are both produced in the complete Si-burning region of a star, which makes the the [Ni/Fe] ratio unaffected by multiple periods of star formation \citep{2020ApJ...900..179K}. The production of Mn and Cr on the other hand, are produced in the incomplete Si-burning region, and could thus have been enhanced relative to Fe during bursts of star formation, as observed in NGC 2865, due to material ejection during SNe  \citep{2020ApJ...900..179K}.

\begin{figure}
    \centering
    \includegraphics[scale=0.4]{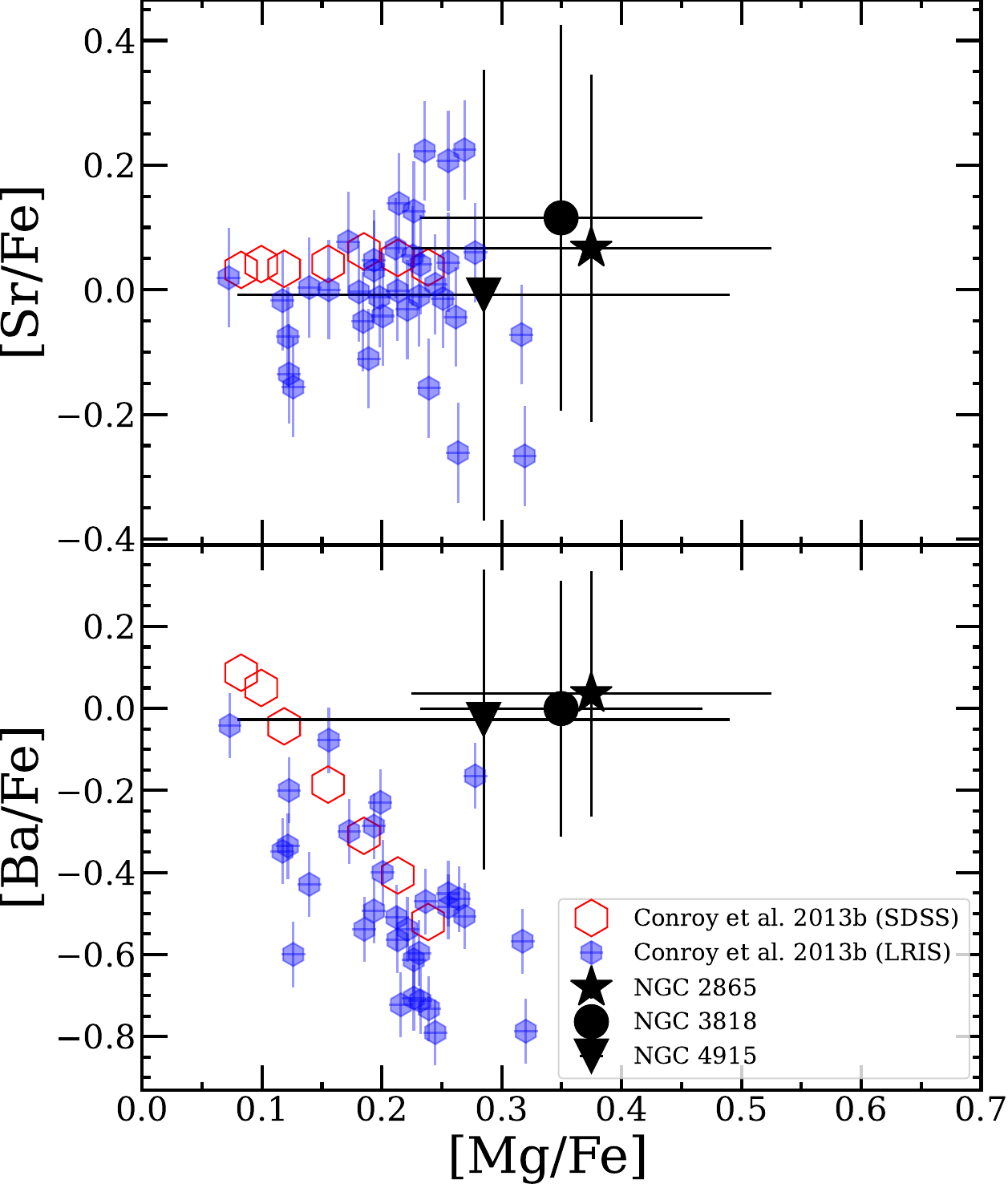}
    \caption{[Sr/Fe] (\textit{upper panel}) and [Ba/Fe] (\textit{bottom panel}) abundance ratios of this work (NGC 2865 - stars, NGC 3818 - circles, and NGC 4915 - triangles) compared to results presented in \citet{Conroy2013:1301.0321v2} (filled, light-blue hexagons for LRIS individual galaxies, and unfilled, red hexagons for stacked SDSS data) for early-type galaxies. We see agreement for the [Sr/Fe] trend, but we do not observe a decreasing [Ba/Fe] trend found in \citet{Conroy2013:1301.0321v2}.}
    \label{fig:srba}
\end{figure}

\begin{figure*}
    \centering
    \includegraphics[scale=0.55]{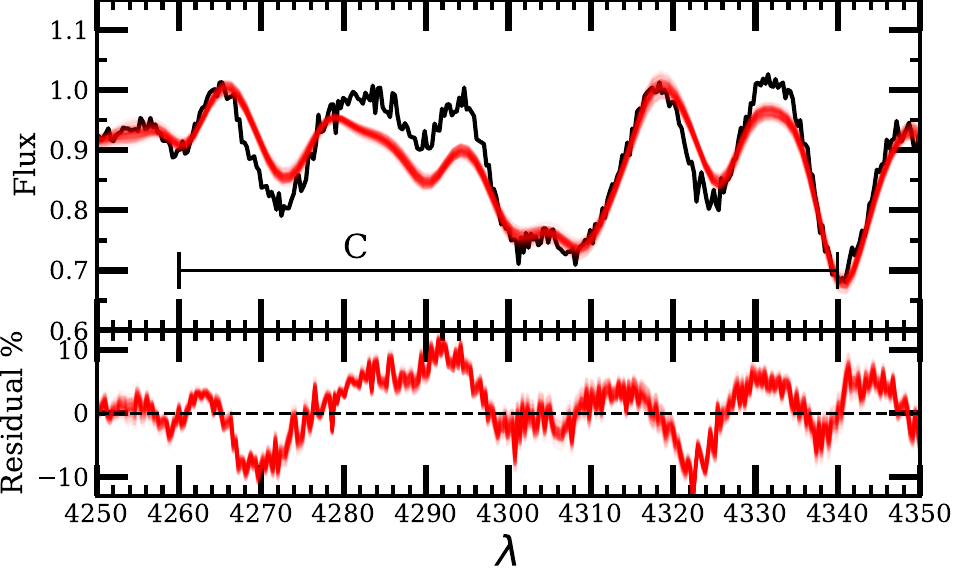}
    \includegraphics[scale=0.55]{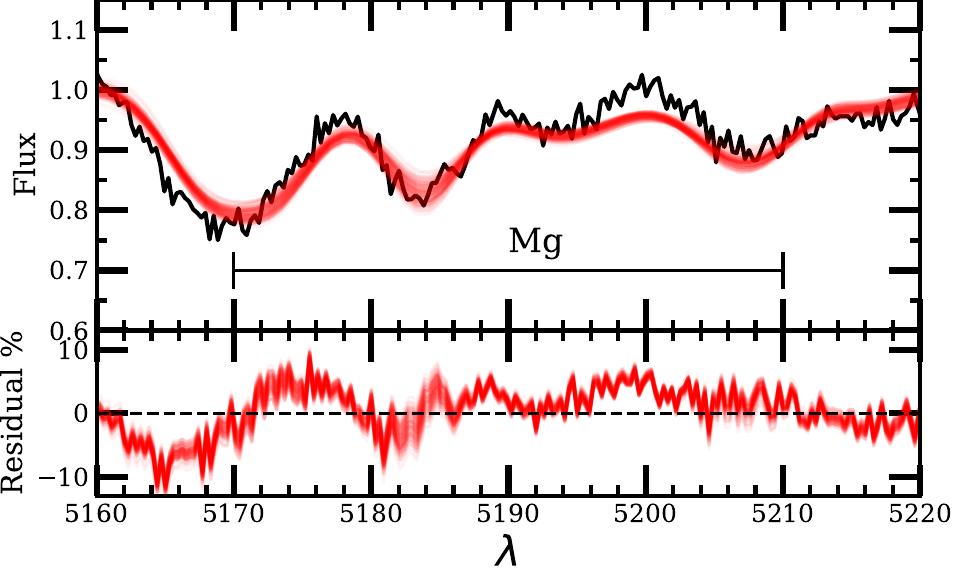}
    \includegraphics[scale=0.55]{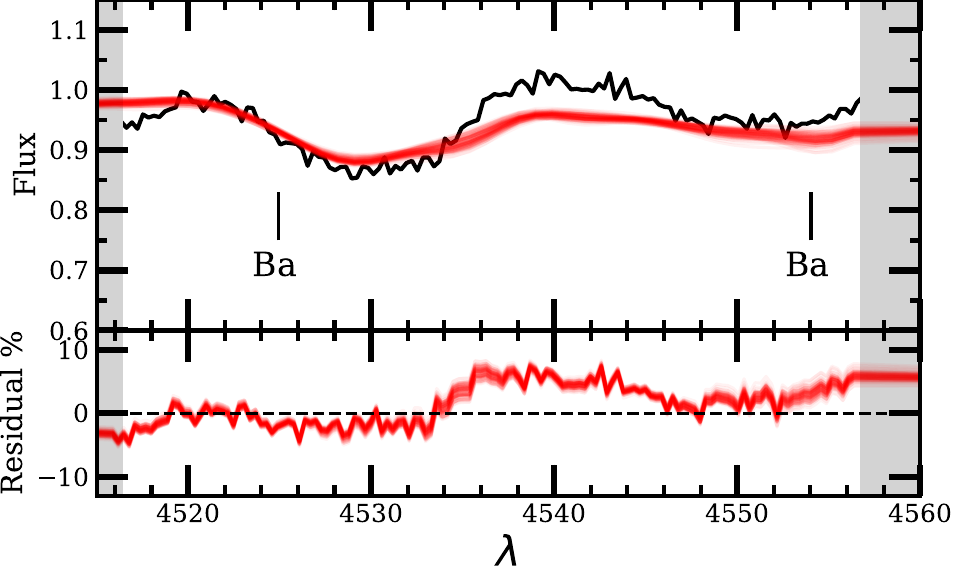}
    \includegraphics[scale=0.55]{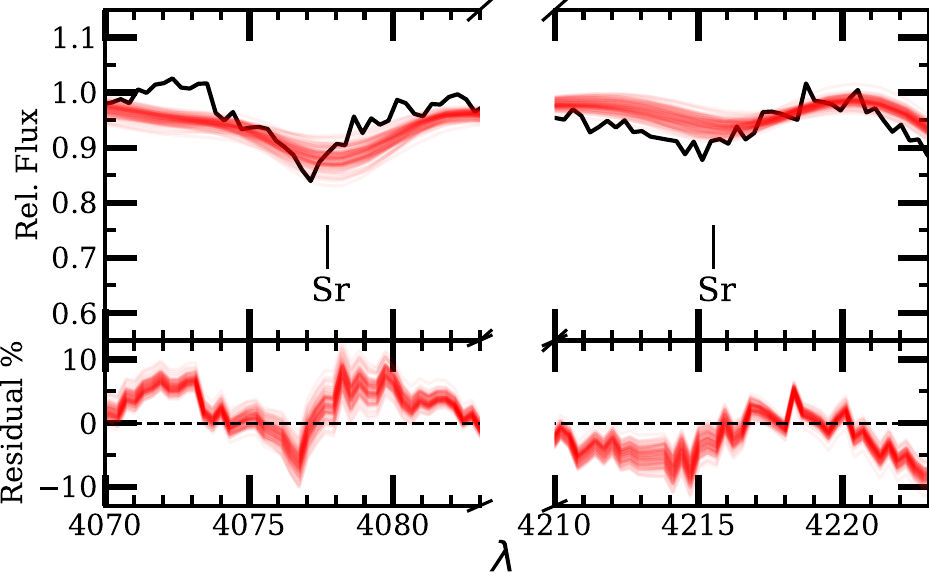}
    \includegraphics[scale=0.55]{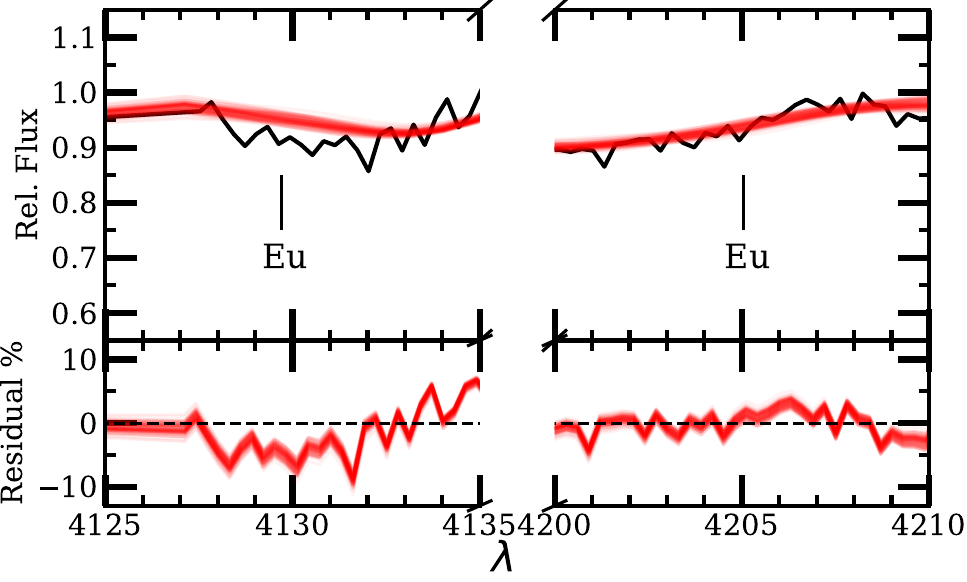}
    \caption{Fits from our bootstrap analysis are overlayed for NGC 2865 for sections of the spectrum that are influenced by C (\textit{top left panel}), Mg (\textit{top right panel}), Ba (\textit{middle left panel}), Sr (\textit{middle right panel}), and Eu (\textit{bottom panel}). The top graph of each panel displays a random sample of 100 of the 1000 model spectra fits are the red lines, while the original data spectrum of NGC 2865 is in black. The bottom graph of each panel shows the residual percentage from each model spectrum and the data spectrum.}
    \label{fig:ncap}
\end{figure*}

\citet{Conroy2013:1301.0321v2} reports abundances for Sr and Ba in ETGs for both stacked SDSS spectra and low-resolution spectra for individual galaxies. We show our results compared to \citet{Conroy2013:1301.0321v2} in Figure \ref{fig:srba}, where the Sr and Ba abundance ratios are plotted for ETGs with respect to their [Mg/Fe] ratios. The decreasing [Ba/Fe] trend with respect to the [Mg/Fe] abundance probes the timescale of star formation within the ETG, with shorter timescales leading to higher [Mg/Fe] values and lower [Ba/Fe] values \citep{Conroy2013:1301.0321v2}. The three ETGs studied in our work are determined to have solar abundances, which agrees with the Sr results of \citet{Conroy2013:1301.0321v2}, but not the decreasing trend of Ba with increasing Mg. 


\section{Discussion} \label{sec:disc}

The abundance ratios shown in Figure \ref{fig:allEls} for previous ETG studies show the observed chemical abundances for quiescent galaxies for a given mass. We explore how the merger history of ETGs could have altered the chemical evolution history in these systems by modeling different merger scenarios. For example, a late ($z\le1$), disk-disk merger would experience a large fraction (up to 40\%) in situ star formation in the central region of the merger remnant (\citealt{1996ApJ...471..115B}, \citealt{2014MNRAS.444.3357N}). As the two progenitor galaxies merged, early dissipation and inflows of gas from the parent galaxies cause an increased star formation rate during the first pass of the merger \citep{1996ApJ...464..641M}. The stars formed in this phase of the merger are in general massive and hence enrich the interstellar medium (ISM) gas on a timescale of $\sim20$ Myr. These massive stars then quickly inject $\alpha$-elements and Fe-peak elements such as Cr, Mn, Co into the ISM.

\subsection{NGC 2865}

NGC 2865 contains the most fine structure out of our 3 ETGs, with a fine structure constant of $\Sigma=6.4$. We note that NGC 2865, which is the only galaxy to show statistically higher Ca, Mn, and Cr abundances compared to ETGs without fine structure, is the youngest, least massive, and most recently perturbed system out of our 3 ETGs. We explore possible physical explanations that would lead to enhancement of certain $\alpha$-, and Fe-peak elements.

\begin{figure}
    \centering
    \includegraphics[scale=0.45]{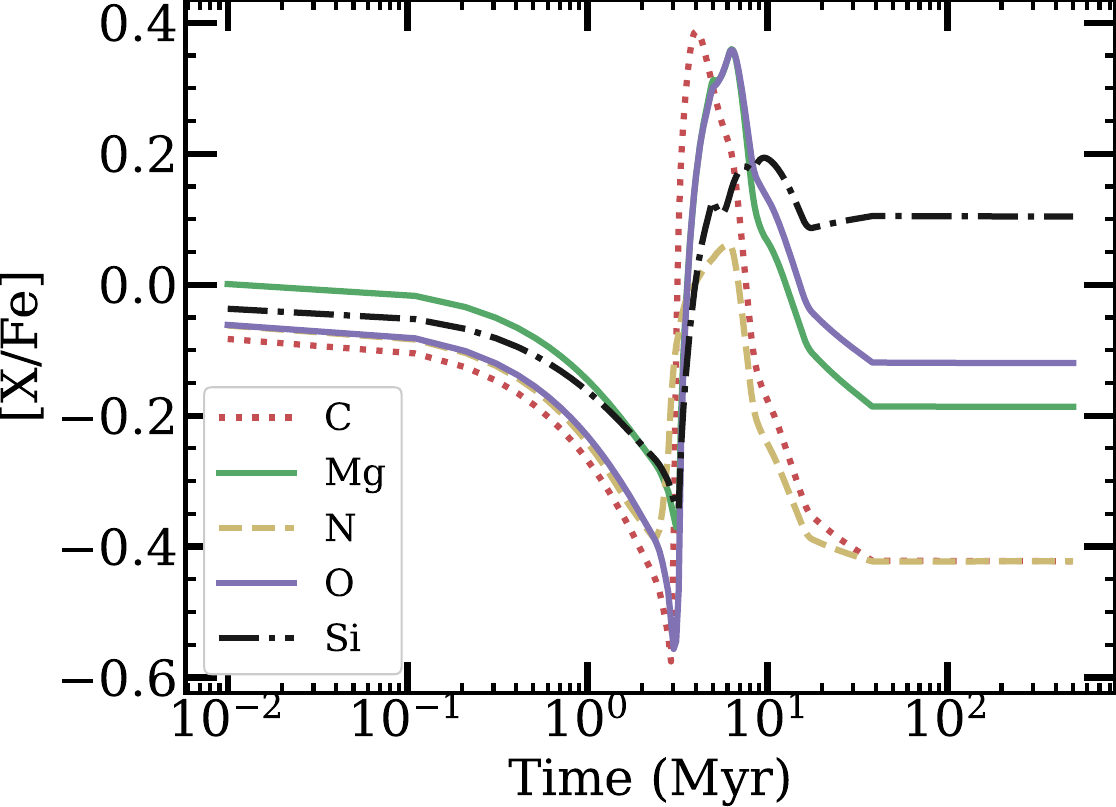}
    \caption{Chemical abundances as a function of time for a gas cloud with an initial solar metallicity using the Starburst99 program (\citealt{1999ApJS..125..489G}, \citealt{2005ApJ...621..695V}, \citealt{2010ApJS..189..309L}, \citealt{2014ApJS..212...14L}). The abundances displayed are derived using the yields modeled by Starburst99 for a gas cloud with a mass comparable to NGC 2865.}
    \label{fig:sb99}
\end{figure}

Simulations of major mergers suggest that there are two periods of intense star formation during the galaxy-pair interaction, one occurring during the first-pass of the progenitor galaxies, and the second occurring during coalescence (\citealt{2019MNRAS.485.1320M}, \citealt{2021MNRAS.503.3113M}). We test if a simple two burst model would be sufficient in explaining the differences in abundances for NGC 2865 compared to other ETGs without fine structure using the Starburst99 program (\citealt{1999ApJS..125..489G}, \citealt{2005ApJ...621..695V}, \citealt{2010ApJS..189..309L}, \citealt{2014ApJS..212...14L}). Starburst99 outputs yields of some elements as a function of gas cloud mass, IMF parameters, and time. We use Starburst99 to model the predicted elemental yields as a function of time after a single burst star formation event. We model this by assuming an initial ISM at solar metallicity with default Starburst99 parameters [IMF, cutoff, etc.] and plot the resulting metallicity of the gas, assuming fill mixing of the enriched gas predicted by SB99 yields. We show the [X/Fe] abundance ratios of gas cloud modeled by Starburst99 in Figure \ref{fig:sb99} for C, N, O, Mg, and Si. We find that there is not enough time for stars born during the starbursts periods of a merger to both significantly contribute to the light output of the galaxy, and to have abundances similar to what we determine for NGC 2865. The discrepancy between our abundances for the $\alpha$-elements and the abundances modeled in Starburst99 suggests that a simple two-burst model scenario cannot fully capture the chemical evolution taking place before or after a major merger event. Additional environmental components such as other periods of starburst activity, chemically peculiar progenitor galaxies, or accretion of chemically enriched satellite galaxies into the central regions of merger remnants could be the cause of the abundance offsets seen in NGC 2865.

\subsection{NGC 3818}

NGC 3818 does not show any abundance differences to ETGs without fine structure. NGC 3818 experienced its major merger earlier than NGC 2865, and which could explain the agreement of [$\alpha$/Fe] ratios to ETGs without fine structure. The fine structure constant of NGC 3818 is the lowest of the 3 ETGs studied in this work at $\Sigma=1.3$ In the case of NGC 3818, its main fine structure feature is a boxy isophote with an embedded disk, which are both structures associated with high mass, gas poor (dry) mergers \citep{2023MNRAS.520.5870G}. If the fine structure was formed in a dry merger, then we would expect a lack of chemical differences to ETGs without fine structure, as there would have been little-to-no star formation during the merger. The abundances we observe in NGC 3818 demonstrate that its dry merger event did not chemically alter the system.

\subsection{NGC 4915}

Finally, in the case of NGC 4915, we find chemical abundance ratios similar to NGC 3818 and other ETGs, however NGC 4915 is fit with a younger stellar population (relative to other ETGs of a similar mass) at its center and it possesses a fine structure constant of $\Sigma=5.5$. \citet{1990ApJ...364L..33S} suggests that NGC 4915 was a product of a disk-disk merger, and if NGC 4915 experienced enough gas inflow to the center of the galaxy spurring star formation, it would explain the presence of a younger stellar population. Assuming that the stellar and gaseous abundance patterns are the same in the progenitor galaxies that formed NGC 4915, the gas present at the time of the merger was likely chemically similar to the gas found in ETGs without fine structure at the same epoch. As a result, the resulting stellar population that formed during the merger would then be chemically similar to that of ETGs without fine structure.

\subsection{Chemical Abundances and Fine Structure}

While the fine structure constant, $\Sigma$, is a useful measurement of how disturbed a system is, the 3 ETGs here show that its value does not indicate a level of chemical enrichment. NGC 3818 and NGC 4915, the two ETGs whose fine structure mostly comes from their boxy isophotes, show little chemical differences to ETGs without fine structure. However, NGC 2865, which possesses tidal tails and extended shells, shows offsets in the $\alpha$, and Fe-peak elements when compared to ETGs without fine structure. As mentioned in Section \ref{sec:intro}, NGC 2865's shells resemble disk galaxies in color as a result of disk-disk galaxy merger, therefore the younger, blue populations of the shells may be linked to the younger stellar populations observed at the center of NGC 2865. NGC 3818 and NGC 4915's lack of abundance to ETGs without fine structure, compared to the chemical offset seen in NGC 2865 could indicate that the type of fine structure (i.e. boxy isophotes versus stellar streams and shells) present around a merger remnant correlates with the chemical composition of the merger remnant.

\section{Conclusions} \label{sec:conclusion}

We present elemental abundance ratios for three early-type elliptical galaxies that have fine structure. The chemical compositions derived from their nuclear spectrum offers a unique look into if and how the stellar populations at the center of ETGs are affected by major mergers. After comparing our ETGs with fine structure to literature ETGs without detected fine structure, we offer the following conclusions:
\begin{itemize}
    \item {We present the derived abundance ratios for a wide array of elements including C, N, O, Mg, Si, Ca, Ti, Cr, Mn, Co, Ni, Cu, Sr, Ba, and Eu. We conduct a bootstrap analysis to probe the reliability of the determined abundances and find that the heavier elements like Cu, Sr, Ba, and Eu are subject to larger uncertainties due to their small response functions and ability to be blended with stronger, nearby elements. Therefore, it is imperative to maximize SNR in spectra and improve our understanding of the response functions of heavy elements and refining our stellar library models if we want to better constrain their abundances.}
    \item {Among the three galaxies we analyzed in this paper, only NGC 2865 shows a chemical difference to literature galaxies without fine structure. We determine that an ETG having fine structure does not necessarily mean its abundance pattern will differ from other ETGs of similar mass.}
    \item {We attribute the abundances differences found in NGC 2865 to the star formation that occurred during its last, major merger. The higher derived abundance ratios in NGC 2865 for Ca, Cr, and Mn as compared to ETGs without fine structure could suggest a more complex star formation history than a simple, two-burst scenario. A more complex scenario would be necessary to characterize NGC 2865's SFH, and would have to consider other merger variables such as the effects of gas inflows and outflows during the merger. Additionally, a more complex SFH could explain why some of the elements such as Ca, Cr, and Mn are enriched, while other elements in the same nucleosynthesis family are not.}
\end{itemize}

Our study of the chemical abundances at the center of ETGs with fine structure has revealed that the presence of fine structure alone does not imply a chemical difference for the stellar populations, however, there is evidence that the \textit{type} of fine structure and the variations in abundance patterns may be a consequence of the same event, the merger and associated star formation. Future chemical abundance studies, especially Integral Field Spectroscopy (IFS) to derive abundance gradients from the center to the outer regions of ETGs with fine structure, will be a valuable tool in determining the extent that the type of fine structure alter the chemical abundances. IFS spectra of ETG regions associated with types of fine structure seen in NGC 2865 (shells, tidal streams), can be compared to the spectra obtained from regions without fine structure in the same galaxy. This comparison will show how the type of fine structure correlates with abundances, and furthermore the chemical evolution history associated with different types of fine structure.

\begin{acknowledgements}
The authors acknowledge University of Florida Research Computing for providing computational resources and support that have contributed to the research results reported in this publication. URL: \url{http://researchcomputing.ufl.edu}

We thank Charlie Conroy not only for his Absorption Line Fitter program, but also his assistance on improving the fit using the high resolution MIKE data. 

Rana Ezzeddine acknowledges support from NSF grant AST-2206263.

The National Radio Astronomy Observatory is a facility of the National Science  Foundation operated under cooperative agreement by Associated Universities, Inc.

Ezequiel Treister acknowledges support from: ANID through Millennium Science Initiative Program - NCN19\_058, CATA-BASAL - ACE210002 and FB210003, FONDECYT Regular - 1190818 and 1200495.
\end{acknowledgements}

%

\vspace{5mm} 
\facility{Magellan: Clay (MIKE spectrograph)}


\software{Absorption Line Fitter (ALF) -  (\citealt{2012ApJ...760...71C}, \citealt{2018ApJ...854..139C}), Python \citep{10.5555/1593511}, CarPy (\citealt{2000ApJ...531..137K}, \citealt{2003PASP..115..688K}), Astropy \citep{2022ApJ...935..167A}, Numpy \citep{harris2020array}, Scipy \citep{2020SciPy-NMeth}, Matplotlib \citep{Hunter:2007}
          }


\bibliography{bib}{}
\bibliographystyle{aasjournal}

\newpage

\appendix

\section{Detailed Results} \label{sec:app}

In order to evaluate the fit of each individual element by ALF, we produce plots that display characteristics of the fitting routine. For each element we display the median, fitted abundance ratio compared to other ETGs (top left panel in each figure), the distribution of fitted values for that element from our bootstrap analysis (top right panel), the parameter covariance between the element and other fitted parameters of ALF (middle row panel), and finally the response function (what wavelength regions the element affects) of the element (bottom row panel).

In the subfigures of Figure 8, we compare our abundance ratios to ETG studies by \citet{2012ApJ...760...71C}, \citet{Conroy2013:1303.6629v2}, \citet{Conroy2013:1301.0321v2}, \citet{2015ApJ...807...11G}, \citet{2021ApJ...923...65F}, and \citet{2021MNRAS.502.5508P}. The colors and symbols used for the literature samples are consistent with Figure \ref{fig:allEls}. We compare literature sources that also use full spectrum fitting methods in order to provide a consistent comparison sample.

The parameter covariance is shown by the correlation coefficients graph. Parameters that show a strong correlation indicate that when ALF produces a model with the selected parameter at a certain value, other parameters that correlate are typically fit with similar trends (e.g. a large O abundance model often has a large C abundance as well).

We include the response function of each element to show the areas of the spectra that are most affected by changes in the specified element. The response function is constructed by taking two models generated by ALF, with all of the same parameter values, except for that of the element of interest. The difference between these two models thus show where the spectra changes when there is a difference abundance ratio present. We use the response function to determine if we are fitting the strongest regions of the spectra for that particular element.

\figsetstart

\figsetnum{8}

\figsettitle{Individual Abundances}

\figsetgrpstart
\figsetgrpnum{8.1}
\figsetgrptitle{Carbon}
\figsetplot{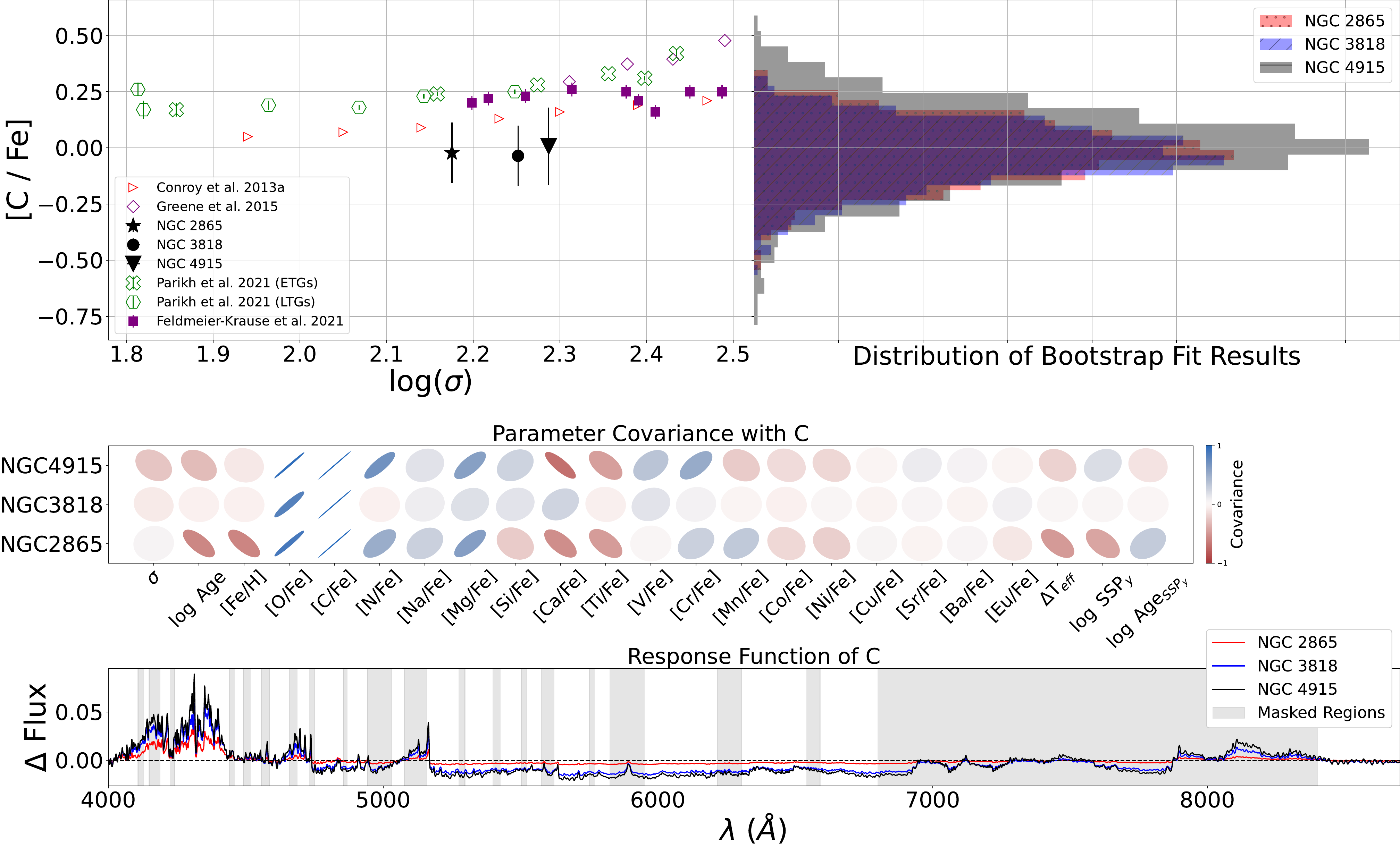} \label{fig:SigC}
\figsetgrpnote{Determinations of carbon abundance in each ETG. \textit{Top left panel:} Comparison of [C/Fe] abundance ratios to literature ETGs plotted against the velocity dispersion. \textit{Top right panel:} Distribution for the carbon abundance from 1000 fitting routines run over spectra sampled from our observed spectrum. The MCMC routine utilized 1024 walkers and a chain of 10,000 after the burn-in. \textit{Middle panel:} Parameter covariance of carbon compared to the other parameters in the fit. Blue, right-leaning ovals indicate a strong correlation, red, left-leaning ovals indicate anti-correlation, and round, pale ellipses indicate no correlation. \textit{Bottom panel:} Difference between two models synthesized with [C/Fe] ratios $\pm$ 0.5 dex above and below the median of the [C/Fe] posterior distribution. Areas of the spectrum that deviate  from 0 show where carbon has an effect on the integrated light.}
\figsetgrpend

\figsetgrpstart
\figsetgrpnum{8.2}
\figsetgrptitle{Nitrogen}
\figsetplot{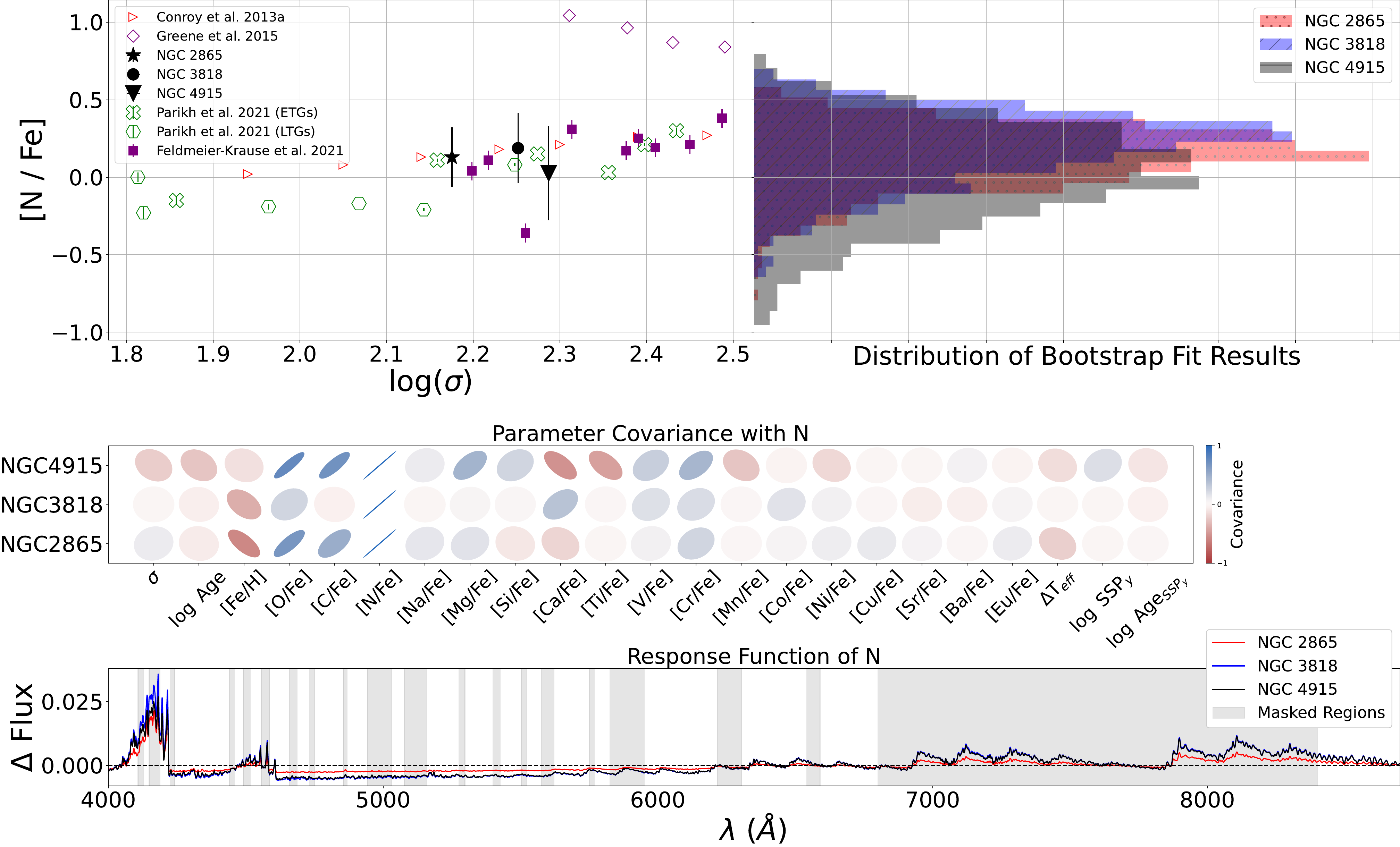}
\figsetgrpnote{Determinations of nitrogen abundance in each ETG. \textit{Top left panel:} Comparison of [N/Fe] abundance ratios to literature ETGs plotted against the velocity dispersion. \textit{Top right panel:} Distribution for the nitrogen abundance from 1000 fitting routines run over spectra sampled from our observed spectrum. The MCMC routine utilized 1024 walkers and a chain of 10,000 after the burn-in. \textit{Middle panel:} Parameter covariance of nitrogen compared to the other parameters in the fit. Blue, right-leaning ovals indicate a strong correlation, red, left-leaning ovals indicate anti-correlation, and round, pale ellipses indicate no correlation. \textit{Bottom panel:} Difference between two models synthesized with [N/Fe] ratios $\pm$ 0.5 dex above and below the median of the [N/Fe] posterior distribution. Areas of the spectrum that deviate  from 0 show where nitrogen has an effect on the integrated light.}
\figsetgrpend

\figsetgrpstart
\figsetgrpnum{8.3}
\figsetgrptitle{Oxygen}
\figsetplot{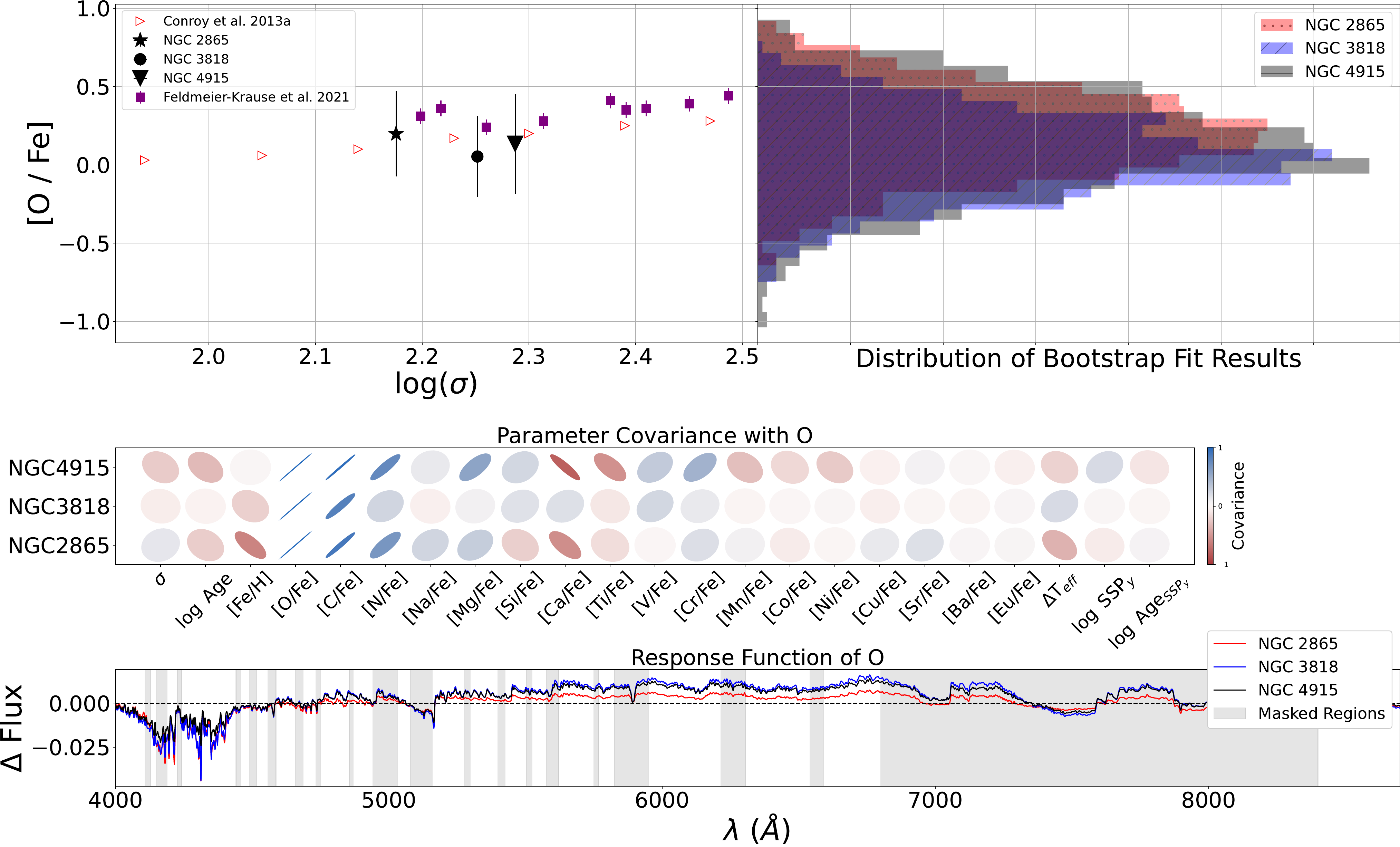}
\figsetgrpnote{Determinations of oxygen abundance in each ETG. \textit{Top left panel:} Comparison of [O/Fe] abundance ratios to literature ETGs plotted against the velocity dispersion. \textit{Top right panel:} Distribution for the oxygen abundance from 1000 fitting routines run over spectra sampled from our observed spectrum. The MCMC routine utilized 1024 walkers and a chain of 10,000 after the burn-in. \textit{Middle panel:} Parameter covariance of oxygen compared to the other parameters in the fit. Blue, right-leaning ovals indicate a strong correlation, red, left-leaning ovals indicate anti-correlation, and round, pale ellipses indicate no correlation. \textit{Bottom panel:} Difference between two models synthesized with [O/Fe] ratios $\pm$ 0.5 dex above and below the median of the [O/Fe] posterior distribution. Areas of the spectrum that deviate  from 0 show where oxygen has an effect on the integrated light.}
\figsetgrpend

\figsetgrpstart
\figsetgrpnum{8.4}
\figsetgrptitle{Magnesium}
\figsetplot{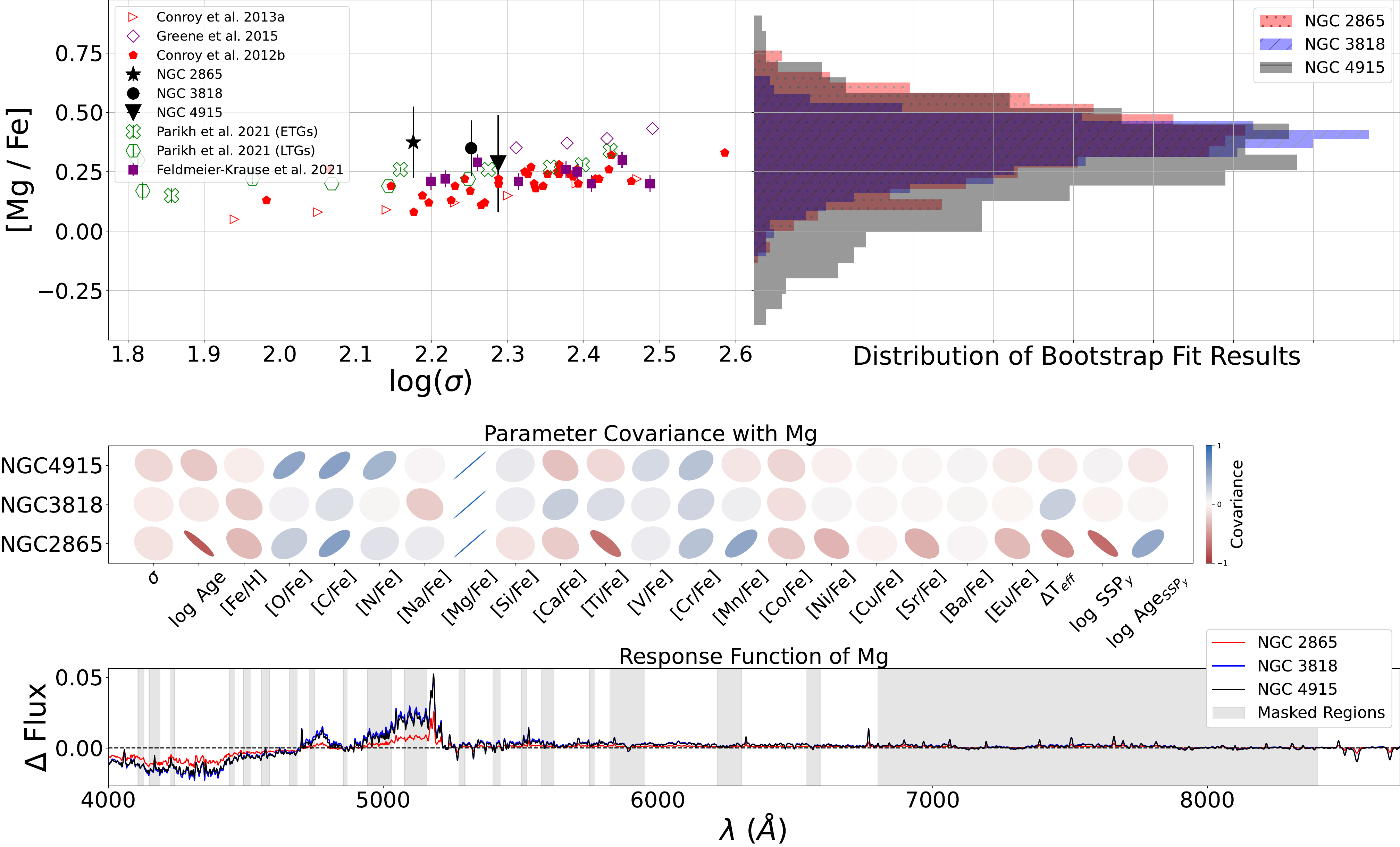}
\figsetgrpnote{Determinations of magnesium abundance in each ETG. \textit{Top left panel:} Comparison of [Mg/Fe] abundance ratios to literature ETGs plotted against the velocity dispersion. \textit{Top right panel:} Distribution for the magnesium abundance from 1000 fitting routines run over spectra sampled from our observed spectrum. The MCMC routine utilized 1024 walkers and a chain of 10,000 after the burn-in. \textit{Middle panel:} Parameter covariance of magnesium compared to the other parameters in the fit. Blue, right-leaning ovals indicate a strong correlation, red, left-leaning ovals indicate anti-correlation, and round, pale ellipses indicate no correlation. \textit{Bottom panel:} Difference between two models synthesized with [Mg/Fe] ratios $\pm$ 0.5 dex above and below the median of the [Mg/Fe] posterior distribution. Areas of the spectrum that deviate  from 0 show where magnesium has an effect on the integrated light.}
\figsetgrpend

\figsetgrpstart
\figsetgrpnum{8.5}
\figsetgrptitle{Silicon}
\figsetplot{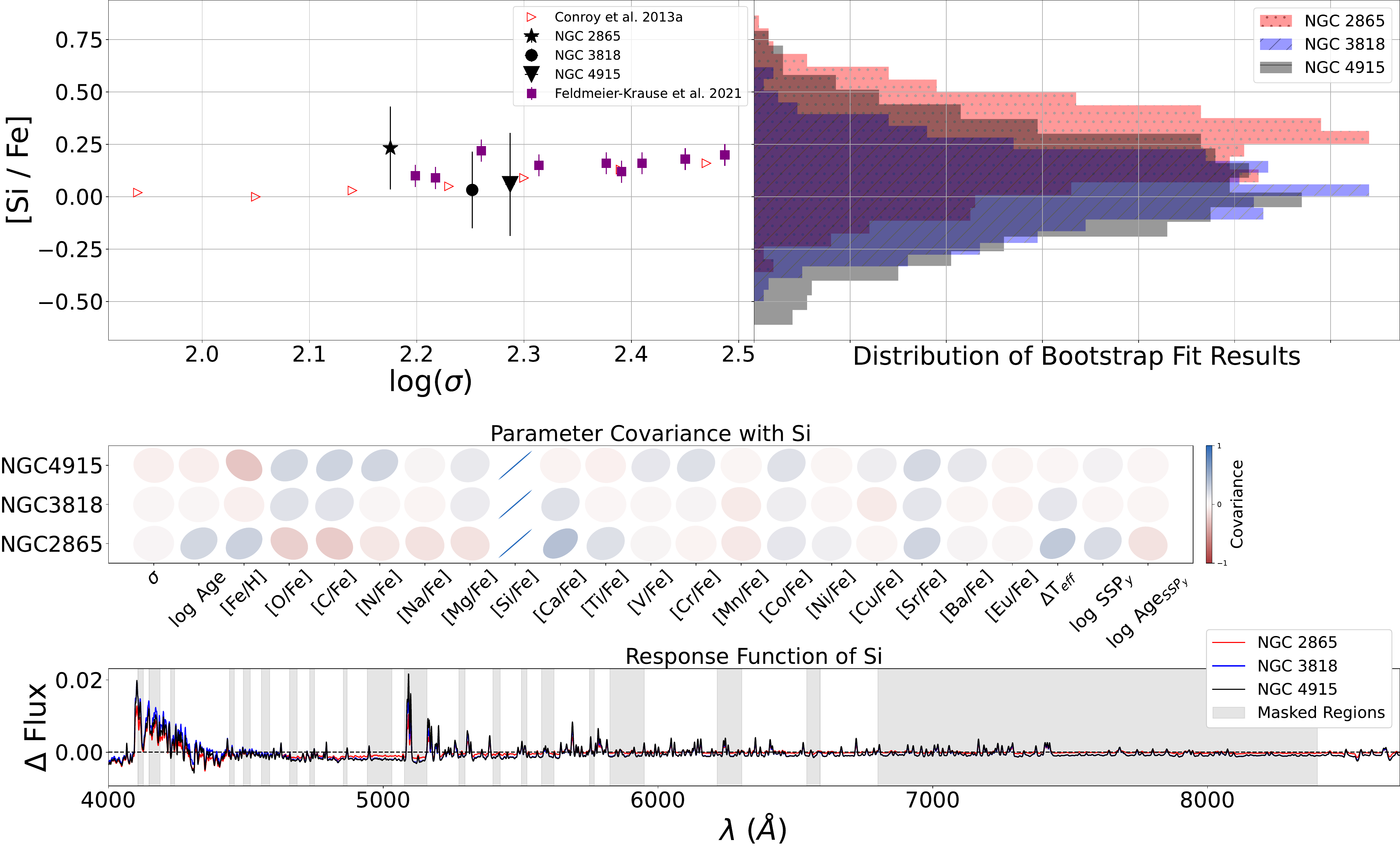}
\figsetgrpnote{Determinations of silicon abundance in each ETG. \textit{Top left panel:} Comparison of [Si/Fe] abundance ratios to literature ETGs plotted against the velocity dispersion. \textit{Top right panel:} Distribution for the silicon abundance from 1000 fitting routines run over spectra sampled from our observed spectrum. The MCMC routine utilized 1024 walkers and a chain of 10,000 after the burn-in. \textit{Middle panel:} Parameter covariance of silicon compared to the other parameters in the fit. Blue, right-leaning ovals indicate a strong correlation, red, left-leaning ovals indicate anti-correlation, and round, pale ellipses indicate no correlation. \textit{Bottom panel:} Difference between two models synthesized with [Si/Fe] ratios $\pm$ 0.5 dex above and below the median of the [Si/Fe] posterior distribution. Areas of the spectrum that deviate  from 0 show where silicon has an effect on the integrated light.}
\figsetgrpend

\figsetgrpstart
\figsetgrpnum{8.6}
\figsetgrptitle{Calcium}
\figsetplot{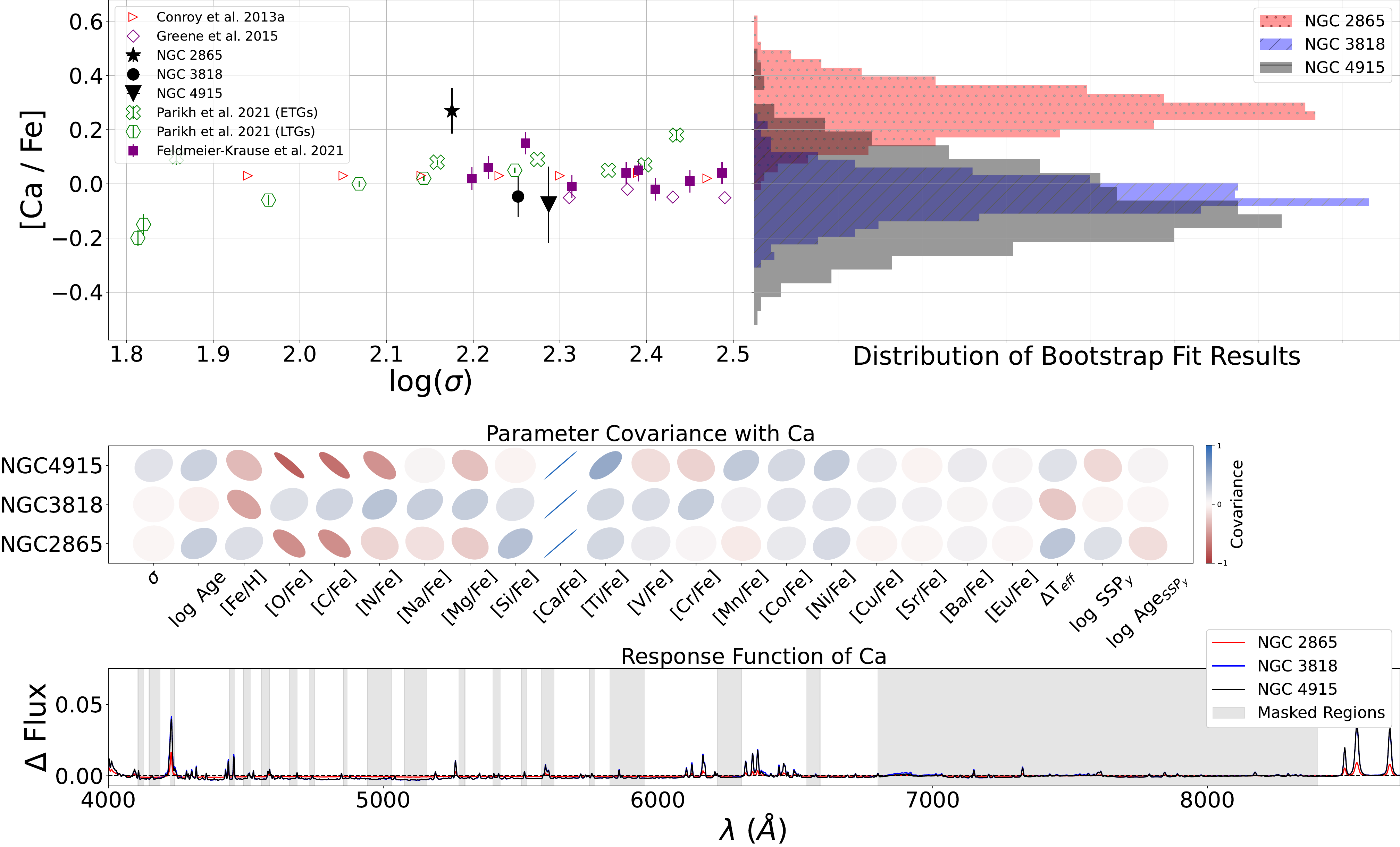}
\figsetgrpnote{Determinations of calcium abundance in each ETG. \textit{Top left panel:} Comparison of [Ca/Fe] abundance ratios to literature ETGs plotted against the velocity dispersion. \textit{Top right panel:} Distribution for the calcium abundance from 1000 fitting routines run over spectra sampled from our observed spectrum. The MCMC routine utilized 1024 walkers and a chain of 10,000 after the burn-in. \textit{Middle panel:} Parameter covariance of calcium compared to the other parameters in the fit. Blue, right-leaning ovals indicate a strong correlation, red, left-leaning ovals indicate anti-correlation, and round, pale ellipses indicate no correlation. \textit{Bottom panel:} Difference between two models synthesized with [Ca/Fe] ratios $\pm$ 0.5 dex above and below the median of the [Ca/Fe] posterior distribution. Areas of the spectrum that deviate  from 0 show where calcium has an effect on the integrated light.}
\figsetgrpend

\figsetgrpstart
\figsetgrpnum{8.7}
\figsetgrptitle{Titanium}
\figsetplot{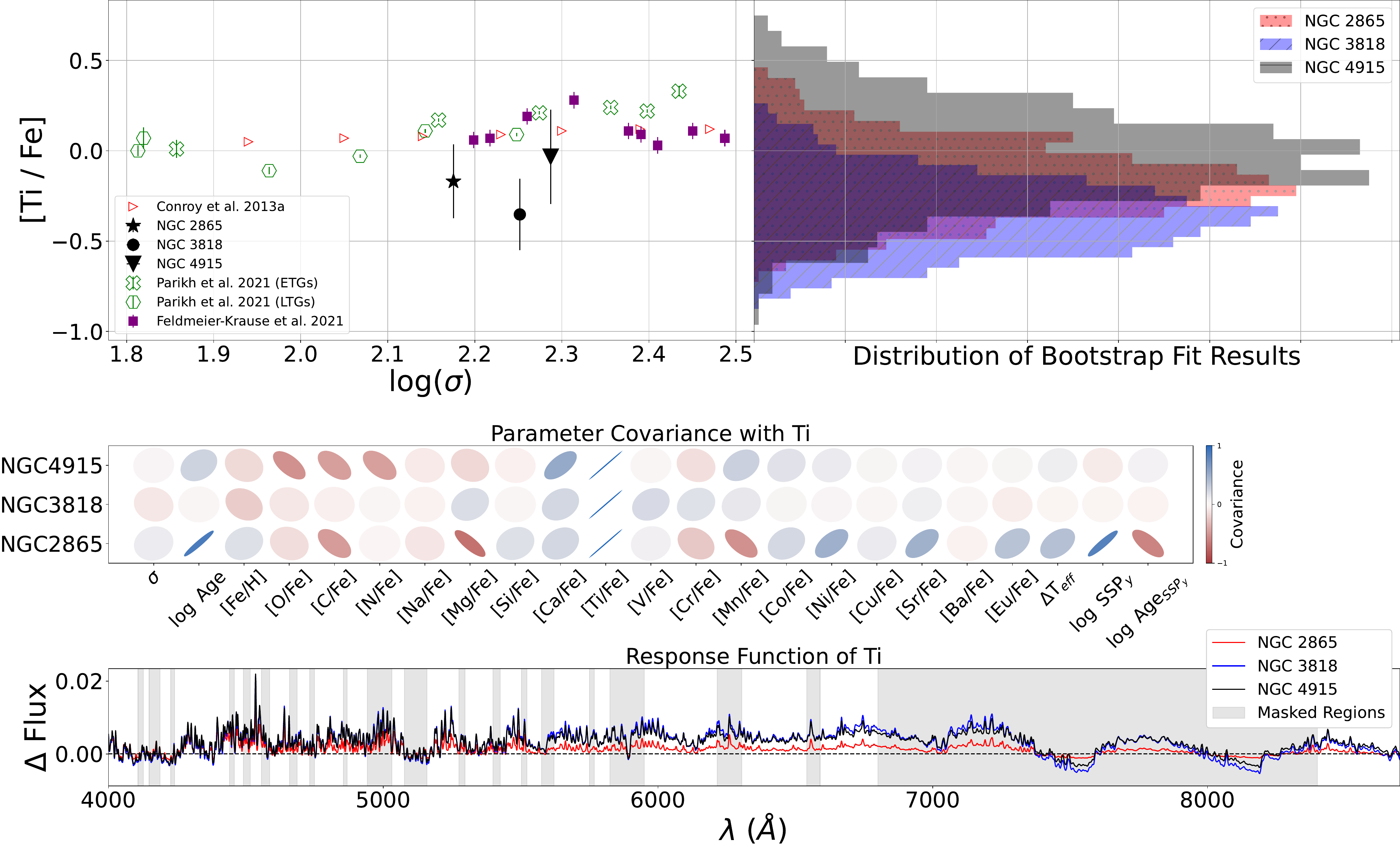}
\figsetgrpnote{Determinations of titanium abundance in each ETG. \textit{Top left panel:} Comparison of [Ti/Fe] abundance ratios to literature ETGs plotted against the velocity dispersion. \textit{Top right panel:} Distribution for the titanium abundance from 1000 fitting routines run over spectra sampled from our observed spectrum. The MCMC routine utilized 1024 walkers and a chain of 10,000 after the burn-in. \textit{Middle panel:} Parameter covariance of titanium compared to the other parameters in the fit. Blue, right-leaning ovals indicate a strong correlation, red, left-leaning ovals indicate anti-correlation, and round, pale ellipses indicate no correlation. \textit{Bottom panel:} Difference between two models synthesized with [Ti/Fe] ratios $\pm$ 0.5 dex above and below the median of the [Ti/Fe] posterior distribution. Areas of the spectrum that deviate  from 0 show where titanium has an effect on the integrated light.}
\figsetgrpend

\figsetgrpstart
\figsetgrpnum{8.8}
\figsetgrptitle{Chromium}
\figsetplot{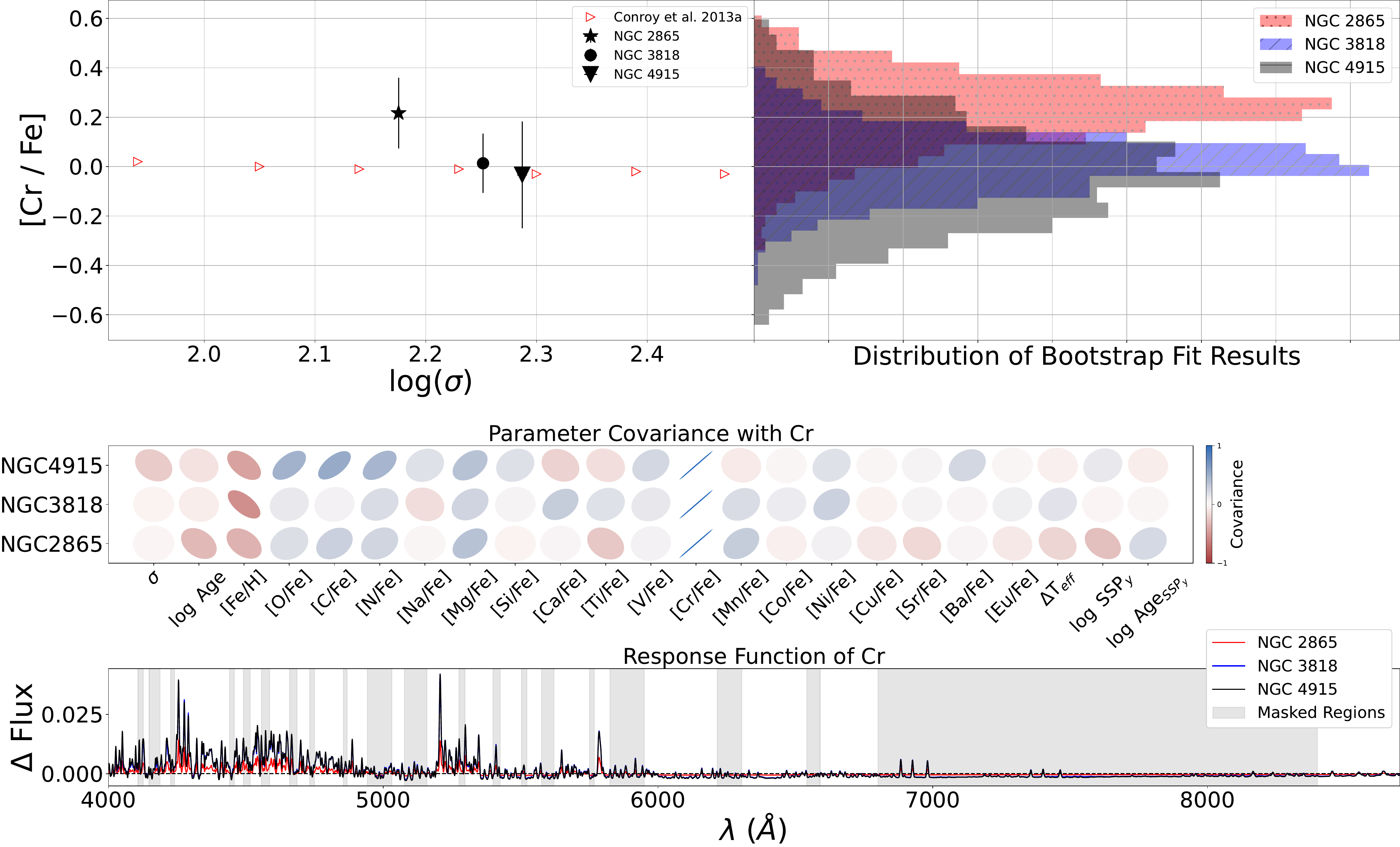}
\figsetgrpnote{Determinations of chromium abundance in each ETG. \textit{Top left panel:} Comparison of [Cr/Fe] abundance ratios to literature ETGs plotted against the velocity dispersion. \textit{Top right panel:} Distribution for the chromium abundance from 1000 fitting routines run over spectra sampled from our observed spectrum. The MCMC routine utilized 1024 walkers and a chain of 10,000 after the burn-in. \textit{Middle panel:} Parameter covariance of chromium compared to the other parameters in the fit. Blue, right-leaning ovals indicate a strong correlation, red, left-leaning ovals indicate anti-correlation, and round, pale ellipses indicate no correlation. \textit{Bottom panel:} Difference between two models synthesized with [Cr/Fe] ratios $\pm$ 0.5 dex above and below the median of the [Cr/Fe] posterior distribution. Areas of the spectrum that deviate  from 0 show where chromium has an effect on the integrated light.}
\figsetgrpend

\figsetgrpstart
\figsetgrpnum{8.9}
\figsetgrptitle{Manganese}
\figsetplot{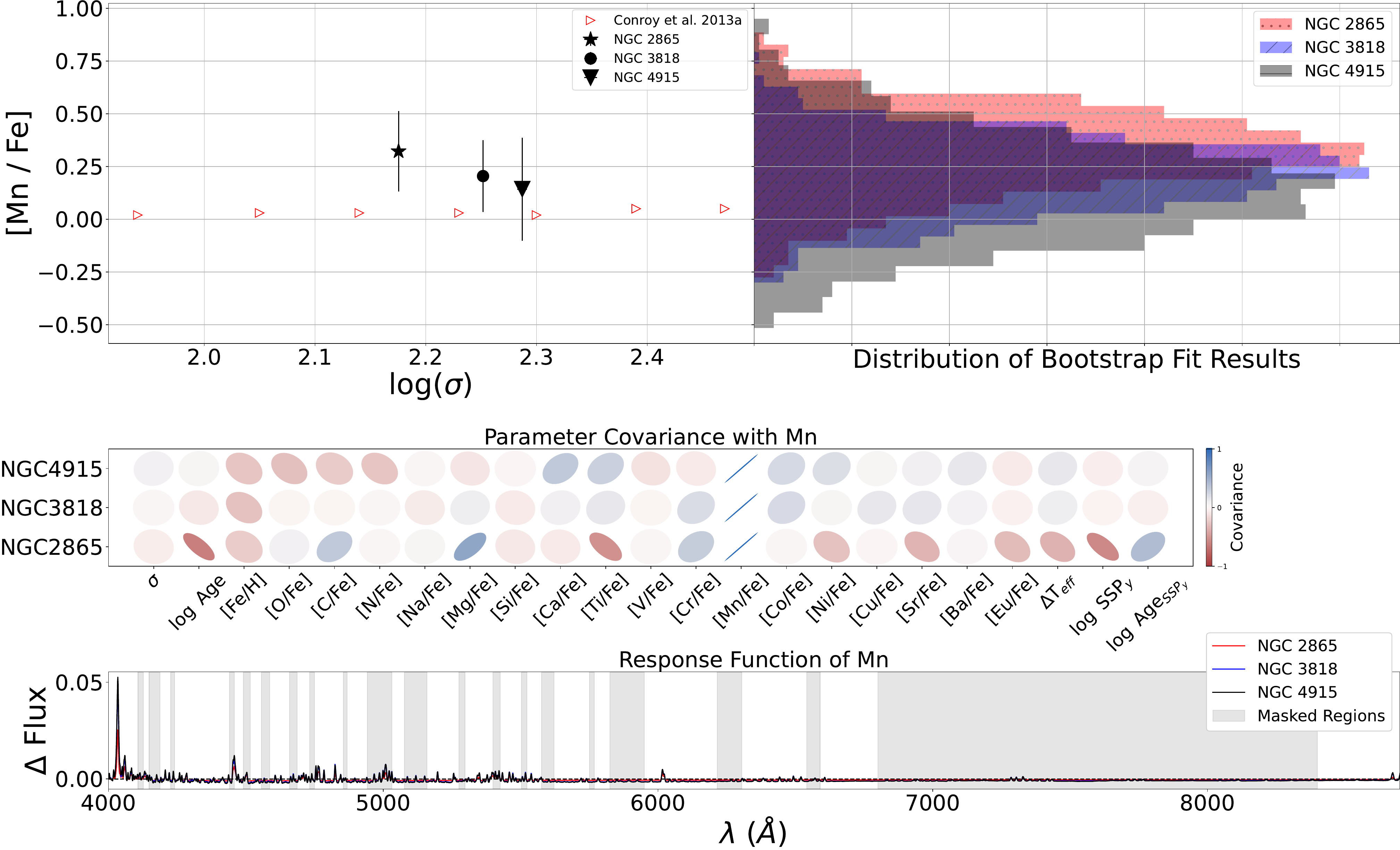}
\figsetgrpnote{Determinations of manganese abundance in each ETG. \textit{Top left panel:} Comparison of [Mn/Fe] abundance ratios to literature ETGs plotted against the velocity dispersion. \textit{Top right panel:} Distribution for the manganese abundance from 1000 fitting routines run over spectra sampled from our observed spectrum. The MCMC routine utilized 1024 walkers and a chain of 10,000 after the burn-in. \textit{Middle panel:} Parameter covariance of manganese compared to the other parameters in the fit. Blue, right-leaning ovals indicate a strong correlation, red, left-leaning ovals indicate anti-correlation, and round, pale ellipses indicate no correlation. \textit{Bottom panel:} Difference between two models synthesized with [Mn/Fe] ratios $\pm$ 0.5 dex above and below the median of the [Mn/Fe] posterior distribution. Areas of the spectrum that deviate  from 0 show where manganese has an effect on the integrated light.}
\figsetgrpend

\figsetgrpstart
\figsetgrpnum{8.10}
\figsetgrptitle{Cobalt}
\figsetplot{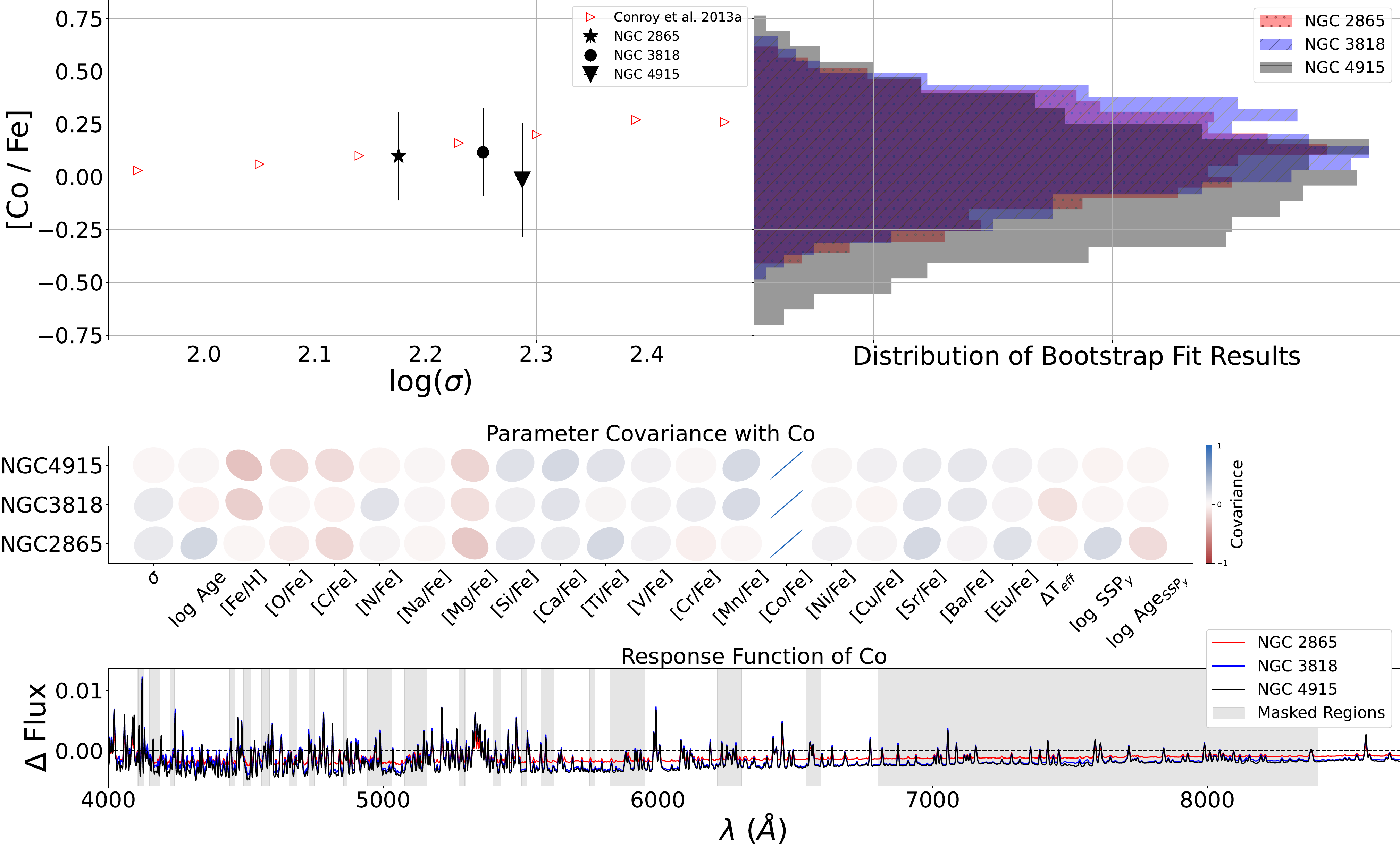}
\figsetgrpnote{Determinations of cobalt abundance in each ETG. \textit{Top left panel:} Comparison of [Co/Fe] abundance ratios to literature ETGs plotted against the velocity dispersion. \textit{Top right panel:} Distribution for the cobalt abundance from 1000 fitting routines run over spectra sampled from our observed spectrum. The MCMC routine utilized 1024 walkers and a chain of 10,000 after the burn-in. \textit{Middle panel:} Parameter covariance of cobalt compared to the other parameters in the fit. Blue, right-leaning ovals indicate a strong correlation, red, left-leaning ovals indicate anti-correlation, and round, pale ellipses indicate no correlation. \textit{Bottom panel:} Difference between two models synthesized with [Co/Fe] ratios $\pm$ 0.5 dex above and below the median of the [Co/Fe] posterior distribution. Areas of the spectrum that deviate  from 0 show where cobalt has an effect on the integrated light.}
\figsetgrpend

\figsetgrpstart
\figsetgrpnum{8.11}
\figsetgrptitle{Nickel}
\figsetplot{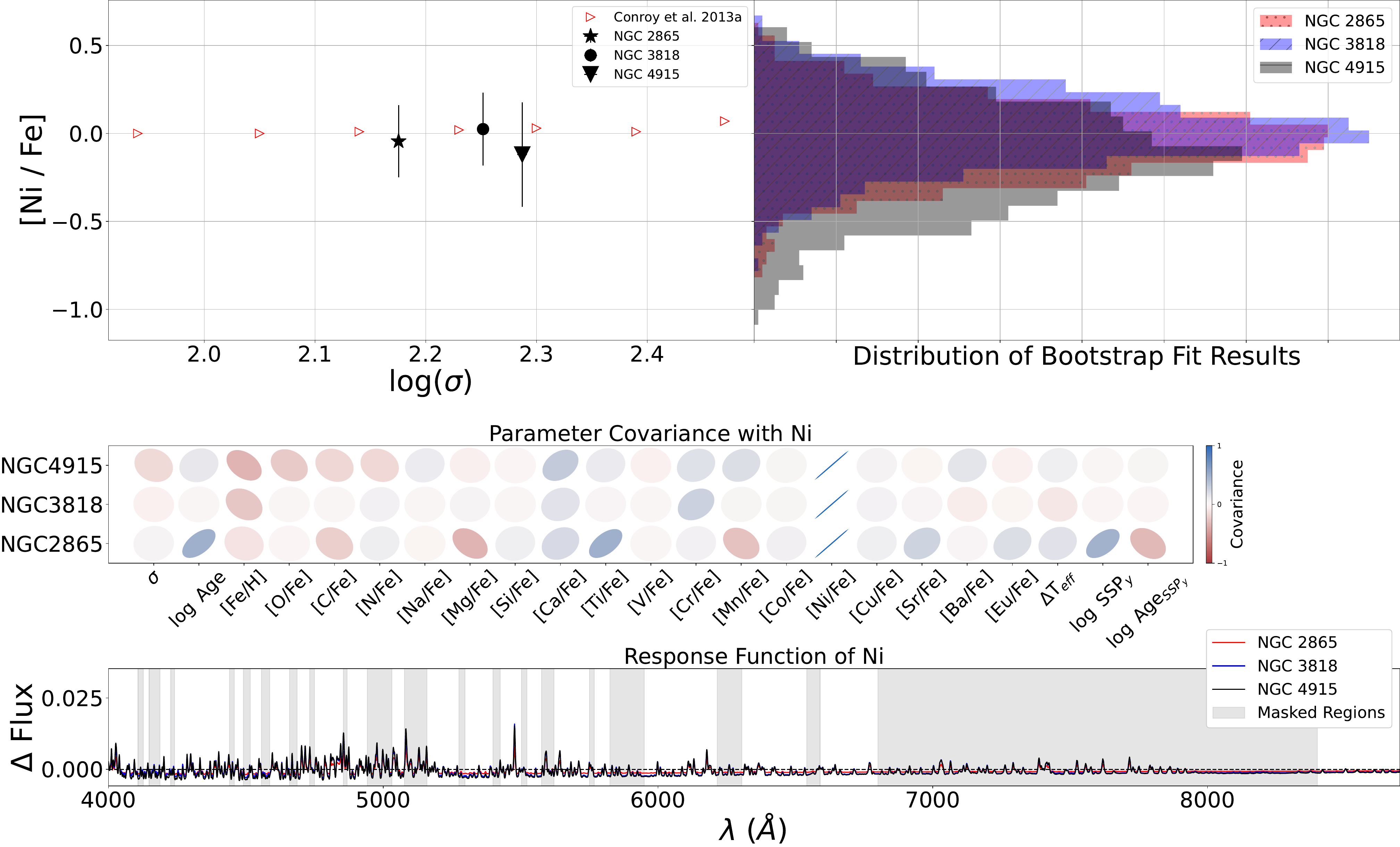}
\figsetgrpnote{Determinations of nickel abundance in each ETG. \textit{Top left panel:} Comparison of [Ni/Fe] abundance ratios to literature ETGs plotted against the velocity dispersion. \textit{Top right panel:} Distribution for the Nickel abundance from 1000 fitting routines run over spectra sampled from our observed spectrum. The MCMC routine utilized 1024 walkers and a chain of 10,000 after the burn-in. \textit{Middle panel:} Parameter covariance of Nickel compared to the other parameters in the fit. Blue, right-leaning ovals indicate a strong correlation, red, left-leaning ovals indicate anti-correlation, and round, pale ellipses indicate no correlation. \textit{Bottom panel:} Difference between two models synthesized with [Ni/Fe] ratios $\pm$ 0.5 dex above and below the median of the [Ni/Fe] posterior distribution. Areas of the spectrum that deviate  from 0 show where Nickel has an effect on the integrated light.}
\figsetgrpend

\figsetgrpstart
\figsetgrpnum{8.12}
\figsetgrptitle{Copper}
\figsetplot{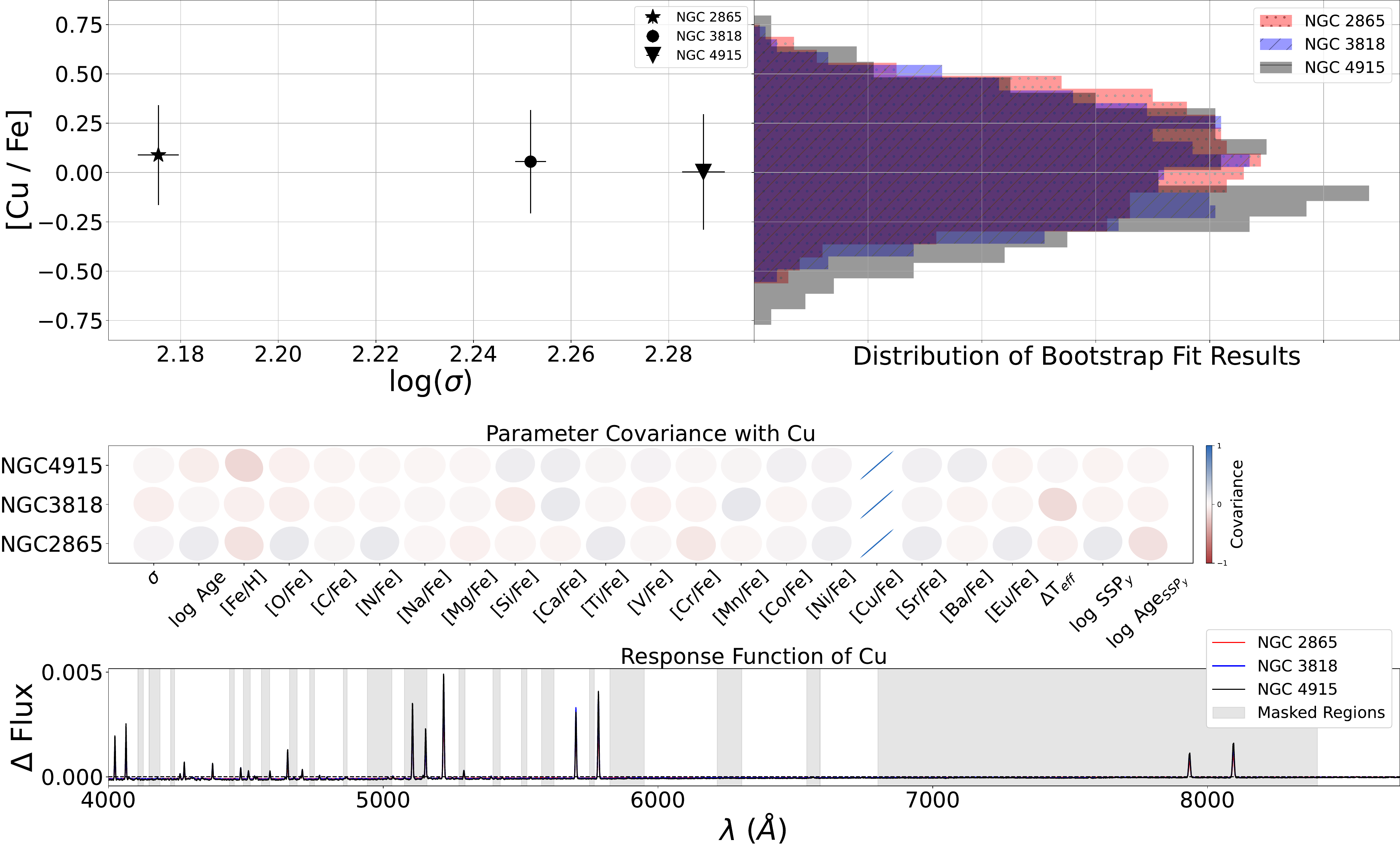}
\figsetgrpnote{Determinations of copper abundance in each ETG. \textit{Top left panel:} Comparison of [Cu/Fe] abundance ratios to literature ETGs plotted against the velocity dispersion. \textit{Top right panel:} Distribution for the copper abundance from 1000 fitting routines run over spectra sampled from our observed spectrum. The MCMC routine utilized 1024 walkers and a chain of 10,000 after the burn-in. \textit{Middle panel:} Parameter covariance of copper compared to the other parameters in the fit. Blue, right-leaning ovals indicate a strong correlation, red, left-leaning ovals indicate anti-correlation, and round, pale ellipses indicate no correlation. \textit{Bottom panel:} Difference between two models synthesized with [Cu/Fe] ratios $\pm$ 0.5 dex above and below the median of the [Cu/Fe] posterior distribution. Areas of the spectrum that deviate  from 0 show where copper has an effect on the integrated light.}
\figsetgrpend

\figsetgrpstart
\figsetgrpnum{8.13}
\figsetgrptitle{Strontium}
\figsetplot{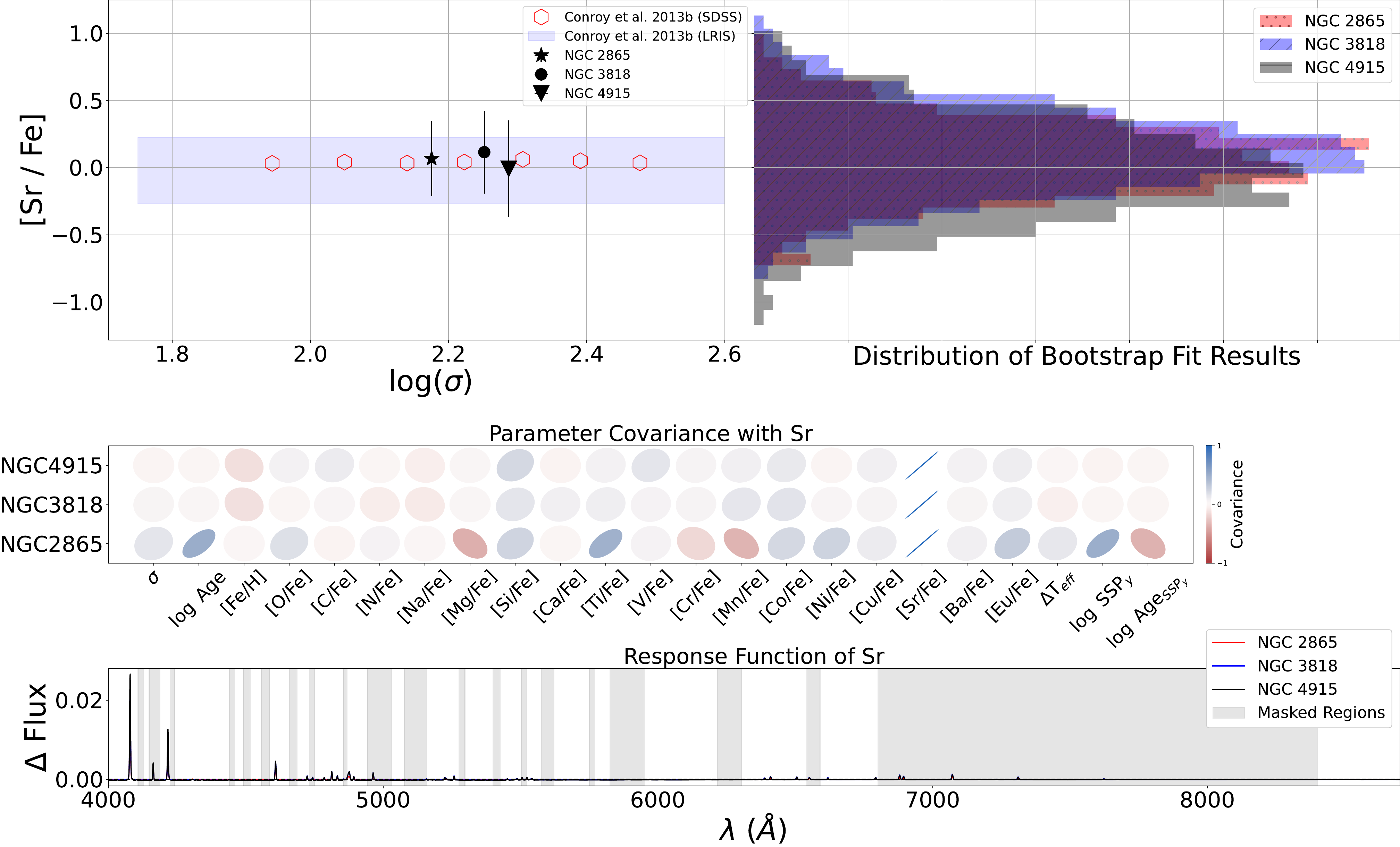}
\figsetgrpnote{Determinations of strontium abundance in each ETG. \textit{Top left panel:} Comparison of [Sr/Fe] abundance ratios to literature ETGs plotted against the velocity dispersion. \textit{Top right panel:} Distribution for the strontium abundance from 1000 fitting routines run over spectra sampled from our observed spectrum. The MCMC routine utilized 1024 walkers and a chain of 10,000 after the burn-in. \textit{Middle panel:} Parameter covariance of strontium compared to the other parameters in the fit. Blue, right-leaning ovals indicate a strong correlation, red, left-leaning ovals indicate anti-correlation, and round, pale ellipses indicate no correlation. \textit{Bottom panel:} Difference between two models synthesized with [Sr/Fe] ratios $\pm$ 0.5 dex above and below the median of the [Sr/Fe] posterior distribution. Areas of the spectrum that deviate  from 0 show where strontium has an effect on the integrated light.}
\figsetgrpend

\figsetgrpstart
\figsetgrpnum{8.14}
\figsetgrptitle{Barium}
\figsetplot{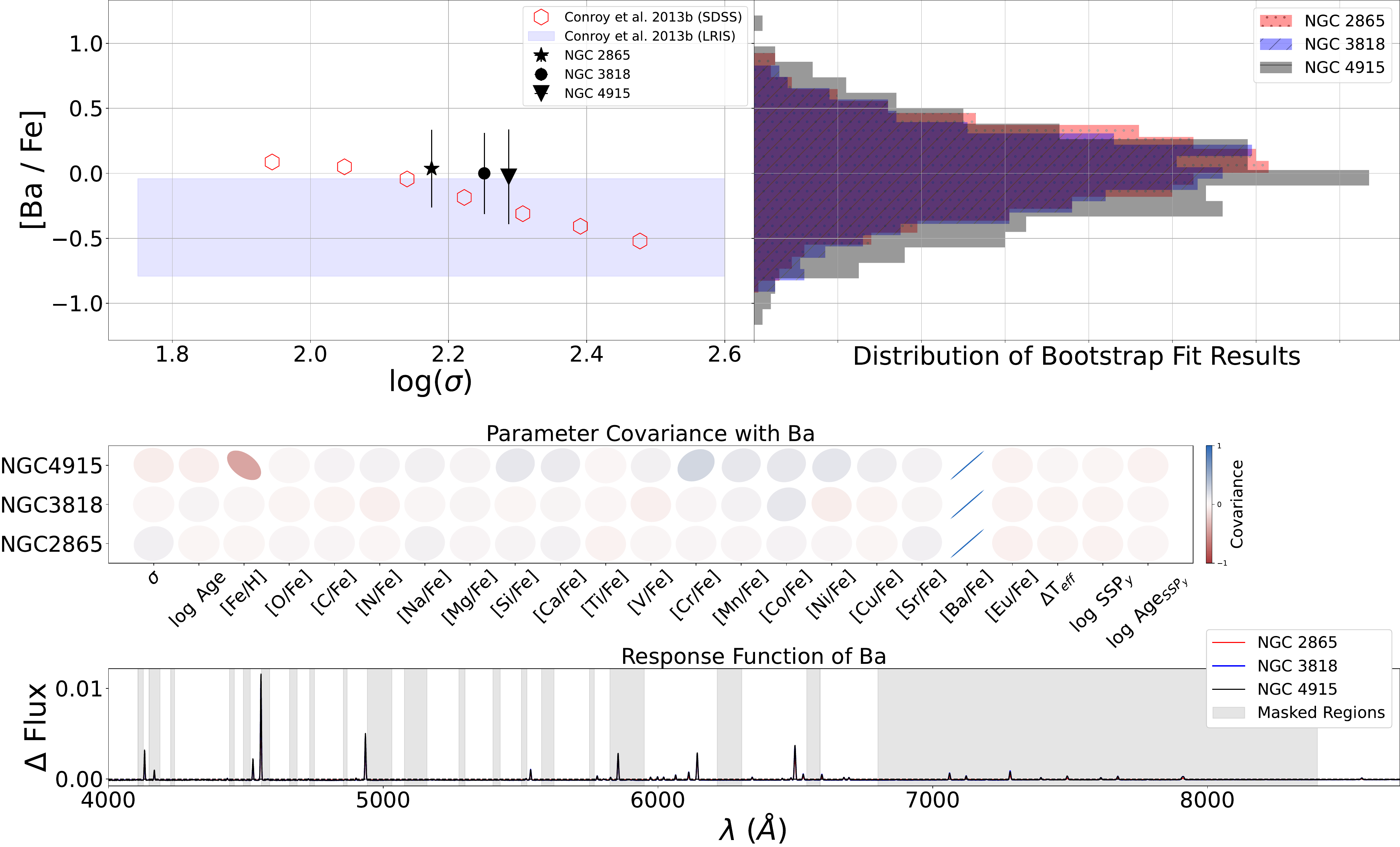}
\figsetgrpnote{Determinations of barium abundance in each ETG. \textit{Top left panel:} Comparison of [Ba/Fe] abundance ratios to literature ETGs plotted against the velocity dispersion. \textit{Top right panel:} Distribution for the barium abundance from 1000 fitting routines run over spectra sampled from our observed spectrum. The MCMC routine utilized 1024 walkers and a chain of 10,000 after the burn-in. \textit{Middle panel:} Parameter covariance of barium compared to the other parameters in the fit. Blue, right-leaning ovals indicate a strong correlation, red, left-leaning ovals indicate anti-correlation, and round, pale ellipses indicate no correlation. \textit{Bottom panel:} Difference between two models synthesized with [Ba/Fe] ratios $\pm$ 0.5 dex above and below the median of the [Ba/Fe] posterior distribution. Areas of the spectrum that deviate  from 0 show where barium has an effect on the integrated light.}
\figsetgrpend

\figsetgrpstart
\figsetgrpnum{8.15}
\figsetgrptitle{Europium}
\figsetplot{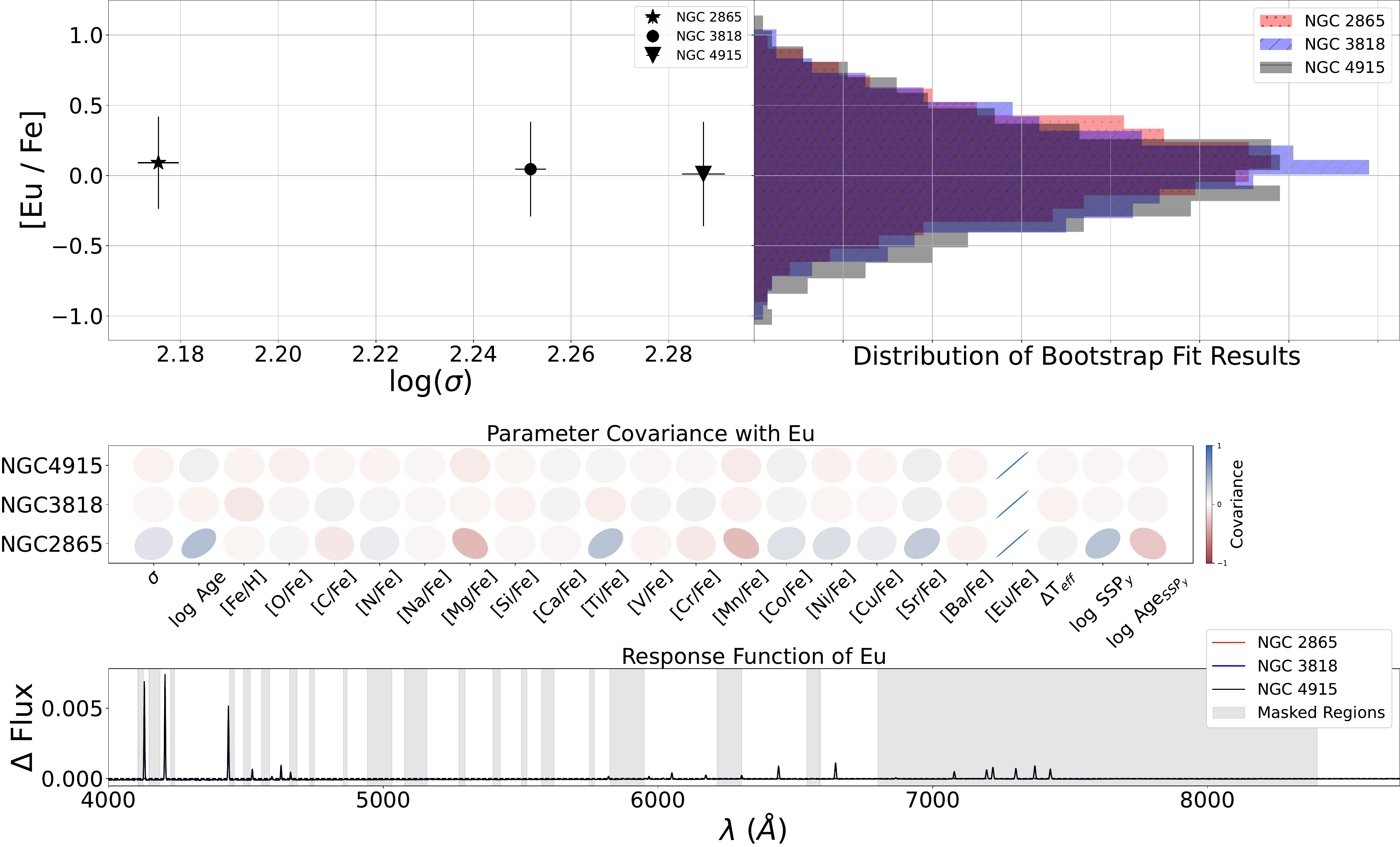} \label{fig:SigEu}
\figsetgrpnote{Determinations of europium abundance in each ETG. \textit{Top left panel:} Comparison of [Eu/Fe] abundance ratios to literature ETGs plotted against the velocity dispersion. \textit{Top right panel:} Distribution for the europium abundance from 1000 fitting routines run over spectra sampled from our observed spectrum. The MCMC routine utilized 1024 walkers and a chain of 10,000 after the burn-in. \textit{Middle panel:} Parameter covariance of europium compared to the other parameters in the fit. Blue, right-leaning ovals indicate a strong correlation, red, left-leaning ovals indicate anti-correlation, and round, pale ellipses indicate no correlation. \textit{Bottom panel:} Difference between two models synthesized with [Eu/Fe] ratios $\pm$ 0.5 dex above and below the median of the [Eu/Fe] posterior distribution. Areas of the spectrum that deviate  from 0 show where europium has an effect on the integrated light.}
\figsetgrpend

\figsetend

\begin{subfigures}
    \begin{figure}
        \centering
        \includegraphics[scale=0.25]{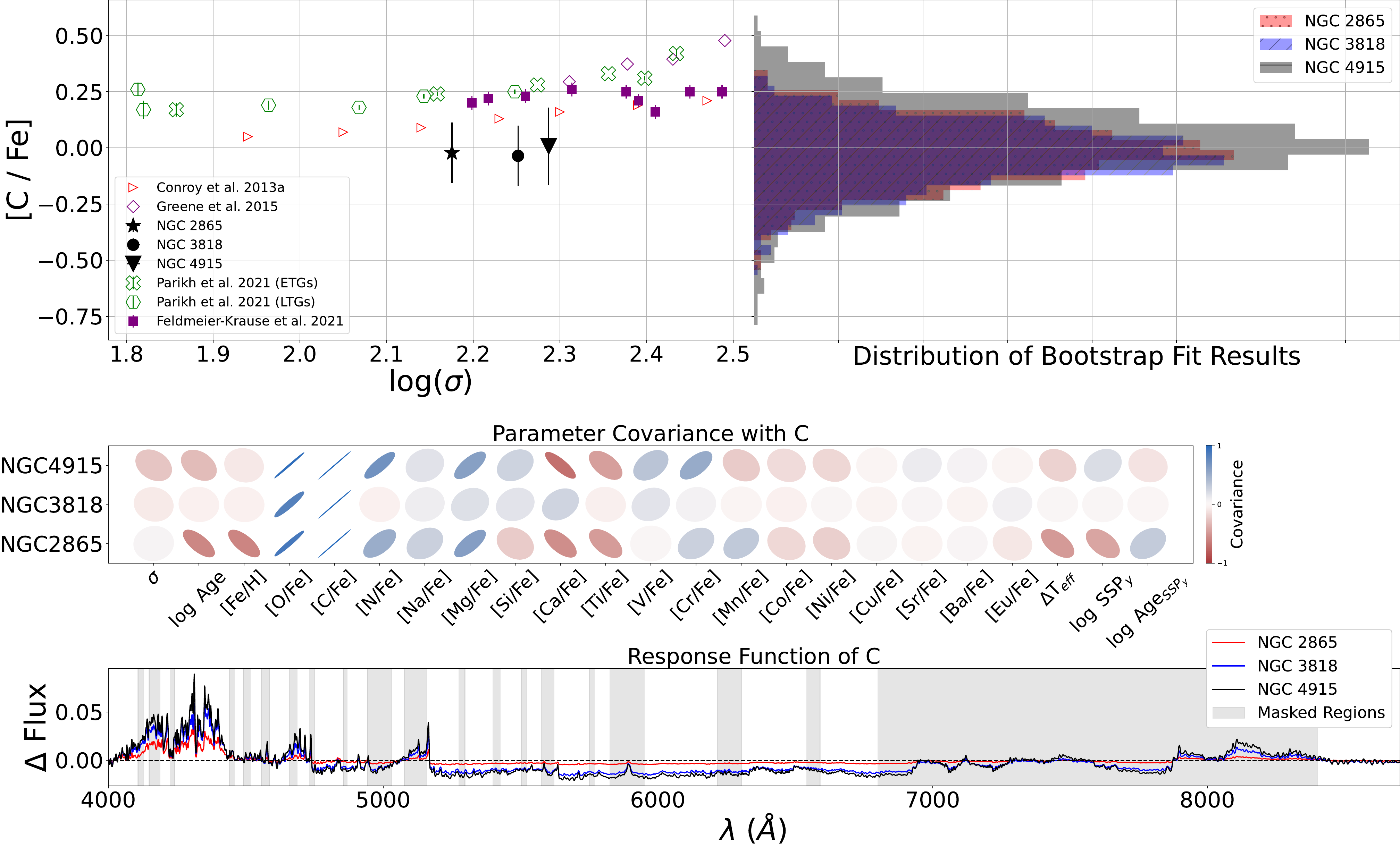}
        \caption{Determinations of carbon abundance in each ETG. \textit{Top left panel:} Comparison of [C/Fe] abundance ratios to literature ETGs plotted against the velocity dispersion. \textit{Top right panel:} Distribution for the carbon abundance from 1000 fitting routines run over spectra sampled from our observed spectrum. The MCMC routine utilized 1024 walkers and a chain of 10,000 after the burn-in. \textit{Middle panel:} Parameter covariance of carbon compared to the other parameters in the fit. Blue, right-leaning ovals indicate a strong correlation, red, left-leaning ovals indicate anti-correlation, and round, pale ellipses indicate no correlation. \textit{Bottom panel:} Difference between two models synthesized with [C/Fe] ratios $\pm$ 0.5 dex above and below the median of the [C/Fe] posterior distribution. Areas of the spectrum that deviate  from 0 show where carbon has an effect on the integrated light.}
    \end{figure}

    \begin{figure}
        \centering
        \includegraphics[scale=0.25]{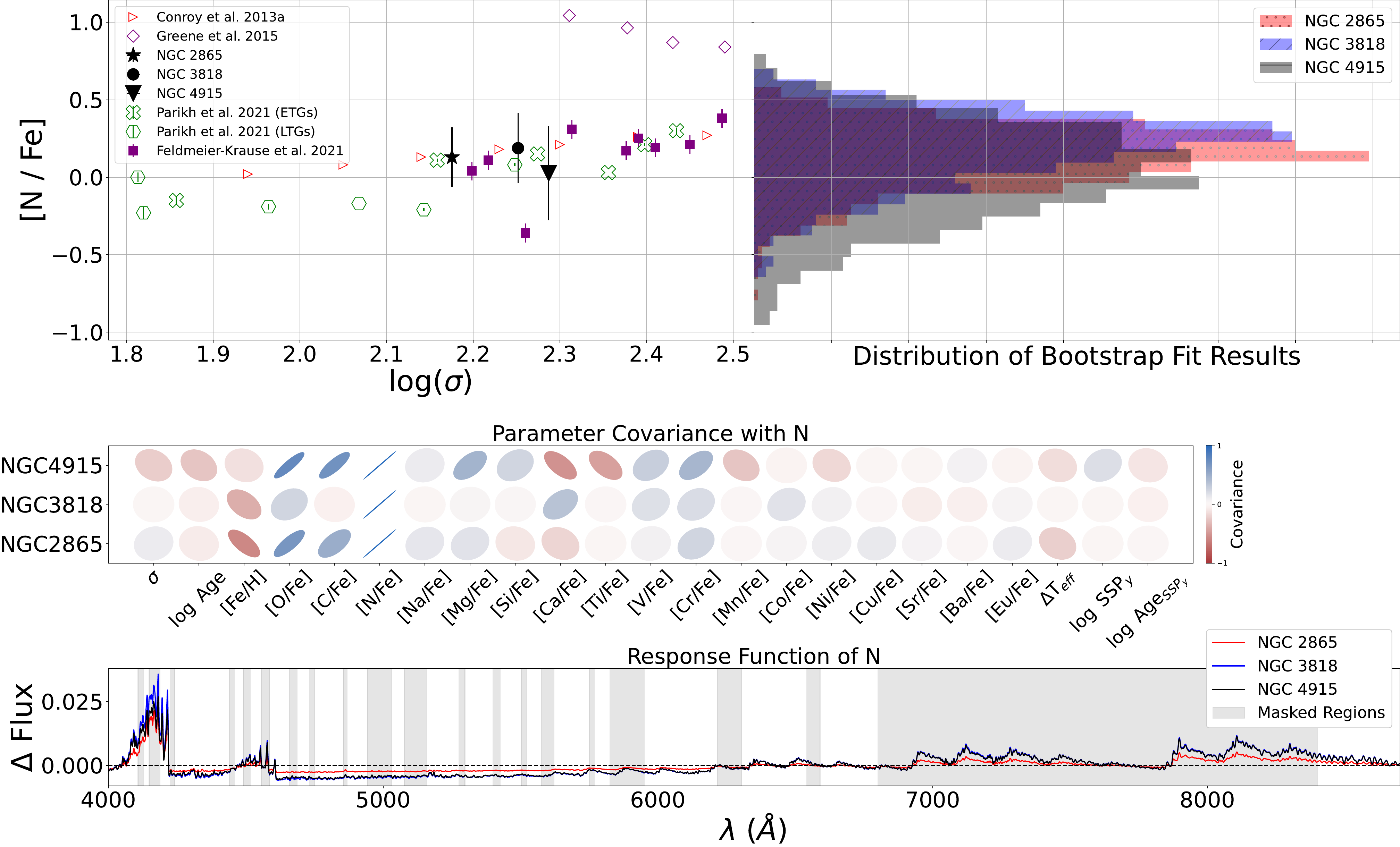}
        \caption{Same as Figure 8a, but for nitrogen.}
    \end{figure}

    \begin{figure}
        \centering
        \includegraphics[scale=0.25]{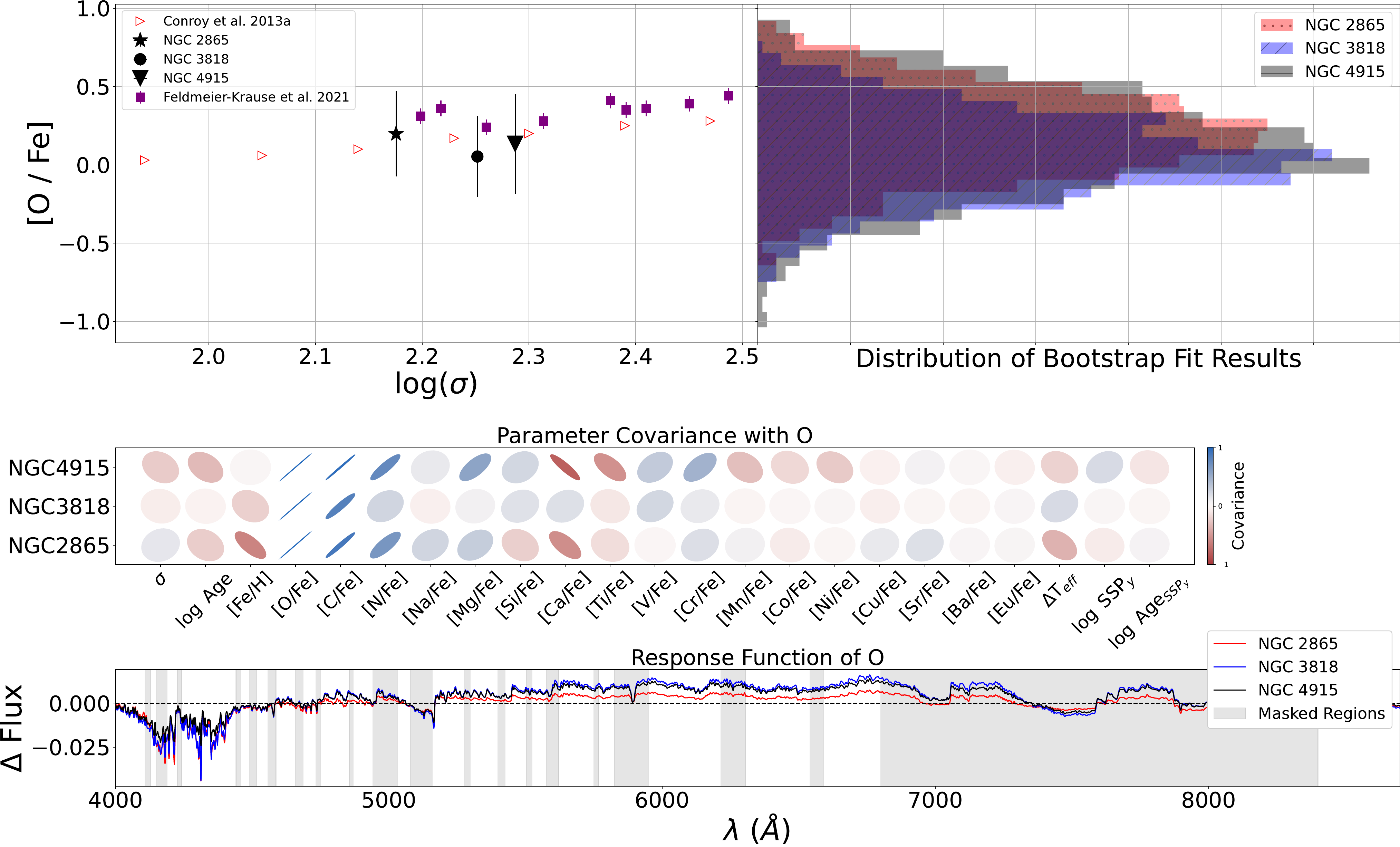}
        \caption{Same as Figure 8a, but for oxygen.}
    \end{figure}

    \begin{figure}
        \centering
        \includegraphics[scale=0.25]{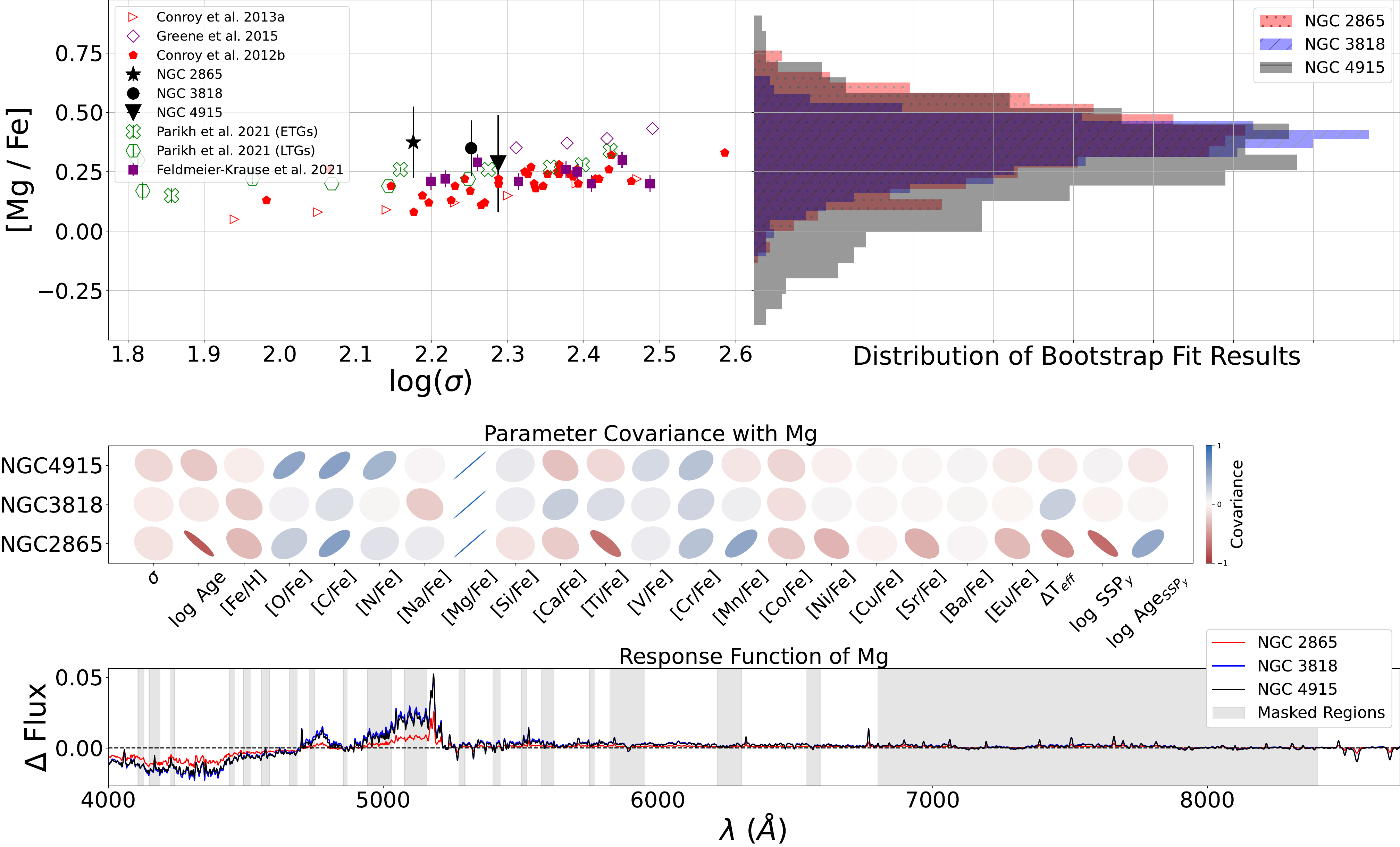}
        \caption{Same as Figure 8a, but for magnesium.}
    \end{figure}

    \begin{figure}
        \centering
        \includegraphics[scale=0.25]{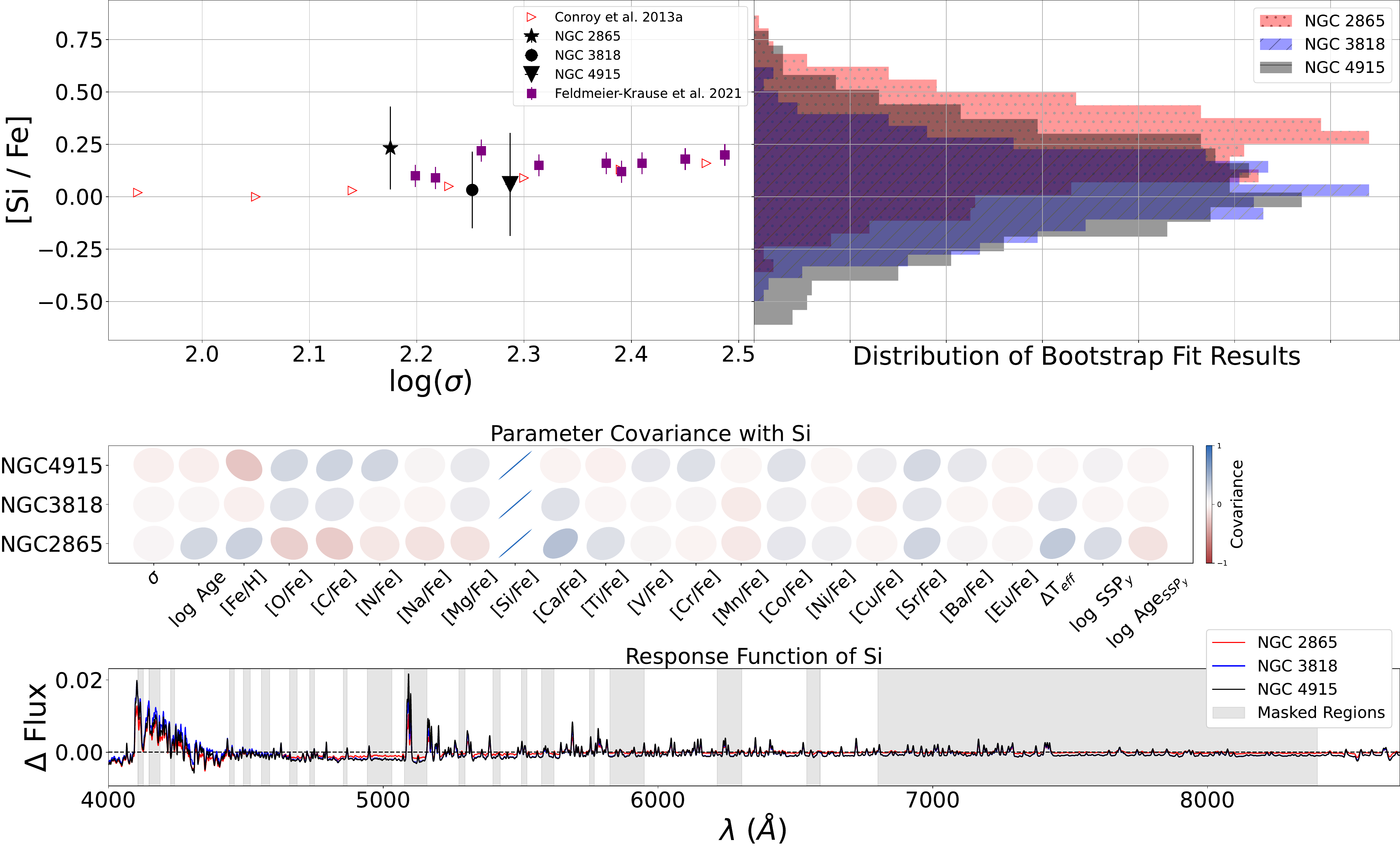}
        \caption{Same as Figure 8a, but for silicon.}
    \end{figure}

    \begin{figure}
        \centering
        \includegraphics[scale=0.25]{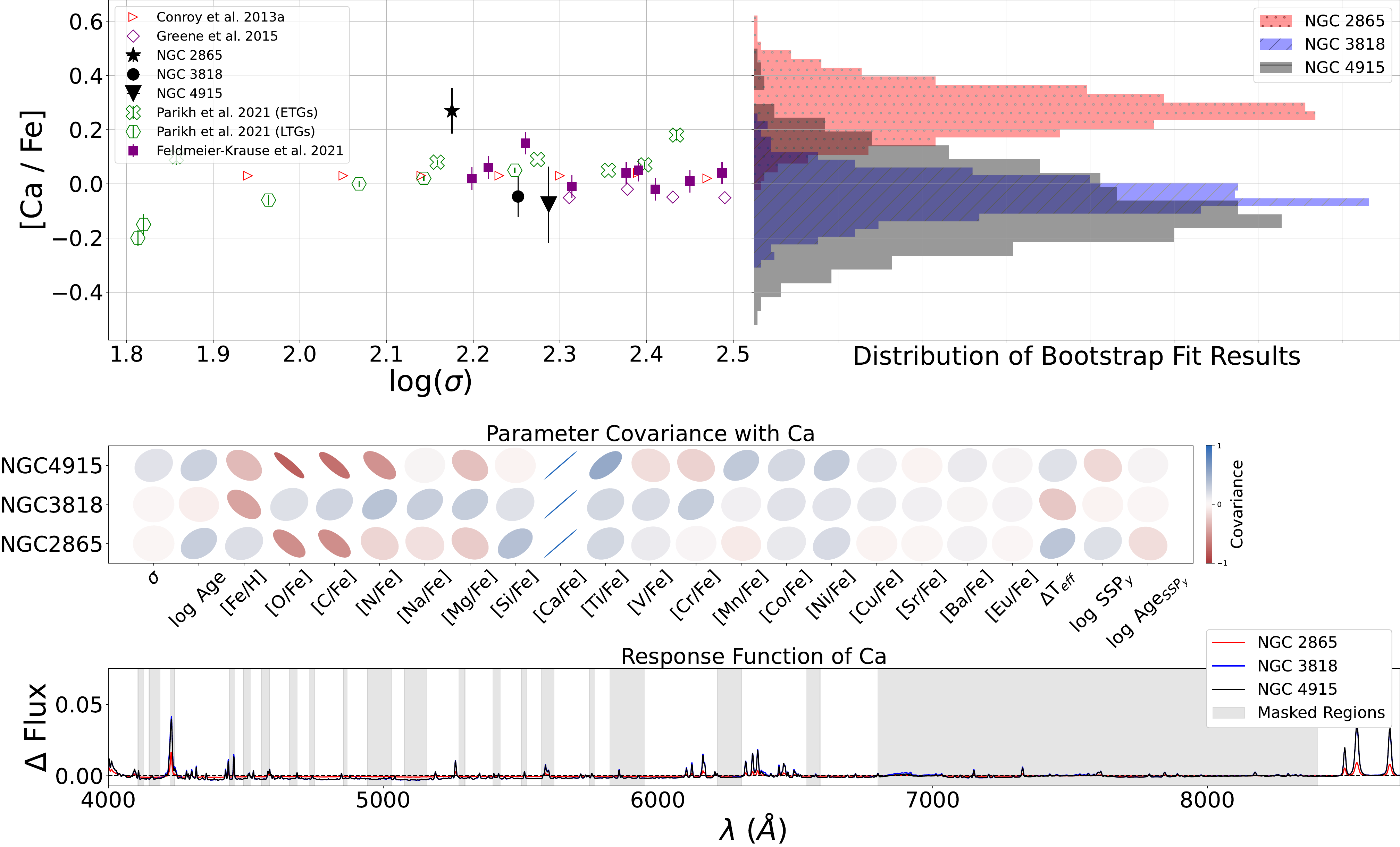}
        \caption{Same as Figure 8a, but for calcium.}
    \end{figure}

    \begin{figure}
        \centering
        \includegraphics[scale=0.25]{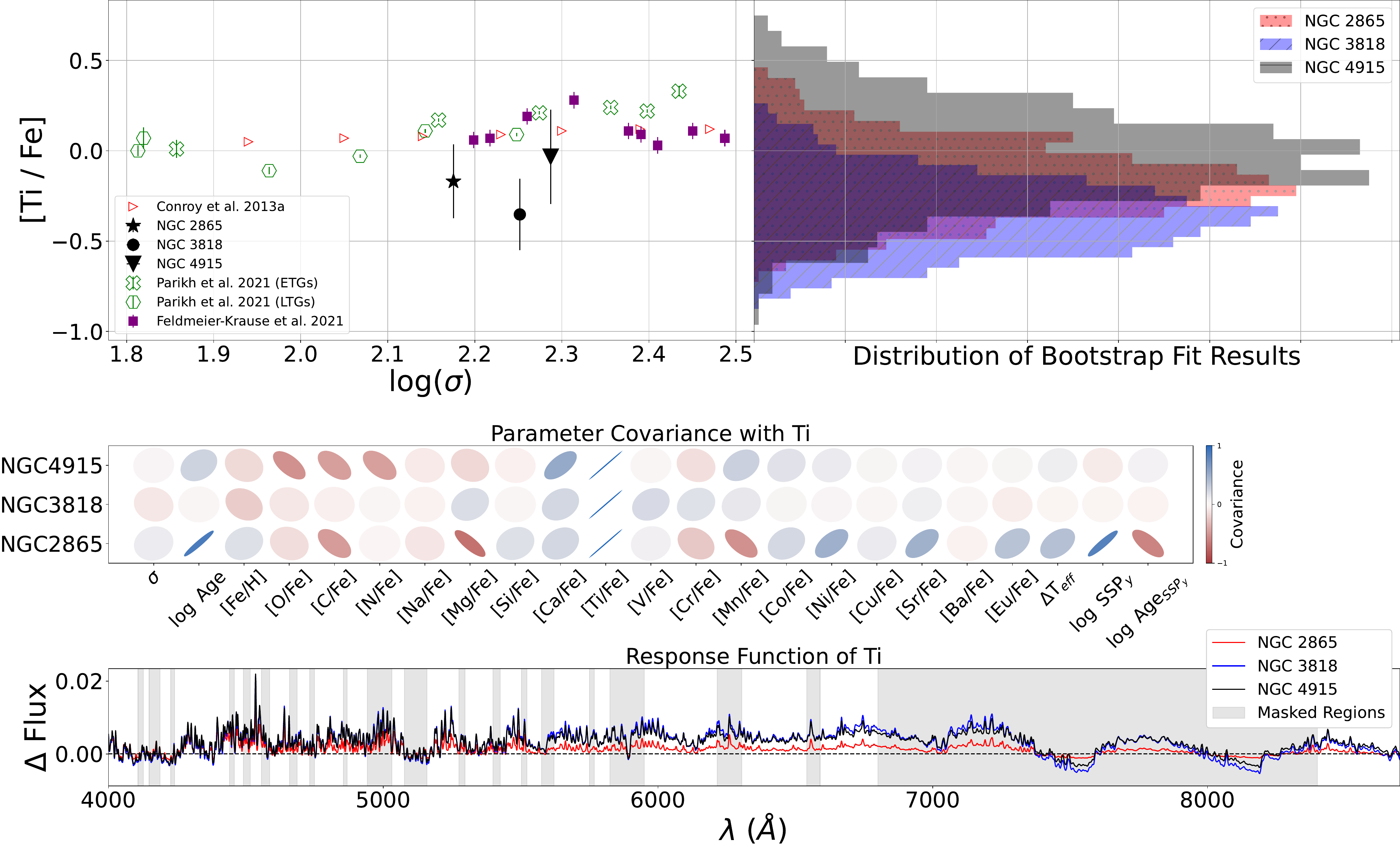}
        \caption{Same as Figure 8a, but for titanium.}
    \end{figure}

    \begin{figure}
        \centering
        \includegraphics[scale=0.25]{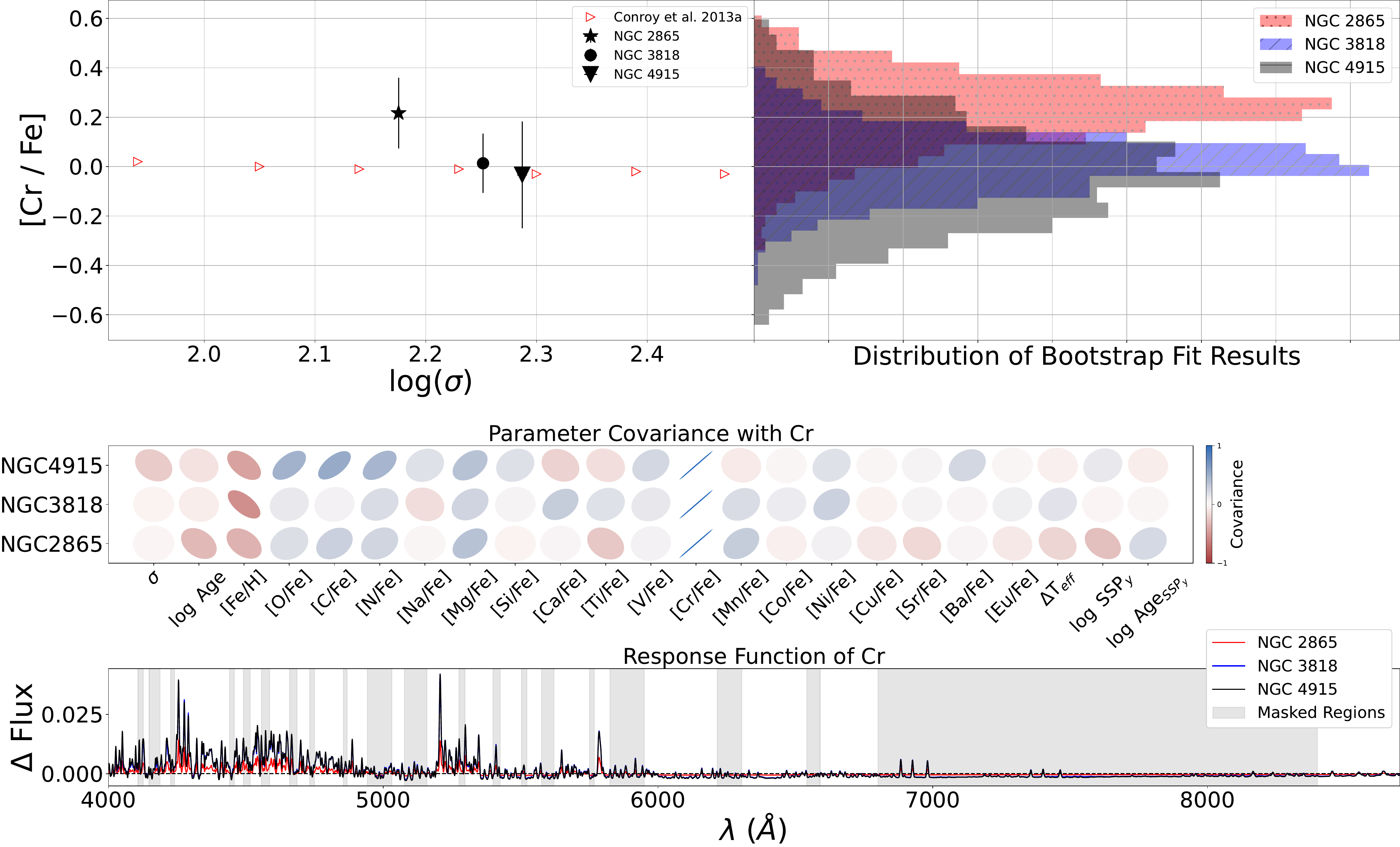}
        \caption{Same as Figure 8a, but for chromium.}
    \end{figure}

    \begin{figure}
        \centering
        \includegraphics[scale=0.25]{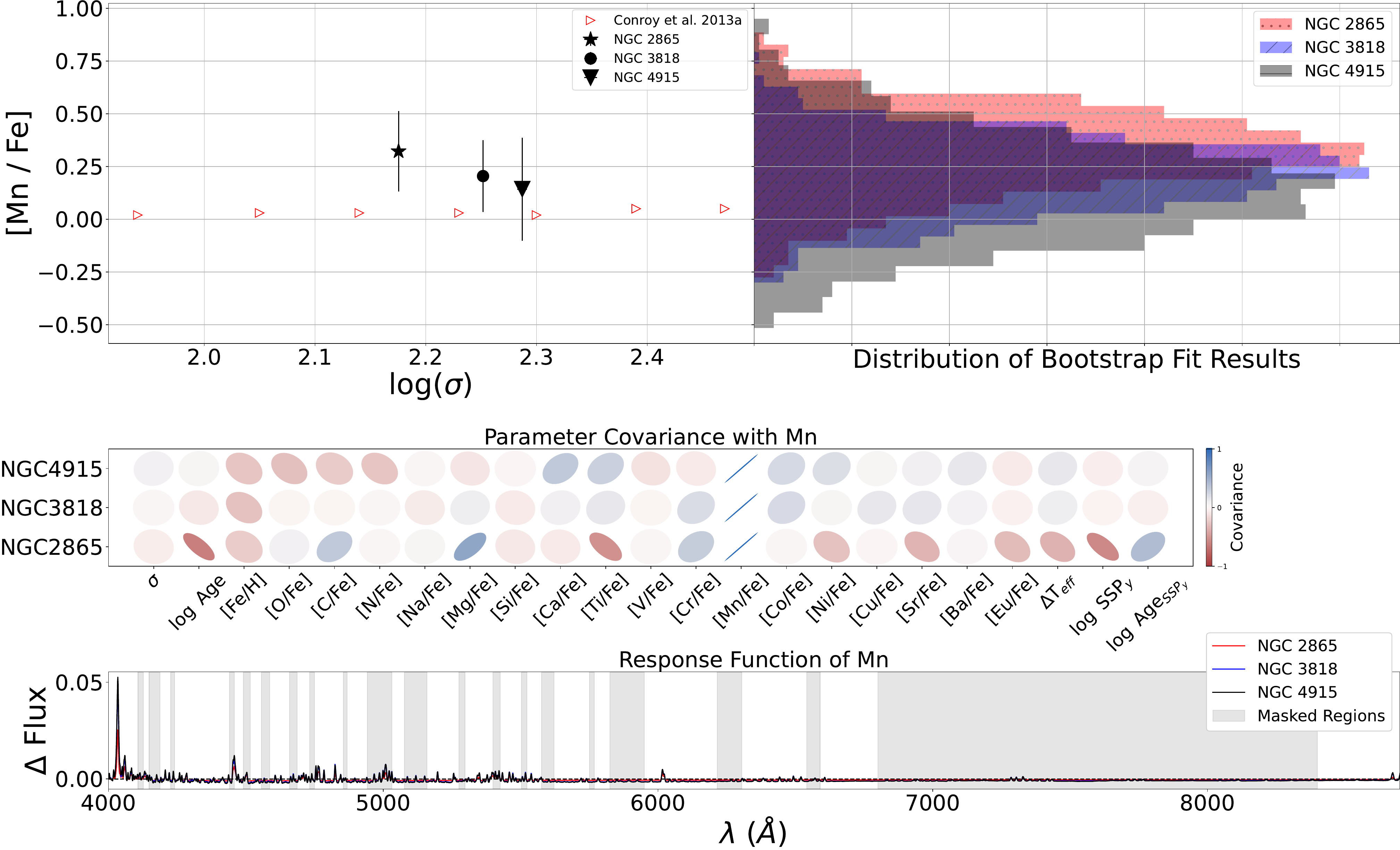}
        \caption{Same as Figure 8a, but for manganese.}
    \end{figure}

    \begin{figure}
        \centering
        \includegraphics[scale=0.25]{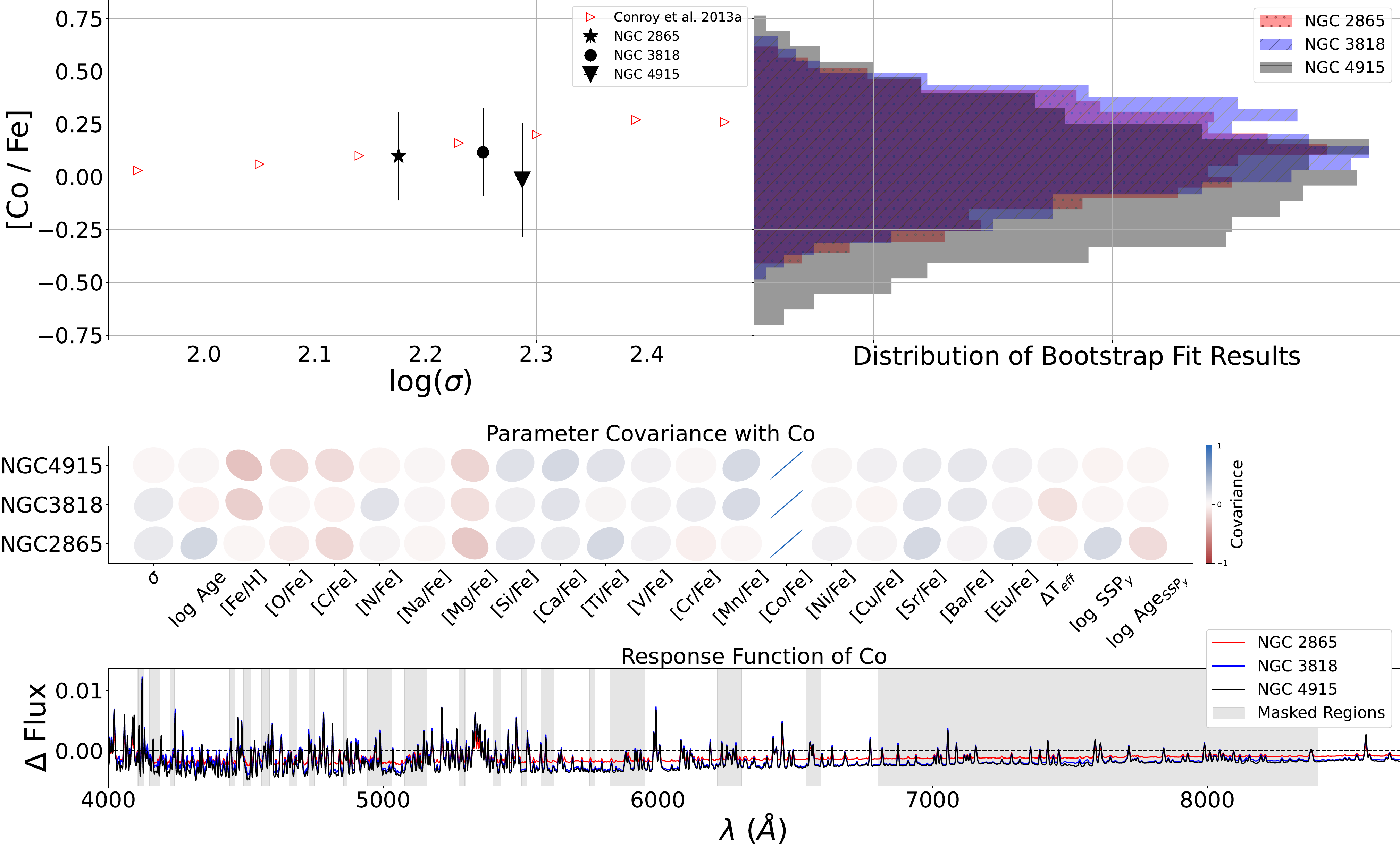}
        \caption{Same as Figure 8a, but for cobalt.}
    \end{figure}

    \begin{figure}
        \centering
        \includegraphics[scale=0.25]{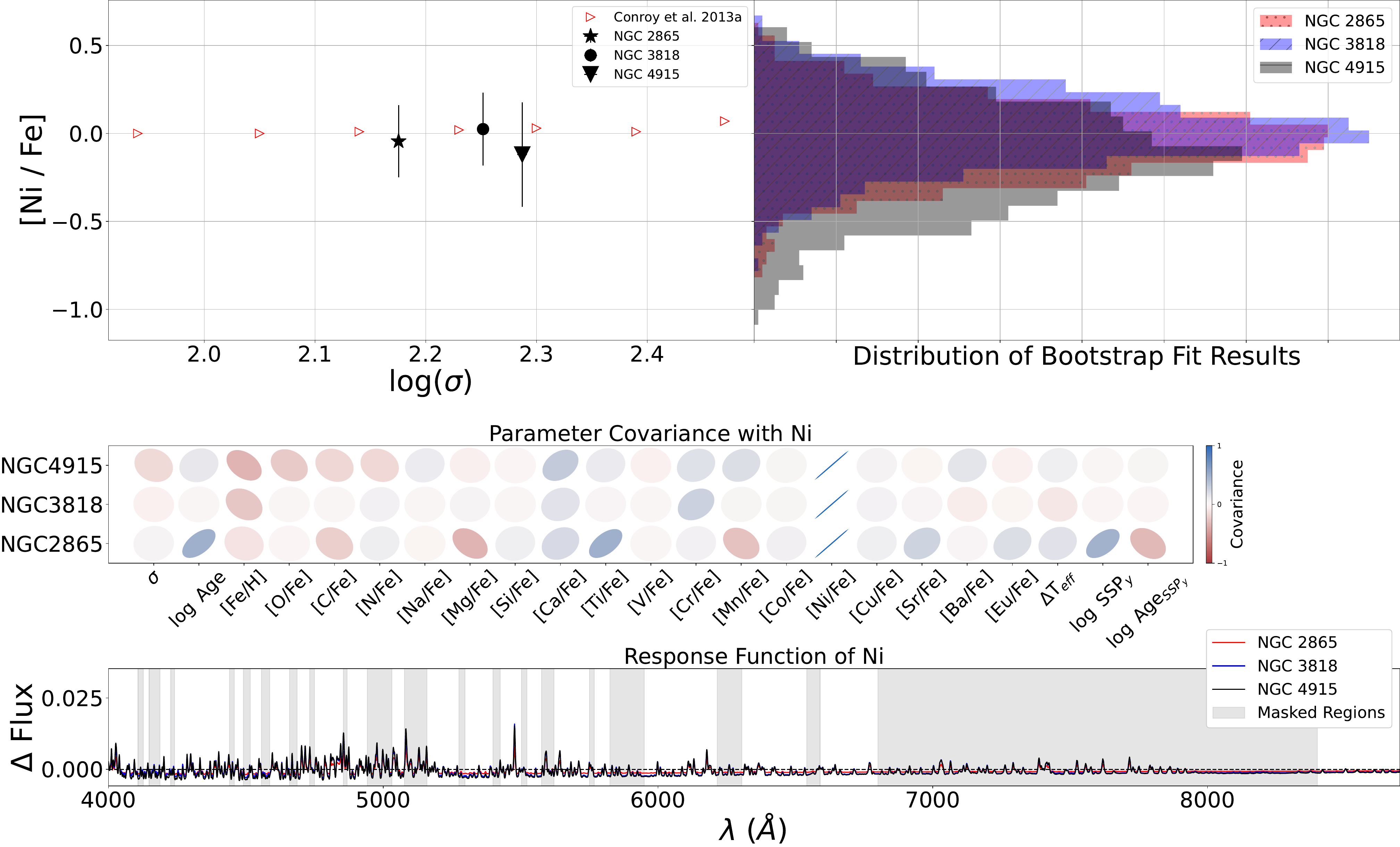}
        \caption{Same as Figure 8a, but for nickel.}
    \end{figure}

    \begin{figure}
        \centering
        \includegraphics[scale=0.25]{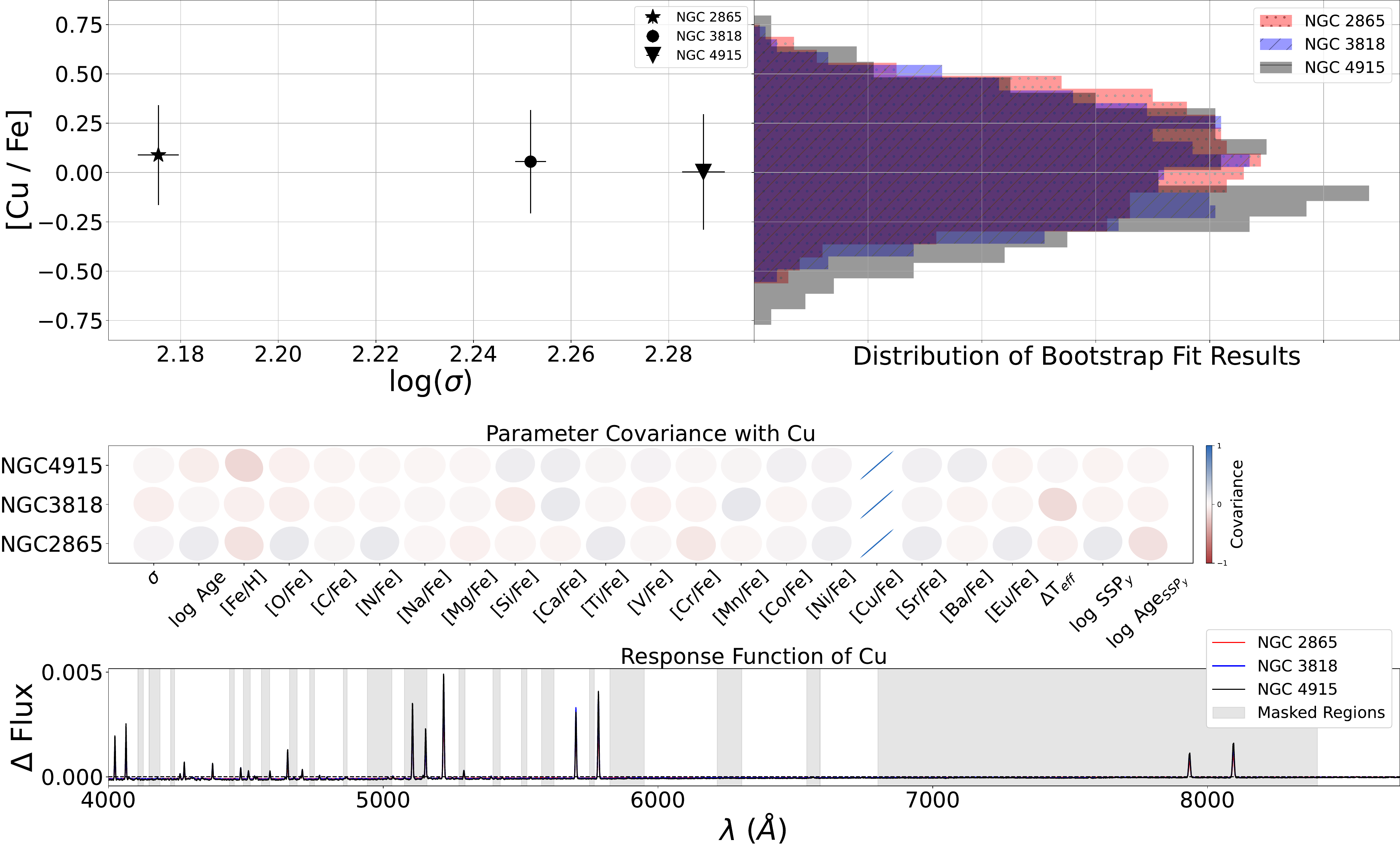}
        \caption{Same as Figure 8a, but for copper.}
    \end{figure}

    \begin{figure}
        \centering
        \includegraphics[scale=0.25]{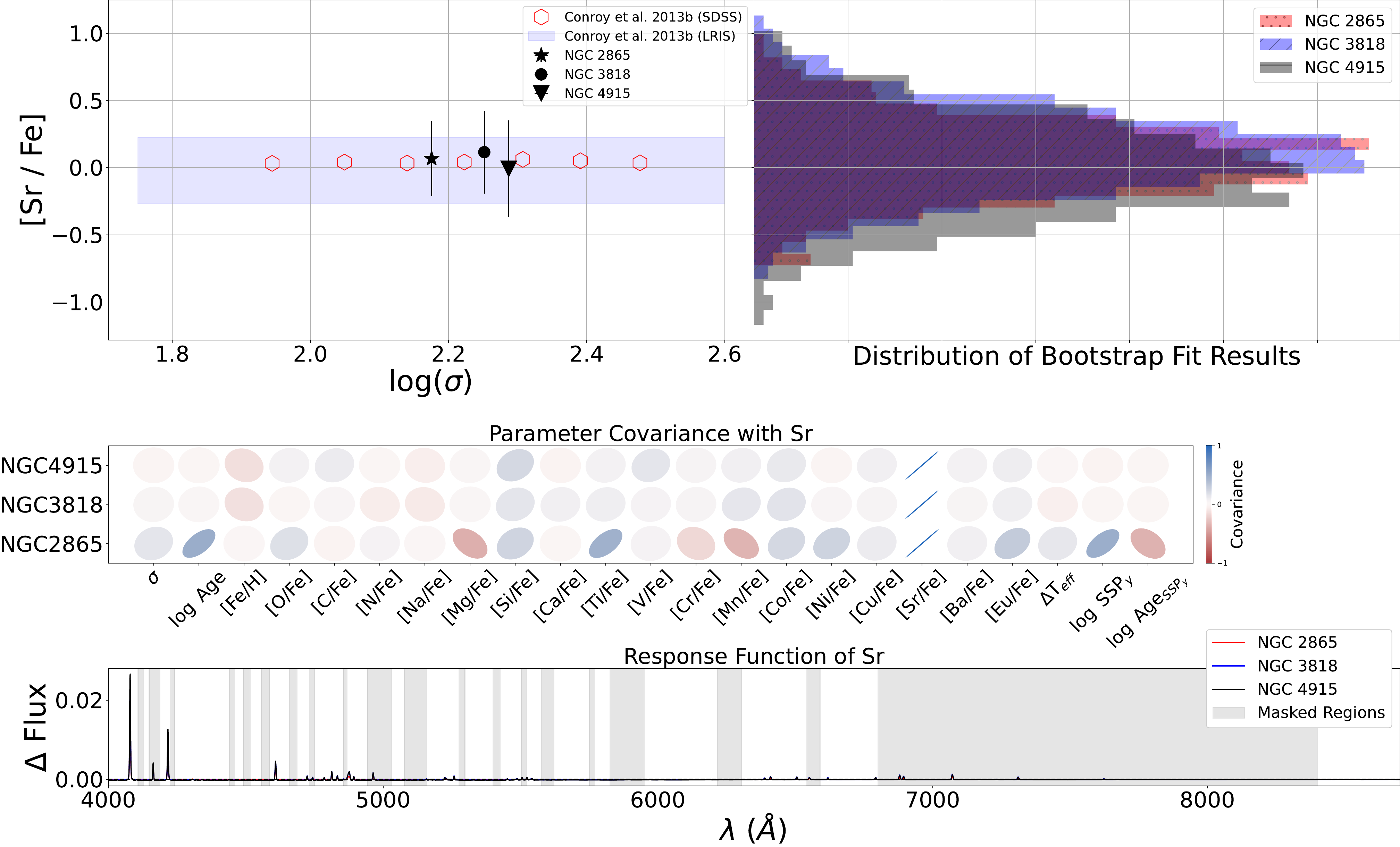}
        \caption{Same as Figure 8a, but for strontium.}
    \end{figure}

    \begin{figure}
        \centering
        \includegraphics[scale=0.25]{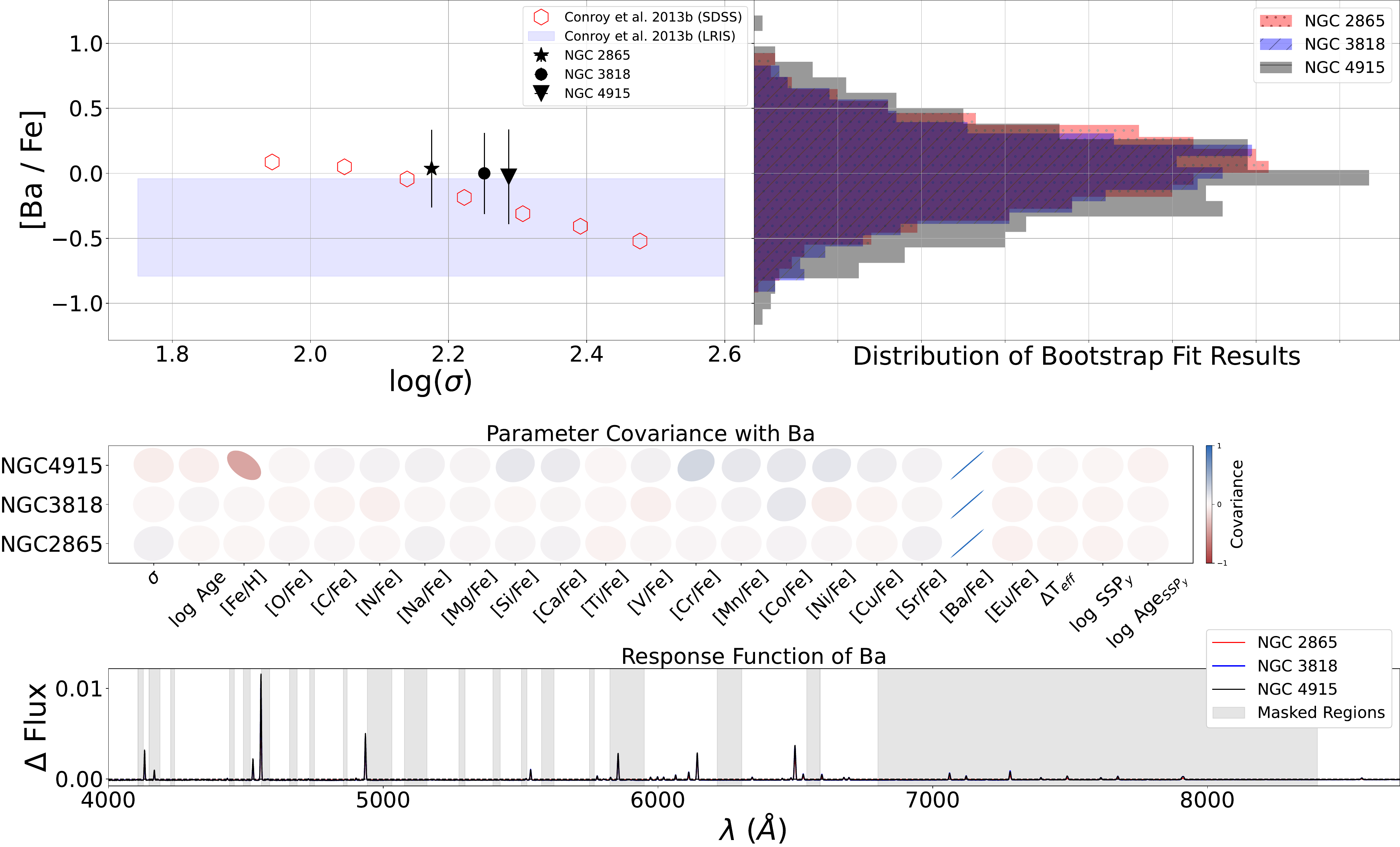}
        \caption{Same as Figure 8a, but for barium.}
    \end{figure}

    \begin{figure}
        \centering
        \includegraphics[scale=0.25]{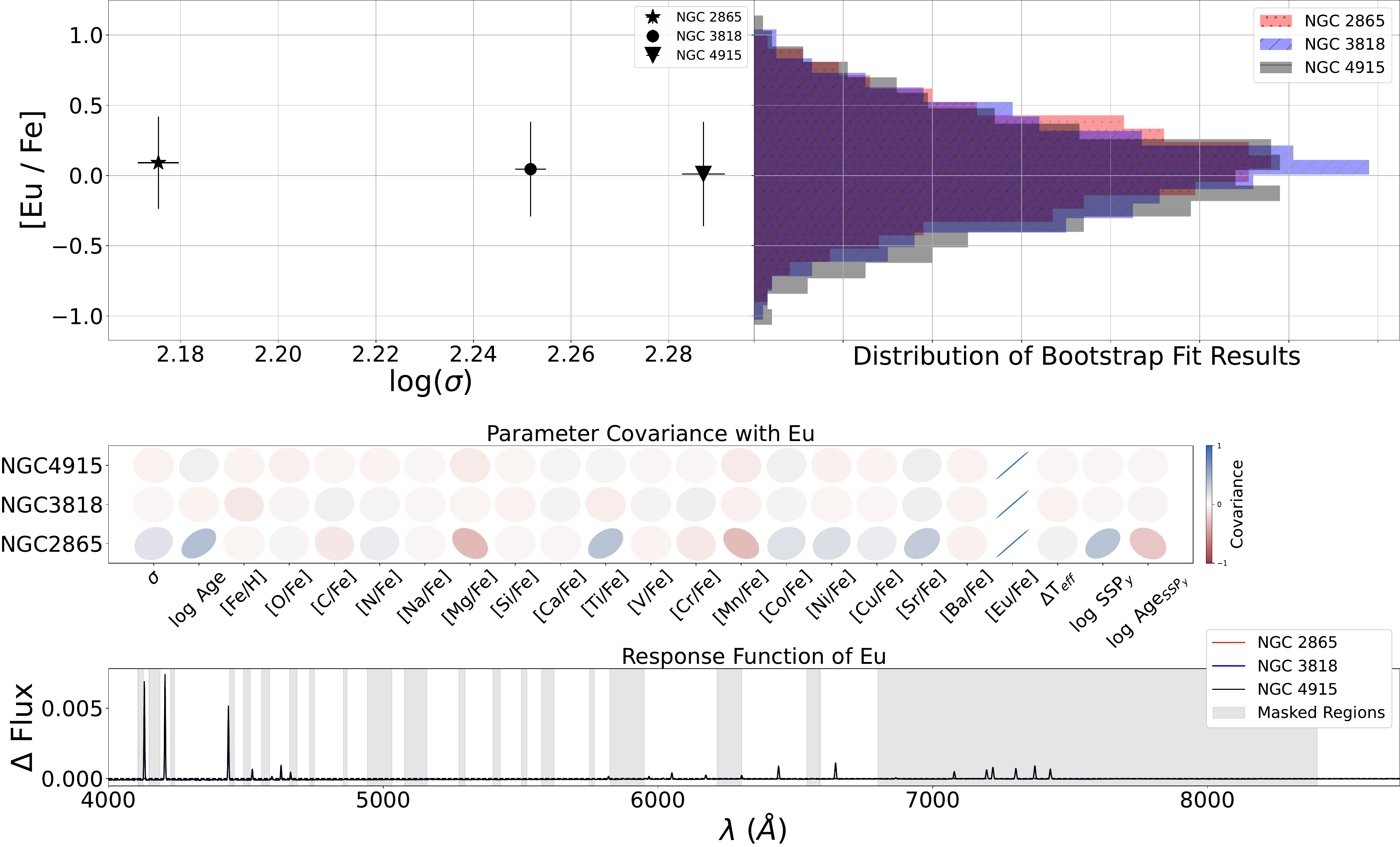}
        \caption{Same as Figure 8a, but for europium.}
    \end{figure}
\end{subfigures}





\end{document}